\documentclass[12pt] {amsart}
\usepackage{amsfonts,amssymb,amsmath,epsfig,graphics,graphicx,rotating}
\usepackage{a4wide}
\usepackage[english]{babel}
\newcommand{\halmos}{\rule{1ex}{1.4ex}}
\makeatletter \@addtoreset{equation}{section} \makeatother

\newtheorem{ittheorem}{Theorem}
\newtheorem{itlemma}{Lemma}
\newtheorem{itproposition}{Proposition}
\newtheorem{itcorollary}{Corollary}
\newtheorem{itdefinition}{Definition}
\newtheorem{itremark}{Remark}
\newtheorem{itexamples}{Examples}

\newenvironment{theorem}{\addtocounter{equation}{1}
\begin{ittheorem}}{\end{ittheorem}}
\newenvironment{lemma}{\addtocounter{equation}{1}
\begin{itlemma}}{\end{itlemma}}
\newenvironment{proposition}{\addtocounter{equation}{1}
\begin{itproposition}}{\end{itproposition}}
\newenvironment{corollary}{\addtocounter{equation}{1}
\begin{itcorollary}}{\end{itcorollary}}
\newenvironment{definition}{\addtocounter{equation}{1}
\begin{itdefinition}}{\end{itdefinition}}
\newenvironment{remark}{\addtocounter{equation}{1}
\begin{itremark}}{\end{itremark}}

\newenvironment{examples}{\addtocounter{equation}{1}
\begin{itexamples}}{\end{itexamples}}

\newenvironment{proofs}{\noindent {\em Proof}.\,\,\,}
{\hspace*{\fill}$\halmos$\smallskip}

\newenvironment{proofsm}{\noindent}
{\hspace*{\fill}$\halmos$\smallskip}

\newcommand{\beq}{\begin{eqnarray}}
\newcommand{\eeq}{\end{eqnarray}}

\newcommand{\beqq}{\begin{eqnarray*}}
\newcommand{\eeqq}{\end{eqnarray*}}

\newcommand{\be}{\begin{equation}}
\newcommand{\ee}{\end{equation}}

\newcommand{\bl}{\begin{lemma}}
\newcommand{\el}{\end{lemma}}

\newcommand{\br}{\begin{remark}}
\newcommand{\er}{\end{remark}}

\newcommand{\bex}{\begin{examples}}
\newcommand{\eex}{\end{examples}}

\newcommand{\bt}{\begin{theorem}}
\newcommand{\et}{\end{theorem}}

\newcommand{\bd}{\begin{definition}}
\newcommand{\ed}{\end{definition}}

\newcommand{\bp}{\begin{proposition}}
\newcommand{\ep}{\end{proposition}}

\newcommand{\bc}{\begin{corollary}}
\newcommand{\ec}{\end{corollary}}

\newcommand{\bpr}{\begin{proofs}}
\newcommand{\epr}{\end{proofs}}

\newcommand{\bprm}{\begin{proofsm}}
\newcommand{\eprm}{\end{proofsm}}

\newcommand{\bi}{\begin{itemize}}
\newcommand{\ei}{\end{itemize}}

\newcommand{\ben}{\begin{enumerate}}
\newcommand{\een}{\end{enumerate}}

\newcommand{\abs}[1]{\left|#1\right|}
\newcommand{\Z}{\mathbb Z}
\newcommand{\R}{\mathbb R}
\newcommand{\N}{\mathbb N}

\newcommand{\Llra}{\Longleftrightarrow}
\newcommand{\Lra}{\Longrightarrow}

\newcommand{\s}{\ensuremath{\mathcal{S}}}
\newcommand{\calL}{\ensuremath{\mathcal{L}}}
\newcommand{\calP}{\ensuremath{\mathcal{P}}}

\newcommand{\ii}{\ensuremath{\mathcal{I}}}
\newcommand{\al}{\ensuremath{\alpha}}
\newcommand{\ga}{\ensuremath{\gamma}}
\newcommand{\Ga}{\ensuremath{\Gamma}}
\newcommand{\de}{\ensuremath{\delta}}

\newcommand{\Om}{\ensuremath{\Omega}}
\newcommand{\si}{\ensuremath{\sigma}}
\newcommand{\Si}{\ensuremath{\Sigma}}
\newcommand{\QED}{\hspace*{\fill}$\halmos$\smallskip}
\newcommand{\ve}{\ensuremath{\varepsilon}}

\newcommand{\dt}{\partial_t}
\newcommand{\dx}{\partial_x}

\newcommand{\Prob}{{\rm I\hspace{-0.8mm}P}}
\begin{document}
\title[Couplings and attractiveness]{Couplings, Attractiveness and Hydrodynamics for 
Conservative Particle Systems}
\author{Thierry Gobron}
\address{ CNRS, UMR 8089, LPTM\\
Universit\'e de Cergy-Pontoise; Site de Saint-Martin 2,
2 avenue Adolphe Chauvin, Pontoise
\\95031 Cergy-Pontoise cedex, France}
\author{Ellen Saada}
\address{CNRS, UMR 6085, LMRS\\
Universit\'e de Rouen;
Avenue de l'Universit\'e, BP.12,
Site du Madrillet,\\
76801 Saint Etienne du Rouvray cedex, France}
\email{Thierry.Gobron@u-cergy.fr}
\email{Ellen.Saada@univ-rouen.fr}
\thanks{}
\begin{abstract}
Attractiveness is a fundamental tool to study interacting 
particle systems and the basic coupling 
construction is a usual route to prove this property, as for instance in simple exclusion.
The derived  Markovian coupled 
process $(\xi_t,\zeta_t)_{t\geq 0}$
satisfies:
 
 (A)  if $\xi_0\leq\zeta_0$ 
 (coordinate-wise), then for all $t\geq 0$, $\xi_t\leq\zeta_t$ a.s.

In this paper, we consider generalized misanthrope models which are conservative particle
systems on $\Z^d$ such that, in each transition, $k$ particles 
 may jump from a site $x$ to another site $y$, with  $k\geq 1$. 
These  models include 
simple exclusion for which $k=1$,  but, beyond that value, 
the basic coupling construction is not possible and a more refined 
one is required.
We give necessary and sufficient conditions on the rates to insure attractiveness; we construct a Markovian 
coupled process which both satisfies (A) and makes
discrepancies between its two 
marginals non-increasing. 
We determine the extremal  
invariant and translation invariant probability measures
under general irreducibility conditions.
We  apply our results to examples including
 a two-species asymmetric exclusion process with charge conservation
(for which $k\le 2$) which arises from a  Solid-on-Solid interface
 dynamics, and a stick process (for which $k$ is unbounded) in correspondence with a generalized discrete Hammersley-Aldous-Diaconis model. We derive the hydrodynamic limit
of these two one-dimensional models.
\end{abstract}
\maketitle
\setcounter{equation}{0}
\section {Introduction}
\label{sec:1}
Attractiveness is a fundamental  
property to study interacting particle systems:  combined with coupling, it is a key tool to
determine the set of invariant (and translation invariant) 
probability measures for the dynamics;  it is also essential in obtaining
hydrodynamic limits for conservative systems under Euler scaling 
(see \cite{L}, \cite{KL}).  Most widely known examples are simple exclusion (SEP), 
zero range (ZRP) or misanthropes (MP) processes (see \cite{A}, \cite{C}). Actually, these examples share two features: first, their infinitesimal transitions 
consist in the jump of a single particle (from a site $x$ to another site $y$); they are  also of {\sl misanthrope type}, in the sense that  their transition
rates depend at most on $x$, $y$, and on occupation numbers at $x$ and $y$. These limitations on the possible variations of the conserved quantity are at the root of the construction of basic coupling, through which attractiveness is shown to hold. 

In this paper, we consider conservative particle systems of misanthrope type but allowing more than one particle to jump at a time 
from a given site to another one. This seemingly slight generalization requires to restart almost from scratch to check attractiveness through a coupling construction. 
We obtain necessary and sufficient conditions on the rates for attractiveness,  giving an explicit construction of an increasing coupling (that is, which shows that the process is attractive).  Using these results, we determine the extremal invariant and translation invariant probability measures  of the process.
Finally, we illustrate our approach on a few examples for which we derive hydrodynamics.

The conservative particle systems we study
are Markov processes $(\eta_t)_{t\geq 0}$ defined on a state space $\Om = X^S$, with $X$ a 
subset of $\Z$, and $S=\Z^d$;
$\eta_t$ is the configuration at time $t$ and
$\eta_t(z)$ its value at site $z\in S$. Depending on $X$, one of the two following interpretations can be used:
for $X\subset\N$, $\eta_t(z)$ is the number of particles  at site $z$;
if $X\subset\Z$, $|\eta_t(z)|$ is the number of unit charges at 
site $z$ (either positive for $\eta_t(z)>0$ or  negative for $\eta_t(z)<0$).
The evolution of the process can be informally described as follows. 
Given a pair $(x,y) \in S^2$ and  an integer $k> 0$, 
with a rate $\Ga_{\eta(x),\eta(y)}^k(y-x)$, $k$ particles (respectively $k$ positive charges) `attempt a jump' 
from site $x$ to site $y$. By this, we mean that the
values at sites $x$ and $y$ are  
changed to $\eta(x)-k$ and $\eta(y)+k$ as long as the resulting 
configuration belongs to the state space. The conserved quantity 
is $\eta(x)+\eta(y)$, the total number of 
particles (resp. the total charge) involved in the transition.

The SEP-ZRP-MP are examples of such processes, for which 
 $X\subset\N$,  $k=1$ is the only allowed move
and the rates take the form $\Ga_{\al,\beta}^1(y-x)=p(y-x)b(\al,\beta)$ for
$\al,\beta\in X,\,x,y\in S$, 
$p(\cdot)$ a translation invariant 
probability kernel on $S$, and $b(\cdot,\cdot)$ a function of the configuration values
at the two involved sites. These processes (see \cite{L}, 
\cite{A}, \cite{C} for details) are 
{\em attractive}, that is, the coordinate-wise (partial) order on 
configurations is preserved (stochastically) by the dynamics (see Section \ref{sec:2} for precise 
definitions), if and only if

${\bf(A)}$ The rate $b(\al,\beta)$ is a function of
$\al$, $\beta$ non-decreasing  
in $\al$ and non-increasing in $\beta$.

Attractiveness for interacting particle systems is derived 
by constructing an {\em increasing Markovian 
coupling}. A {\em coupled} Markov process $(\xi_t,\zeta_t)_{t\geq 0}$ 
is a process on $\Om\times\Om$ such that each marginal is a 
copy of the original process $(\eta_t)_{t\geq 0}$. The coupling is {\em increasing} if 
partial order between marginal configurations is preserved
under the coupled dynamics.

The simplest candidate for this purpose is 
the  {\em basic} (or Vasershtein) {\em coupling}. It 
is designed to make the two coupled processes 
$(\xi_t)_{t\geq 0},\,(\zeta_t)_{t\geq 0}$ 
`as similar as possible', and its construction can be obtained by relying on three rules: 
\hfill\break\noindent 
(i) coupled transitions have the same departure and arrival sites for the two
marginals,\hfill\break\noindent
(ii) an identical change in both processes is given the largest possible rate,\hfill\break\noindent
(iii) non coupled transitions are supplemented to get the correct left and right marginals.

In the context of SEP-ZRP-MP, these rules give: by  (i), the same probability $p(y-x)$ is  taken 
for coupled 
jumps from site $x$ to site $y$. Then, if the  values at sites $x$ and $y$ are
$\xi_{t^-}(x)=\al$, $\xi_{t^-}(y)=\beta$, $\zeta_{t^-}(x)=\ga$ and 
$\zeta_{t^-}(y)=\de$, three 
changes may occur  (remember that at most one particle can jump in a transition): \hfill\break\noindent
* with rate $\left(b(\al,\beta)\wedge b(\ga,\de)\right)p(y-x)$, 
a particle jumps from $x$ to $y$ simultaneously in both processes;\hfill\break\noindent 
* with rate $\bigl(b(\al,\beta) - b(\al,\beta)\wedge b(\ga,\de)\bigr)p(y-x)$, 
only a $\xi$-particle jumps from $x$ to $y$;\hfill\break\noindent 
* with rate $\bigl(b(\ga,\de) - b(\al,\beta)\wedge b(\ga,\de)\bigr)p(y-x)$, 
only a $\zeta$-particle jumps from $x$ to $y$.

By rule (ii), the last two changes are mutually exclusive, and their rates are fixed by rule (iii). 
Notice that the resulting Markovian
coupling is not necessarily increasing. In fact, 
there are exactly two 
instances in which the coordinate-wise order $\xi_{t^-}\le\zeta_{t^-}$ 
might be destroyed, namely ($\al\le\ga$ and $\beta=\de$) or ($\al=\ga$ and 
$\beta\le\de$). In the first (resp. second) one, the isolated jump of
a $\xi$-particle (resp. a $\zeta$-particle) would break the  
order between configurations. Attractiveness requires that  the corresponding rates  are equal
to zero, or equivalently 
\beqq
b(\al,\de)\le b(\ga,\de) &&\hbox{ for all } 
\de \hbox{ and }\al\le\ga \cr
b(\ga,\beta)\le b(\ga,\de) &&\hbox{  for all } 
\ga \hbox{ and } \beta\ge\de
\eeqq
which is exactly the content of Condition ${\bf(A)}$.

Beyond attractiveness, the basic coupling construction turns out 
to be essential to characterize the set 
$\left(\mathcal {I}\cap\mathcal{S}\right)_e$ of 
extremal invariant and translation invariant probability measures
of $(\eta_t)_{t\ge 0}$, through a control 
of the evolution of {\em discrepancies} between the marginals of 
$(\xi_t,\zeta_t)_{t\geq 0}$. There is a discrepancy at $x\in S$ at time $t$ if 
$\xi_t(x)\not=\zeta_t(x)$. In SEP-ZRP-MP, in a coupled transition the number of discrepancies  on the involved sites remains constant whenever the values of the two marginal configurations are 
 ordered, but decreases otherwise. 
This is a consequence of the fact that Condition ${\bf(A)}$ (applied twice) leads to inequalities between rates on unordered configurations, 
$$b(\al,\beta)
 \le b(\ga,\de)\hbox{ for all } 
\al<\ga \hbox{ and }\beta>\de$$
This decreasing property is crucial: 
under suitable conditions on  $p(y-x)$ and $b(\al,\beta)$, it permits the identification of $\left(\mathcal {I}\cap\mathcal{S}\right)_e$
as a one parameter family $\{\mu_\rho\}_\rho$ of product  probability measures, 
where the parameter $\rho$ fixes the average particle density per site. 

Beyond these classical
examples, the requirements needed for attractiveness through rules (i-iii) might be so drastic that the basic coupling construction is not possible. 
We therefore have to weaken those rules in order to get attractiveness through a coupling construction imposing
less restrictive conditions on the rates.
At the same time, we have to ask for a decrease of discrepancies, which came for free within basic coupling but could be less obvious in a larger setting.
 
Our starting point will be the more investigated case of pure jump 
processes with denumerable state space, which provides us with a set of necessary conditions on the rates for 
attractiveness. Restricting  ourselves to the class of conservative systems we consider in this paper,
it turns out that it is possible to construct a Markovian coupling,  which is increasing whenever these necessary 
conditions are fulfilled, hence proving also their sufficiency. Another property of this coupling is that  discrepancies are non increasing. 
However, a decrease of discrepancies requires additional irreducibility conditions. 
Under them, we prove that  $\left(\mathcal {I}\cap\mathcal{S}\right)_e$ is a one parameter family of probability measures. This gives the first step in the derivation of hydrodynamic limits under Euler scaling, which we explicitely achieve for two one-dimensional
examples, where an alternative proof of macroscopic stability is needed in order to follow the approach of \cite{BGRS3}.

It is important to notice that the restriction to single particle jumps ($k=1$) is essential in the basic coupling construction and in the derivation of attractiveness,  in particular in the two following instances: it induces implicitely a total order on $S$ and a partial lexicographic order on the state space; it drastically restricts the ways by which the order between configurations can be broken. These two points make the weakening of basic coupling rules useless and hide the interplay between partial order, conservation rule and optimization of coupling rates, though its
understanding seems necessary to deal with more complex systems.  Working with $k>1$ gives  some hints in these directions. In a forthcoming paper, we will consider attractiveness for conservative particle systems with speed change, a  problem which requires a fine tuning of these (yet mostly hidden) properties.

 In the last part of this paper, we will develop our results on various examples: First, in the well known cases where only $k=1$
is possible (thus in particular SEP-ZRP-MP)
our construction reduces to basic coupling. Moreover, our irreducibility conditions relax restrictions on the rates for ZRP-MP and allow to determine of $\left(\mathcal {I}\cap\mathcal{S}\right)_e$ in more general cases than previously treated.\hfill\break
As soon as we depart from $k\le 1$ case, both constructions show very different features: we focus on  two  one-dimensional examples, a  discrete nearest neighbor stick process in correspondence with a generalized  (that is, with nearest neighbor jumps) discrete Hammersley-Aldous-Diaconis (HAD) model, for which $k$ is unbounded, and 
a two-species asymmetric exclusion process with charge 
conservation, for which $k\le 2$.  A continuous stick process with totally asymmetric jumps was introduced in \cite{Se} in relation with the (continuous) HAD model, and the discrete HAD process was studied in particular in \cite{FM}. The two-species asymmetric exclusion process was introduced 
in \cite{CDFG} in the context of solid-on-solid 
interface dynamics and independently in  \cite{TS1}, \cite{TS2}.   Particular versions of this dynamics have been investigated in \cite{FT}, \cite{FN}
in order to derive hydrodynamics in non-attractive cases.

The paper is organised as follows. 
In Section \ref{sec:2}, we first recall the definition and properties of 
attractiveness. Then we introduce the conservative models 
$(\eta_t)_{t\ge 0}$ we are 
interested in, and state our main results: the set of necessary and sufficient conditions on the transition rates of $(\eta_t)_{t\ge 0}$
for attractiveness; the construction 
of the Markovian increasing coupling it involves, and under which discrepancies 
decrease. Proofs are given in Section \ref{sec:4}.
In Section \ref{sec:IcapS}, we obtain $\left(\mathcal {I}\cap\mathcal{S}\right)_e$ 
for $(\eta_t)_{t\ge 0}$ under additional irreducibility conditions. In Section \ref{sec:6}, we analyze the hydrodynamic behavior of $(\eta_t)_{t\ge 0}$ when the dynamics is one-dimensional and nearest neighbor.
We have illustrated our results on examples, the classical ones, SEP-ZRP-MP, as well as on the
 stick process and the two-species asymmetric exclusion 
process: In Section \ref{application_coupling}  and Appendix (Section \ref{sec:A1}), the coupling construction is presented in some detail and in
Section \ref{sec:IcapS_ex} we determine $\left(\mathcal {I}\cap\mathcal{S}\right)_e$ for each example.
In section \ref{sec:6} we 
obtain the hydrodynamic limit of the stick process and the two species asymmetric exclusion; for the latter we prove 
heuristic results given in
\cite{CDFG} and rederived more recently by \cite{TS1}, \cite{TS2}. 
\section { Preliminaries and  results}
\label{sec:2}
\subsection {Background}
\label{sec:2.1}
We first recall the general definitions and theorems about 
attractiveness and couplings that we use in this paper. 

Let $(\eta_t)_{t\geq 0}$ be an interacting 
particle system of state space $\Om = X^S$, with $X$ a 
subset of $\Z$, and $S=\Z^d$. We denote by $T(t)$ the 
semi-group of this Markov process and by ${\calL}$ its 
infinitesimal generator.  We mainly refer to \cite{L}, which deals with 
Feller processes on compact state spaces.
Since we also consider other cases, we assume that
\be \label{oleinik-ea} 
\begin{array}{l} (\eta_t)_{t\geq 0} \mbox{ is a well defined Markov process on a subset }
\Omega_0 \subset \Omega\\ \mbox{such that for any bounded local function $f$ on }\Omega_0, \\
\forall\eta\in\Omega_0,\,\,\displaystyle{\lim_{t\to 0}\frac{T(t)f(\eta)-f(\eta)}t={\calL}f(\eta)<+\infty}
\end{array}\ee 

A {\em coupled process} $(\xi_t,\zeta_t)_{t\geq 0}$ of two copies 
$(\xi_t)_{t\geq 0}$ and $(\zeta_t)_{t\geq 0}$ of $(\eta_t)_{t\geq 0}$
is a Markov process with state space $\Om_0\times\Om_0$, 
such that each marginal is a copy of the original process
$(\eta_t)_{t\geq 0}$. 

There is a partial
order (coordinate-wise) on the state space $\Om$, defined by
\be\label{order}
\forall \xi,\zeta\in \Om ,\quad \xi\leq\zeta \Llra 
( \forall x\in S,\quad\xi (x) \leq\zeta (x) )
\ee
\bd {\rm (see \cite{KKO})}
Let $W$ be a set endowed with a partial order relation.
A subset $V\subset W$ is {\rm increasing} if
\be\label{incset}
\forall l\in V,m\in W,\quad l\leq m \Lra m\in V.
\ee
A subset $V\subset W$ is {\rm decreasing} if 
\be\label{decset}
\forall l\in V,m\in W,\quad l\geq m \Lra m\in V.
\ee
A function $f:W\rightarrow\R$ is {\rm monotone} if
\be\label{incfct}
\forall l,m \in W, \quad l\leq m \Lra f(l) \leq f(m).
\ee
\ed
\bex\label{ex_1} (\cite{KK})
Let $l\in W$. Then $I_l=\{m\in W:l\leq m \}$ is an
increasing set, and $D_l=\{m\in W:l\geq m\}$ a decreasing one.
\eex
\br\label{rk_1}
For any subset $V\subset W$,
\[ V \mbox{ is increasing }\Llra  W\setminus V \mbox{ is decreasing }\Llra 
{\bf 1}_V \mbox{ is monotone}.\]
\er
We denote by $\mathcal{M}$ the set  of all bounded, monotone 
continuous functions on $\Om$. 
The partial order \eqref{order} induces a {\em stochastic order} 
(see \cite{KKO}) on the set ${\calP}$ of probability 
measures on $\Om$ endowed with the weak topology:
\be\label{incprob}
\forall\nu,\nu'\in{\calP},\quad\nu\leq \nu'\Llra 
( \forall  f\in\mathcal{M},\nu(f)\leq \nu'(f)).
\ee
 \br\label{rk_2}
$\nu\leq \nu'\Llra \nu(V)\leq \nu'(V)
\text{ for all increasing subsets } V\subset\Om$.

Indeed, on the one hand use Remark \ref{rk_1}, and on the other hand, notice that for any non-negative  $f\in\mathcal{M}$ and all $s\ge 0$, the set $\{f\ge s\}$ is increasing, and $\nu(f)=\int_0^\infty \nu\{f\ge s\}ds$.
\er
\bt\label{equiv-attractive} {\rm \cite[Theorem II.2.2]{L}}
For the particle system $(\eta_t)_{t\geq 0}$, the 
following two statements are equivalent: 

(a) $f\in\mathcal{M}$ implies $T(t)f\in\mathcal{M}$ 
for all $t\geq 0$.

(b) For $\nu,\nu'\in{\calP}$, $\nu\leq \nu'$ implies 
$\nu T(t)\leq \nu'T(t)$ for all $t\geq 0$.
\et
\bd\label{def-attractive} {\rm \cite[Definition II.2.3]{L}}
The particle system $(\eta_t)_{t\geq 0}$ is {\rm attractive} if the 
equivalent conditions of Theorem \ref{equiv-attractive} are satisfied.
\ed
By Remarks \ref{rk_1} and \ref{rk_2}, 
it is enough to check Theorem \ref{equiv-attractive}(a) 
for indicator functions of all increasing sets.

In the following our main object is the infinitesimal generator  ${\calL}$ of the process
since we look for necessary and sufficient conditions on the transition rates 
that yield attractiveness of the particle system. 
As mentioned in \cite{KKO}, this was investigated 
by many authors for pure jump processes (see also \cite{K},  \cite[Chapter 4]{S}) on a countable state space. 
In such cases, sufficiency consists in writing the semigroup as 
an exponentiation of the generator to check Theorem \ref{equiv-attractive}(b).
This follows from Strassen's Theorem, which links stochastic order 
to coupling (for a thorough analysis of this link, see 
\cite{KKO}):
\bt\label{th-attractive} {\rm \cite[Theorem II.2.4]{L}}
For  $\nu,\nu'\in{\calP}$, a necessary and sufficient condition for 
$\nu\leq \nu'$ is the existence of a probability measure 
$\mu$ on $\Om\times\Om$, called a coupled probability measure,
which satisfies
 
(a) $\mu\{(\xi,\zeta):\xi\in A\}=\nu(A),\,
\mu\{(\xi,\zeta):\zeta\in A\}=\nu'(A)$,  for all Borel sets 
$A\in\Om$, and
 
(b) $\mu\{(\xi,\zeta):\xi\leq\zeta\}=1$.
\et
Here we take the usual but longer route to derive attractiveness 
for interacting particle systems (see \cite{L}), and check 
Theorem \ref{equiv-attractive}(a) by constructing a {\em Markovian 
increasing coupling}, that is a  coupled process 
$(\xi_t,\zeta_t)_{t\geq 0}$ with the property that 
$\xi_0\leq\zeta_0$ implies
\be\label{coupl-attr}
P^{(\xi_0,\zeta_0)}\{\xi_t\leq\zeta_t\}=1
\ee
for all $t\geq 0$, where $P^{(\xi_0,\zeta_0)}$ denotes the 
distribution of $(\xi_t,\zeta_t)_{t\geq 0}$ with initial state 
$(\xi_0,\zeta_0)$.

Our strategy for proving and using attractiveness will consist in three steps: \hfill\break\noindent
(a) obtain necessary conditions on the transition rates of the model; \hfill\break\noindent
(b) construct a Markovian coupling and show that it is increasing under conditions obtained in (a), thus proving that they are also sufficient; \hfill\break\noindent
(c) verify that discrepancies cannot increase. When supplemented with some conditions 
on the rates,
this will allow to determine  $\left(\mathcal {I}\cap\mathcal{S}\right)_e$.
\subsection { The model, attractiveness and coupling}
\label{sec:2.2}
{}From now on, we restrict our analysis to the following class of models.
 
Let $S=\Z^d$ be the set of sites and $X\subset \Z$ be the set of admissible values on each site. In our examples, 
$X$ will be either a finite subset of $\Z$ or $\N$. The 
infinitesimal generator $\calL$ of the process $(\eta_t)_{t\ge 0}$ 
on $\Om = X^S$ is given, for a local function $f$, by
\be\label{gene}
\mathcal L f(\eta) = \sum_{x,y\in S} \sum_{\al,\beta\in X}
\chi_{x,y}^{\al,\beta}(\eta)\sum_{k\in\N} \Ga_{\al,\beta}^k (y-x)
(f(S_{x,y}^k\eta) -f(\eta))
\ee
where $\chi_{x,y}^{\al,\beta}$ is  the indicator function of 
configurations with values
$(\al,\beta)$ on $(x,y)$,
\be
\chi_{x,y}^{\al,\beta}(\eta)=
\begin{cases}
1 &\hbox{ if } \eta(x)=\al \hbox{ and } \eta(y)=\beta\cr
0& \hbox{ otherwise}\cr
\end{cases}
\ee
$S_{x,y}^k$ is a local operator performing the
transformation whenever possible (the value $k=0$ is not excluded)
\be
\bigl(S_{x,y}^k\eta\bigr)(z) =
\begin{cases}
\eta(x)-k &\hbox{ if }  z=x \hbox{ and } \eta(x)-k \in X , \eta(y)+k \in X\cr
\eta(y)+k &\hbox{ if } z=y \hbox{ and } \eta(x)-k \in X, \eta(y)+k \in X\cr
\eta(z) &\hbox{ otherwise}\cr
\end{cases}
\ee
This particle system is conservative, with $\eta(x)+\eta(y) $ the  {\sl conserved quantity} in a transition
 between sites $x$ and $y$.

Unless $X$ is finite and the rates $\Ga_{.,.}^.(.)$'s have finite range
(in which case we refer to \cite[Chapter 1]{L}), additional conditions on the transition rates, and/or a reduction of the state space, are required to ensure that \eqref{gene} is the infinitesimal generator of a well defined Markov process. Since such conditions differ for the specific model one deals with, we assume \eqref{oleinik-ea} and state here only a common necessary restriction on the rates (complete precise assumptions being given on examples).  For all $z\in S,\,\al,\beta\in X$, the rates $\Ga_{\al,\beta}^k(z)$ satisfy
\be\label{test_H1'}
\sum_{k\in\N}\Ga_{\al,\beta}^k(z)<\infty  
\ee

{}For notational convenience, we will often drop the 
explicit dependence on $z$ of  $\Ga_{\al,\beta}^k(z)$ ; we also set 
\be
\Ga_{\al,\beta}^k = 0 \hbox{ if } \al-k \not\in X,  
\beta + k \not\in X,  \hbox{ or } k=0.
\ee
\bd\label{def:ordered} The notation 
$(\al,\beta)\leq(\ga,\de)$ is equivalent to $\al\le\ga,\,\beta\le\de$; the two pairs $(\al,\beta),(\ga,\de)$
are {\rm ordered} if $(\al,\beta)\leq(\ga,\de)$ or $(\al,\beta)\geq(\ga,\de)$; they are {\rm not ordered} otherwise, 
that is when $(\al<\ga,\beta>\de)$ or $(\al>\ga,\beta<\de)$.\ed
The main result of this section is
\bt\label{new-main}
The particle system $(\eta_t)_{t\geq 0}$ is attractive if 
and only if for all $(\al,\beta,\ga,\de)\in X^4$ with $(\al,\beta) \le (\ga,\de)$,
for all $(x,y)\in S^2$, 
\beq
\forall l\ge 0,  \sum_{k'>\de-\beta+l} \ \Ga_{\al,\beta}^{k'}(y-x) &\le&
\sum_{l'> l}\ \Ga_{\ga,\de}^{l'}(y-x)\label{inc2} \\
\forall k\ge 0, \qquad \sum_{k'> k}  \ \Ga_{\al,\beta}^{k'}(y-x) &\ge&
\sum_{l'> \ga - \al +k}   \ \Ga_{\ga,\de}^{l'} (y-x)
 \label{dec2}
\eeq 
\et
Note that by \eqref{test_H1'} the above sums are finite.
We now give the succession of steps which leads to this theorem.
Proofs  are postponed to Section \ref{sec:4}.

First, we prove necessary conditions for attractiveness:
\bp\label{ff0}
If the particle system $(\eta_t)_{t\geq 0}$ is attractive, then
 for all 
$(\al,\beta,\ga,\de)\in X^4$ with $(\al,\beta) \le (\ga,\de)$,
for all $(x,y)\in S^2$, inequalities 
\eqref{inc2}--\eqref{dec2} hold.
\ep
The other part of Theorem \ref{new-main} is harder and is obtained 
by constructing explicitly a Markovian coupling, 
which appears to be increasing under conditions  \eqref{inc2}--\eqref{dec2}.
The evolution of the coupled process 
$(\xi_t,\zeta_t)_{t\geq 0}$ we look for is defined through its
infinitesimal generator $\overline{\calL}$, whose rates 
will be derived from those of ${\calL}$. 
Here,  as in the basic coupling construction,
\emph{the same} departure and arrival sites $x$ 
and $y$  are chosen for jumps of coupled particles.
\bp\label{princ}
The operator $\overline{\calL}$ defined on $\Om_0\times\Om_0$ as 
\beq\nonumber
\overline{\calL}f(\xi,\zeta)&=&
\sum_{x,y\in S}
\sum_{\al,\beta\in X} \sum_{\ga,\de \in X} 
\chi_{x,y}^{\al,\beta}(\xi)\chi_{x,y}^{\ga,\de}(\zeta)\times\\
\label{gene2}
&&\qquad\qquad\sum_{k,l}
G_{\al,\beta;\, \ga,\de}^{k;\, l} (y-x)
\bigl(f(S_{x,y}^k \xi, S_{x,y}^{l} \zeta) -f(\xi,\zeta)\bigr)
\eeq
with coupling rates $ G_{\al,\beta;\, \ga,\de}^{k;\, l}$ 
given by \eqref{cou2}--\eqref{cou2-3} below
for all 
$(\al,\beta,\ga,\de)\in X^4$ and all non-negative $k,\,l$
as functions of the initial rates $\Ga_{\al,\beta}^{k'}$ and 
$\Ga_{\ga,\de}^{l'}$, is the generator of a Markovian coupling 
between two copies of the
Markov process  defined by \eqref{gene}.

{}For all  positive $k,\,l$, we set
\beq\label{cou2}
\hbox{        }G_{\al,\beta;\, \ga,\de}^{k;\, l}&=&
\Bigl(\Ga_{\al,\beta}^{k} - \Ga_{\al,\beta}^{k} \wedge
\bigl( \Si_{\ga,\de}^l- \Si_{\al,\beta}^k \wedge \Si_{\ga,\de}^l
\bigr)\Bigr)\nonumber\\
&& \qquad\qquad\qquad\qquad\wedge
\Bigl( \Ga_{\ga,\de}^{l}  -  \Ga_{\ga,\de}^{l} \wedge
\bigl(\Si_{\al,\beta}^k - \Si_{\al,\beta}^k \wedge  \Si_{\ga,\de}^l
\bigr)\Bigr)
\\
\hbox{        } G_{\al,\beta;\, \ga,\de}^{0;\, l}&=&
 \Ga_{\ga,\de}^{l}  -   \Ga_{\ga,\de}^{l} \wedge
\bigl(\Si_{\al,\beta}^0 - \Si_{\al,\beta}^0 \wedge  \Si_{\ga,\de}^l
\bigr)
\label{cou2-1}\\
\hbox{        } G_{\al,\beta;\, \ga,\de}^{k;\, 0}&=&
\Ga_{\al,\beta}^{k} - \Ga_{\al,\beta}^{k} \wedge
\bigl( \Si_{\ga,\de}^0- \Si_{\al,\beta}^k \wedge \Si_{\ga,\de}^0
\bigr)\label{cou2-2}
\\
\hbox{        } G_{\al,\beta;\, \ga,\de}^{0;\, 0} &=& 0
\label{cou2-3}\
\eeq
where the $\Si_{\al,\beta}^k$'s denote partial sums of rates 
for $k\ge 0$,
\be\label{Si_al-beta}
\Si_{\al,\beta}^k = \sum_{k'>k} \ \Ga_{\al,\beta}^{k'}.
\ee
\ep
 In other words, at time $t$,  configurations
$\xi_{t^-},\,\zeta_{t^-}$ are changed into 
$\xi_{t}=S_{x,y}^k\xi_{t^-},\,\zeta_{t}=S_{x,y}^l\zeta_{t^-}$
with rate $G_{\xi_{t^-}(x),\xi_{t^-}(y);
\,\zeta_{t^-}(x),\zeta_{t^-}(y)}^{k;\, l}(y-x)$. 
Rates with superscript `$k; 0$' (resp. `$0; l$')
correspond
to uncoupled changes in the $\xi_.$ (resp. $\zeta_.$) process.

Since the coupling rates are constructed for a fixed pair of sites $(x,y)$, 
we dropped from the notation the explicit  dependence of the rates on $(x,y)$ 
and wrote $\Ga_{\al,\beta}^k$ for $\Ga_{\al,\beta}^k(y-x)$ 
and $G_{\al,\beta;\, \ga,\de}^{k;\, l}$ for $G_{\al,\beta;\, \ga,\de}^{k;\, l}(y-x)$.
\br\label{sigma-ok}
The partial sums $\Si_{\al,\beta}^k$'s in \eqref{Si_al-beta}
are well defined by \eqref{test_H1'}. As a direct consequence of 
expressions \eqref{cou2}--\eqref{cou2-3} for the coupling rates,
the coupled process is well defined on $\Om_0\times\Om_0$ 
(cf. \eqref{oleinik-ea}). 
\er
\br\label{cou2-symm} Formulas \eqref{cou2}--\eqref{cou2-3}
 are symmetric under exchange of marginals:
for all $(\al,\beta,\ga,\de)$, $k$ and $l$
\[  G_{\al,\beta;\, \ga,\de}^{k;\, l}=  G_{\, \ga,\de;\al,\beta}^{l;\, k}\]
\er
\br\label{cou2-diag}
Like in the basic coupling construction,
the diagonal coupling rates are equal to the original
rates of the marginal processes: for  $(\al,\beta,k)=(\ga,\de,l)$, 
\eqref{cou2} becomes
\[
 G_{\al,\beta;\,\al,\beta}^{k;\,k}=
\Bigl(\Ga_{\al,\beta}^k - \Ga_{\al,\beta}^k \wedge
\bigl( \Si_{\al,\beta}^k- \Si_{\al,\beta}^k \wedge \Si_{\al,\beta}^k
\bigr)\Bigr)=\Ga_{\al,\beta}^k
\]
\er
In the next two propositions, we give some hints on the construction and 
structure of the coupling rates; for any fixed $(\al,\beta,\ga,\de) \in X^4$,
non zero coupling rates are located on a ``staircase-shaped" path in the $(k,l)$ quadrant: most of the coupling rates are zero 
 and for $n>0$ there is at most one pair $(k,l)$ such that $k+l=n$ and 
 $G_{\al,\beta;\, \ga,\de}^{k;\, l} >0 $. Alternatively, the coupling rates can be defined by induction.
\bp\label{altprinc2}
For any $(\al,\beta,\ga,\de) \in X^4$, all $z\in S$, let $(k_0,l_0) = (0,0)$ and  $\mathcal P_{\al,\beta;\, \ga,\de}=\{(k_i,l_i)\}_{i\ge 0}$ be the sequence of points in $\N\times \N$
defined for all $i> 0$ as
\beq\label{eq2.40}
\begin{cases}
{k_{i}=k_{i-1} +1, \quad l_{i}=l_{i-1} }&\hbox{ if }  \quad \Si_{\al,\beta}^{k_{i-1}} \ge \Si_{\ga,\de}^{l_{i-1}}\cr
{k_{i}=k_{i-1},  \quad l_{i}=l_{i-1} +1}& \hbox{ otherwise}\cr
\end{cases}
\eeq
The coupling rates defined by equations  \eqref{cou2}--\eqref{cou2-3} satisfy

(i) For all  $(k_i,l_i)\in \mathcal P_{\al,\beta;\, \ga,\de}$ 
\beq\label{altprinc2-sola}
G_{\al,\beta;\, \ga,\de}^{k_i;\, l_i} =
\Si_{\al,\beta}^{k_{i-1}} \vee  \Si_{\ga,\de}^{l_{i-1}} 
-  \Si_{\al,\beta}^{k_i} \vee  \Si_{\ga,\de}^{l_i} 
\eeq
(ii) For all $(k,l)\notin \mathcal P_{\al,\beta;\, \ga,\de}$
\beq\label{altprinc2-solb}
G_{\al,\beta;\, \ga,\de}^{k;\, l} =0
\eeq
\ep
\bp\label{altprinc}
The set of coupling rates defined by equations  \eqref{cou2}--\eqref{cou2-3}
is the unique solution of the recursion relations:
 
For all  $(\al,\beta,\ga,\de) \in X^4$, all $z\in S$ and all positive $k$,$l$, 
\beq
\qquad G_{\al,\beta;\, \ga,\de}^{k;\, l} &=&
\bigl(\Ga_{\al,\beta}^k -\sum_{l'>l} G_{\al,\beta;\, \ga,\de}^{k;\, l'}\bigr) \wedge
\bigl(\Ga_{\ga,\de}^l -\sum_{k'>k} G_{\al,\beta;\, \ga,\de}^{k';\, l}\bigr)
\label{coupl2}\\
\qquad  G_{\al,\beta;\, \ga,\de}^{0;\, l} &=& \Ga_{\ga,\de}^l 
- \sum_{k'>0} G_{\al,\beta;\, \ga,\de}^{k';\, l} 
\label{coupl3}\\
\qquad G_{\al,\beta;\, \ga,\de}^{k;\, 0} &=& \Ga_{\al,\beta}^k  
- \sum_{l'>0} G_{\al,\beta;\, \ga,\de}^{k;\, l'} 
\label{coupl4}\\
G_{\al,\beta;\, \ga,\de}^{0;\, 0} &=& 0
\label{coupl5}
\eeq
\ep
We end this subsection with a proposition giving the last step to Theorem \ref{new-main}:
\bp\label{main}
Under conditions \eqref{inc2}--\eqref{dec2}, the 
Markovian coupling defined in Proposition \ref{princ} is increasing.
More precisely, if $(\al,\beta),(\ga,\de)$ are  ordered, then the coupling 
 rate $G_{\al,\beta;\, \ga,\de}^{k;\, l}$ (with non-negative $k,l$) is 
 nonzero only if the increments satisfy
\beq\label{4.10}
\begin{cases}
l-k \in \{-(\de-\beta),\cdots,\ga-\al\}& 
\qquad \hbox{ if } 
(\al,\beta) \le (\ga,\de)\\
l-k \in \{-(\al-\ga),\cdots,\beta-\de\}&
\qquad \hbox{ if }
(\al,\beta) \ge (\ga,\de)
\end{cases}
\eeq 
\ep
\subsection { Evolution of discrepancies}
\label{sec:2.3}
Attractiveness expresses that two ordered configurations remain such (stochastically) under the dynamics.
Many applications require also a control over unordered pairs of configurations. This can either be a consequence of attractiveness or require some additional hypotheses, that we investigate in this subsection.
\bd\label{def-disc}
In the  coupled process  $(\xi_t,\zeta_t)_{t\ge 0}$, there is a {\rm discrepancy} 
 at site $z\in S$ at time $t$ if $\xi_t(z)\ne\zeta_t(z)$. This 
discrepancy is {\rm positive} (resp. {\rm negative}) {\rm
of width} $a>0$ if $\xi_t(z)-\zeta_t(z)=a$ (resp. $\xi_t(z)-\zeta_t(z)=-a$). 
\ed
We first show that
under the Markovian increasing coupling constructed in 
Proposition \ref{princ}, the sum of the widths of the
discrepancies (involved in a transition) between the two 
coupled processes $(\xi_t)_{t\ge 0}$ and $(\zeta_t)_{t\ge 0}$ 
is non-increasing (later on, we will sometimes forget the 
word `width', and speak of non-increasing or decreasing 
discrepancies). 

We fix $x,y$ two sites of $S$.
Denote by $\overline{\calL}_{x,y}$ the generator of jumps between 
them,
\beq\label{local-L}
\overline{\calL}_{x,y}f(\xi,\zeta)=&&
\sum_{\al,\beta\in X} \sum_{\ga,\de \in X} \chi_{x,y}^{\al,\beta}(\xi)
\chi_{x,y}^{\ga,\de}(\zeta)\times\label{3.23}\\
&&\Bigl(\sum_{k,l}
G_{\al,\beta;\, \ga,\de}^{k;\, l} (y-x)
\bigl(f(S_{x,y}^k \xi, S_{x,y}^{l} \zeta) -f(\xi,\zeta)\bigr)\nonumber\\
&&\quad+ \sum_{k,l} G_{\beta,\al;\, \de,\ga}^{l;\, k} (x-y)
\bigl(f(S_{y,x}^l \xi, S_{y,x}^k \zeta) -f(\xi,\zeta)\bigr)
\Bigr)
\nonumber
\eeq 
for a local function $f$, and two configurations $\xi,\zeta\in\Om_0$.

Let $f_{x,y}^{\pm}=f_{y,x}^{\pm}$ be
the functions  
which measure the width of positive or negative discrepancies between $\xi$ 
and $\zeta$ on sites $x$ and $y$
\beq\label{f_xy}
\psi_x^\pm(\xi,\zeta)&:=&[\xi(x)-\zeta(x)]^\pm,\cr
f_{x,y}^\pm(\xi,\zeta)&:=&\psi_x^\pm(\xi,\zeta) + \psi_y^\pm(\xi,\zeta)
\eeq
The quantities 
\be\label{Delta_neg}
\Delta^\pm(\xi(x),\xi(y),\zeta(x),\zeta(y),k,l) := f_{x,y}^\pm
(S_{x,y}^k\xi,S_{x,y}^l\zeta) -f_{x,y}^\pm (\xi,\zeta) 
\ee
measure how, in a coupled transition on sites $x,y$, discrepancies 
evolve.
Since by the conservation rule, 
\[
f_{x,y}^+(\xi,\zeta) - f_{x,y}^-(\xi,\zeta)= \bigl(\xi(x)+\xi(y)\bigr)-\bigl(\zeta(x)+\zeta(y)\bigr)
=f_{x,y}^+ (S_{x,y}^k\xi,S_{x,y}^l\zeta)- f_{x,y}^- (S_{x,y}^k\xi,S_{x,y}^l\zeta)
\]
the two quantities  $\Delta^\pm$ in \eqref{Delta_neg} are equal and we set
\beq\label{Delta+=Delta-} 
\Delta (\xi(x),\xi(y),\zeta(x),\zeta(y),k,l) &=& \Delta^\pm(\xi(x),\xi(y),\zeta(x),\zeta(y),k,l)
\eeq
with the meaning that positive and negative discrepancies disappear (or get reduced)
by pair annihilation.
Notice the symmetry
\be\label{Delta_sym}
\Delta (\xi(y),\xi(x),\zeta(y),\zeta(x),l,k) = \Delta(\xi(x),\xi(y),\zeta(x),\zeta(y),k,l)
\ee
because by exchanging respectively $\xi(x)$ and $\xi(y)$,
$\zeta(x)$ and $\zeta(y)$, $k$ and $l$,
\beqq
[(\xi(y)-l)-(\zeta(y)-k)]^+ &=& [(\xi(y)+k)-(\zeta(y)+l)]^+ \cr
[(\xi(x)+l)-(\zeta(x)+k)]^+ &=& [(\xi(x)-k)-(\zeta(x)-l)]^+ 
\eeqq
Next theorem details how discrepancies decrease.
\bt\label{discrep}
For all $(\xi,\zeta)\in\Om_0\times\Om_0$, $(x,y)\in S^2$, 
\be\label{gen_negatif}
\overline{\calL}_{x,y}f_{x,y}^+(\xi,\zeta)=
\overline{\calL}_{x,y}f_{x,y}^-(\xi,\zeta)\leq 0\qquad
\ee
More precisely, setting $(\al,\beta, \ga,\de)=(\xi(x),\xi(y),\zeta(x),\zeta(y
))$,  the coupling rate $G_{\al,\beta;\, \ga,\de}^{k;\, l}(y-x)$ may be positive
 only for non negative increments $k$, $l$ satisfying:
\be\label{crnz}
-[\al-\ga]^+ - [\de - \beta]^+ \le l-k \le [\ga - \al]^+ + [\beta -\de]^+
\ee
and in that case
\be\label{Delta_neg-detail}
\begin{cases}
\Delta (\al,\beta, \ga,\de,k,l)=0 & \mbox{ for }
(\al,\beta),(\ga,\de) \mbox{  ordered }\\ 
\Delta (\al,\beta, \ga,\de,k,l) =0 & \mbox{ for }
(\al,\beta),(\ga,\de) \mbox{  not ordered} \\ & \mbox{   and  }k-l\in\{0,(\al-\ga)+(\de-\beta)\}\\ 
\Delta (\al,\beta, \ga,\de,k,l)<0 & \mbox{ otherwise }
\end{cases}
\ee
\et
In \eqref{Delta_neg-detail}, the first line 
expresses attractiveness. The second line deals with discrepancies of opposite signs and contains two cases: 
when $k-l=0$, the same number of particles moves together in both processes so that discrepancies do not change (as in basic coupling, for which $k=l=1$);
when $k-l=(\al-\ga)+(\de - \beta)$, the positions of discrepancies 
at sites $x$ and $y$ are exchanged.  The third line 
expresses a decrease of discrepancies, it will be the key tool in the determination of
$\left(\mathcal {I}\cap\mathcal{S}\right)_e$  (see Section \ref{sec:IcapS}). 
Notice that for basic coupling, this case corresponds only to uncoupled jumps,
 while here it may involve non-zero values for both  $k$ and $l$; the jump gets
the two marginals closer by merging together two discrepancies of opposite signs. The resulting pairs of values need not be ordered since discrepancies can reduce their widths without disappearing and possibly exchange their signs (as in the second line). This new phenomenon of (partial) {\sl exchange of discrepancies} does exist  {\sl beyond} basic coupling (it requires $|k-l|\ge 2$) and induces some changes in the behavior of the coupled process. More precisely 
\bd\label{idsc-exchange}
The increasing coupling defined by \eqref{gene2} {\rm allows exchanges of discrepancies} whenever there exist  $(x,y)\in S^2$, $(\al,\beta,\ga,\de)\in X^4$ with $(\al-\ga) (\beta-\de) <0$, non-negative $k,l$ with $|k-l|>|\al-\ga|\vee |\beta-\de|$ such that  $G_{\al,\beta;\, \ga,\de}^{k;\, l}(y-x) > 0$. The exchange of discrepancies is {\rm total} if $k-l=(\al-\ga)+(\de -\beta)$ and {\rm partial} otherwise. \ed

\bl\label{lem:noteod}
The increasing coupling defined by \eqref{gene2} does not allow exchanges of discrepancies  if and only if for any $(x,y)\in S^2$,
$(\al,\beta,\ga,\de)\in X^4$, non-negative $k,l$ such that  $G_{\al,\beta;\, \ga,\de}^{k;\, l}(y-x) >0$,
we have
\be\label{noteod}
-([\al-\ga]^+ \vee  [\de -\beta]^+) \le l-k \le [\ga-\al]^+ \vee [\beta-\de]^+
\ee
\el
\br\label{rk:noteod}
When $(\al,\beta),(\ga,\de)$ are  ordered, \eqref{noteod} reduces to inequalities  \eqref{4.10} for attractiveness. Otherwise \eqref{noteod} is stronger than \eqref{crnz} (which is a consequence of attractiveness).
\er

 A first application of Theorem 
\ref{discrep} is to determine extremal
invariant and translation invariant probability measures for the model (see Theorem \ref{th:(i cap s)_e}) under irreducibility conditions on the rates, that cope with exchanges of discrepancies and other phenomena which appear beyond the case $k\le 1$. 

A second application is  {\sl macroscopic stability}, 
 an essential step to derive the hydrodynamic limit of one-dimensional models (see Section \ref{sec:6}). This property, introduced in \cite[Proposition 3.1]{bm}  for one-dimensional finite-range simple exclusion processes, gives a control of the time evolution of discrepancies in the coupled process. However, its proof does not extend to models with jumps
 of size $k\ge 1$, in particular because the latter induce  
 exchanges of discrepancies.  
  So we prove here
 
\bp\label{macrostab-ppv}
We assume that the particle system defined by \eqref{gene}
is one dimensional ($S=\Z$) with only nearest neighbor transitions, and that the increasing coupling defined by \eqref{gene2} does not allow exchanges of discrepancies. 
Then, if the coupled process $(\xi_s,\zeta_s)_{s\ge 0}$
is such that
$\sum_{x\in\Z}[\xi_0(x)+\zeta_0(x)]<+\infty$, we have for
every $t>0$,
\be\label{eq:macrostab}
S(\xi_t,\zeta_t)\leq S(\xi_0,\zeta_0)\ee
where, for $x\in\Z$, 
\be\label{def:sx}
s_x(\xi_s,\zeta_s):=\sum_{y\leq
x}[\xi_s(y)-\zeta_s(y)],\qquad
S(\xi_s,\zeta_s):=\sup_{x\in\Z}\abs{s_x(\xi_s,\zeta_s)}
\ee 
\ep
\br\label{rk:macrostab}
Both restriction to nearest neighbor interactions and Condition \eqref{noteod} are necessary
to get the macroscopic stability property \eqref{eq:macrostab}. See Section \ref{sec:4} for counter-examples.
\er
\bp\label{macrostab:cns}
The following inequalities 
imply \eqref{noteod}: For all
$(\al,\beta,\ga,\de)\in X^4$, non-negative $k,l$,
\beq\label{cs-sum_l}\Si_{\ga,\de}^l &\ge& \Si_{\al,\beta}^L,\quad L=l+[\al-\ga]^+ \vee [\de-\beta]^+\\\label{cs-sum_k}
\Si_{\al,\beta}^k &\ge& \Si_{\ga,\de}^K,\quad K=k+[\ga-\al]^+ \vee [\beta-\de]^+
\eeq
\ep
\br\label{cs-sum_l=inc2}
When $(\al,\beta), (\ga,\de)$ are ordered, \eqref{cs-sum_l}--\eqref{cs-sum_k} coincide with \eqref{inc2}--\eqref{dec2}  and thus reduce to attractiveness conditions.
\er
\section { Proofs.}
\label{sec:4}

\noindent 
{\bf Proof of Proposition \ref{ff0}.}
 
\bprm
We proceed in two steps. The first one yields an inequality for the generator of an attractive system,
when applied on indicators of increasing sets. Such inequalities were first derived in \cite{M} for Markov jump processes on a countable state space $E$, with an infinitesimal generator bounded in $l_1(E)$. 
The second step specializes the previous inequality to our model on which it reads  \eqref{inc2}--\eqref{dec2}.

{\em Step 1.}
Let $(\xi,\zeta)\in\Om_0\times\Om_0$ be two configurations  such 
that $\xi\le\zeta$. Let $V\subset \Om$ be an  increasing 
cylinder set. If $\xi\in V$ or $\zeta\notin V$, 
\be\label{meme}
{\bf 1}_V(\xi)={\bf 1}_V(\zeta)
\ee
By attractiveness, for all $t\ge 0$,
$$(T(t){\bf 1}_V)(\xi)\le (T(t){\bf 1}_V)(\zeta)$$
 (use Remark \ref{rk_2} and Theorem \ref{equiv-attractive}(a)). Combining 
this with \eqref{meme},
\[
t^{-1}[(T(t){\bf 1}_V)(\xi)-{\bf 1}_V(\xi)]\le 
t^{-1}[(T(t){\bf 1}_V)(\zeta)-{\bf 1}_V(\zeta)],
\]
which gives, by Assumption \eqref{oleinik-ea},
\be\label{ineq_gen}
({\calL}{\bf 1}_V)(\xi)\le ({\calL}{\bf 1}_V)(\zeta)
\ee
We have
\beq\label{value_gen}
({\calL}{\bf 1}_V)(\xi)
&=& \sum_{x,y\in S} \sum_{\al,\beta\in X}
\chi_{x,y}^{\al,\beta}(\xi)\sum_{k} \Ga_{\al,\beta}^k (y-x)
\left({\bf 1}_V(S_{x,y}^k\xi)-{\bf 1}_V(\xi)\right)
\nonumber\\
&=&-{\bf 1}_V(\xi)\sum_{x,y\in S} \sum_{\al,\beta\in X}
\chi_{x,y}^{\al,\beta}(\xi)\sum_{k}
{\bf 1}_{\Om_0 \setminus V}(S_{x,y}^k\xi)\Ga_{\al,\beta}^k (y-x)
\nonumber\\
&& +{\bf 1}_{\Om_0 \setminus V}(\xi)\sum_{x,y\in S} \sum_{\al,\beta\in X}
\chi_{x,y}^{\al,\beta}(\xi)\sum_{k}
{\bf 1}_{V}(S_{x,y}^k\xi)\Ga_{\al,\beta}^k (y-x)
\eeq

{\em Step 2.}
We fix $(x,y)\in S^2$, $x\ne y$,
$(\al,\beta,\ga,\de)\in X^4$, with $(\al,\beta) \le (\ga,\de)$,
and  two configurations $(\xi,\zeta)$ in $\Om_0\times\Om_0$
such that $\xi(x)=\al$, $\xi(y)=\beta$, $\zeta(x)=\ga$, $\zeta(y)=\de$,
and $\xi(z)=\zeta(z)$ for all $z\ne x,y$. Thus $\xi\le\zeta$.

We fix numbers $(p_z\in X, z\in S)$, which satisfy 
$p_x <\xi(x)=\al$, $p_y>\zeta(y)=\de$, and $p_z=\xi(z)=\zeta(z)$ 
for all $z\in S\setminus\{x,y\}$.
We denote by $\|\cdot\|$ the $L_1$-norm on $S$. For $n\ge \|y-x\|$, we define 
\beqq
I_y(n)=\{\eta\in\Om:\,\eta(z)\ge p_z, \; \hbox{ for all } z\not=x \hbox{ such that } \|y-z\|\le n \}
\eeqq
By Example \ref{ex_1},  $I_y(n)$ is 
an increasing cylinder set. Since neither $\zeta$ nor $\xi$ belong to $I_y(n)$, we have using 
\eqref{value_gen}
\beqq
({\calL}{\bf 1}_{I_y(n)})(\xi)&=&
\sum_{x',y'\in S} \sum_{\al',\beta'\in X}
\chi_{x',y'}^{\al',\beta'}(\xi)\sum_{k'}
{\bf 1}_{I_y(n)}(S_{x',y'}^{k'}\xi)\Ga_{\al',\beta'}^{k'} (y'-x')\\
&=&\sum_{k'\in\N}
{\bf 1}_{\{\beta+k'\ge p_y\}}
\Ga_{\al,\beta}^{k'}(y-x)
+\sum_{z:\|y-z\|> n}\, \sum_{k'\in\N}
{\bf 1}_{\{\beta+k'\ge p_y\}}
\Ga_{\xi(z),\beta}^{k'}(y-z)
\eeqq
and similarly
\beqq
({\calL}{\bf 1}_{I_y(n)})(\zeta)&=&\sum_{l'\in\N}
{\bf 1}_{\{\de+l'\ge p_y\}}
\Ga_{\ga,\de}^{l'}(y-x)
+\sum_{z:\|y-z\|> n}\, \sum_{l'\in\N}
{\bf 1}_{\{\de+l'\ge p_y\}}
\Ga_{\zeta(z),\de}^{l'}(y-z)
\eeqq
so that \eqref{ineq_gen} writes
\beqq
&&\sum_{k'\in\N:\beta+k'\ge p_y }
\Ga_{\al,\beta}^{k'}(y-x)
+\sum_{z:\|y-z\|> n}\, \sum_{k'\in\N}
{\bf 1}_{\{\beta+k'\ge p_y\}}
\Ga_{\xi(z),\beta}^{k'}(y-z)\\
&&\qquad \le
\sum_{l'\in\N: \de+l'\ge p_y}
\Ga_{\ga,\de}^{l'}(y-x)
+\sum_{z:\|y-z\|> n}\, \sum_{l'\in\N}
{\bf 1}_{\{\de+l'\ge p_y\}}
\Ga_{\zeta(z),\de}^{l'}(y-z)
\eeqq
Taking the monotone limit $n\to\infty$ gives
\beq\label{incr_xy-y}
\sum_{k'\in\N:\,\beta+k'\ge p_y}\Ga_{\al,\beta}^{k'}(y-x)\le 
\sum_{l'\in\N:\,\de+l'\ge p_y}\Ga_{\ga,\de}^{l'}(y-x)
\eeq
A similar argument can be used with the decreasing set
\beqq
D_{x}(n)=\{\eta\in\Om:\,\eta(z)\le p_z, \; \hbox{ for all } z\not=y \hbox{ such that } \|x-z\|\le n \}
\eeqq 
Configurations  $\xi$ and $\zeta$ belong to its complement $\Om_0\setminus D_{x}(n)$ (which is  increasing by  Remark \ref{rk_1}) and inequality \eqref{ineq_gen} for this set
leads to 
\beq\label{decr_xy-x}
\sum_{k'\in\N:\,\al-k'\le p_x}\Ga_{\al,\beta}^{k'}(y-x)\ge  
\sum_{l'\in\N:\,\ga-l'\le p_x}\Ga_{\ga,\de}^{l'}(y-x)
\eeq
{}Finally, taking 
$p_y=\de+l+1$ in \eqref{incr_xy-y} and   
$p_x=\al-k-1$ in \eqref{decr_xy-x} gives 
\eqref{inc2}--\eqref{dec2}.
\eprm

In order to prove Proposition \ref{princ}, we  
give equivalent expressions for the coupling rates.
\bl\label{lemma-cou-alt}
The following expressions 
are equivalent to 
\eqref{cou2}--\eqref{cou2-2} for all 
$(\al,\beta,\ga,\de)\in X^4$ and  positive $k,l$:
\beq\label{cou-alt1}
G_{\al,\beta;\, \ga,\de}^{k;\, l}=&&
\Ga_{\al,\beta}^{k} \wedge
\bigl( \Si_{\ga,\de}^{l-1}- \Si_{\al,\beta}^k 
\wedge \Si_{\ga,\de}^{l-1}\bigr)
-\Ga_{\al,\beta}^{k} \wedge
\bigl( \Si_{\ga,\de}^l- \Si_{\al,\beta}^k \wedge \Si_{\ga,\de}^l\bigr)
\\
\label{cou-alt2}
G_{\al,\beta;\, \ga,\de}^{k;\, l}=&&
\bigl( \Si_{\al,\beta}^{k-1}- 
\Si_{\al,\beta}^{k-1} \wedge \Si_{\ga,\de}^{l}\bigr)
\wedge \Ga_{\ga,\de}^l
- \bigl( \Si_{\al,\beta}^k- \Si_{\al,\beta}^k \wedge \Si_{\ga,\de}^l\bigr)
\wedge \Ga_{\ga,\de}^l
\\
\label{cou-alt3}
G_{\al,\beta;\, \ga,\de}^{0;\, l}=&&
\bigl( \Si_{\ga,\de}^{l-1}- \Si_{\al,\beta}^0 \wedge \Si_{\ga,\de}^{l-1}\bigr)
-\bigl( \Si_{\ga,\de}^l- \Si_{\al,\beta}^0 \wedge \Si_{\ga,\de}^l\bigr) 
\\
\label{cou-alt4}
G_{\al,\beta;\, \ga,\de}^{k;\, 0}=&&
\bigl( \Si_{\al,\beta}^{k-1}- \Si_{\al,\beta}^{k-1} \wedge \Si_{\ga,\de}^{0}\bigr)
-
\bigl( \Si_{\al,\beta}^k- \Si_{\al,\beta}^k \wedge \Si_{\ga,\de}^0\bigr)
\eeq
\el
\bpr
We prove only \eqref{cou-alt2}, \eqref{cou-alt4}; the two other  equations follow by symmetry from Remark \ref{cou2-symm}.
Let $(\al,\beta,\ga,\de)\in X^4$, $k>0$  and $l\ge 0$.
Using the elementary identity
\be\label{idw}
\forall a,b,c \ge 0,\quad 
a \wedge (c - b\wedge c) =  (a+b) \wedge c - b \wedge c
\ee
with $a= \Ga_{\al,\beta}^k$, $b=\Si_{\al,\beta}^k$, $c=\Si_{\ga,\de}^l$ leads to (we have $\Ga_{\al,\beta}^k=\Si_{\al,\beta}^{k-1}-\Si_{\al,\beta}^k$)
\beq\label{cou-diff}
\Ga_{\al,\beta}^k - \Ga_{\al,\beta}^k \wedge
\bigl( \Si_{\ga,\de}^l- \Si_{\al,\beta}^k \wedge \Si_{\ga,\de}^l\bigr)
&=& \bigl[ (a+b)- (a+b) \wedge c \bigr] - \bigl[ b - b \wedge c \bigr]\nonumber\\
&=& \bigl(\Si_{\al,\beta}^{k-1}- \Si_{\al,\beta}^{k-1} \wedge \Si_{\ga,\de}^l\bigr) -
\bigl( \Si_{\al,\beta}^k - \Si_{\al,\beta}^k \wedge \Si_{\ga,\de}^l\bigr) 
\eeq
Recalling \eqref{cou2-2} and setting $l=0$ in \eqref{cou-diff} gives \eqref{cou-alt4}.
Inserting \eqref{cou-diff} in \eqref{cou2}, and using identity \eqref{idw} again with $a= \bigr(\Si_{\al,\beta}^{k-1}- \Si_{\al,\beta}^{k-1} \wedge \Si_{\ga,\de}^l\bigr) -
\bigl( \Si_{\al,\beta}^k - \Si_{\al,\beta}^k \wedge \Si_{\ga,\de}^l\bigr)$, $b=\Si_{\al,\beta}^k- \Si_{\al,\beta}^k \wedge \Si_{\ga,\de}^l$ and $c=\Ga_{\ga,\de}^l$ gives
\beqq
G_{\al,\beta;\, \ga,\de}^{k;\, l} &=&
\bigl[\bigr(\Si_{\al,\beta}^{k-1}- \Si_{\al,\beta}^{k-1} \wedge \Si_{\ga,\de}^l\bigr) -
\bigl( \Si_{\al,\beta}^k - \Si_{\al,\beta}^k \wedge \Si_{\ga,\de}^l\bigr)\bigr]\\
&&\qquad\qquad\wedge
\bigl[\Ga_{\ga,\de}^l - \Ga_{\ga,\de}^l\wedge (\Si_{\al,\beta}^k- \Si_{\al,\beta}^k \wedge \Si_{\ga,\de}^l)\bigr]\\
&=& \bigr(\Si_{\al,\beta}^{k-1}- \Si_{\al,\beta}^{k-1} \wedge \Si_{\ga,\de}^l\bigr) \wedge \Ga_{\ga,\de}^l -
\bigr(\Si_{\al,\beta}^k- \Si_{\al,\beta}^k \wedge \Si_{\ga,\de}^l\bigr)\wedge \Ga_{\ga,\de}^l
\eeqq
which is \eqref{cou-alt2}.
\epr

\noindent 
{\bf Proof of Proposition \ref{princ}.}
 
\bprm
Let $(\al,\beta, \ga,\de)\in X^4$ and  positive $k,l$.
Using \eqref{cou2-2}, \eqref{cou-alt1} and a telescopic argument,
partial sums of coupling rates read 
\beqq
\sum_{l'=0}^l G_{\al,\beta;\, \ga,\de}^{k;\, l'}=
G_{\al,\beta;\, \ga,\de}^{k;\, 0}+
\sum_{l'=1}^l G_{\al,\beta;\, \ga,\de}^{k;\, l'}=
\Ga_{\al,\beta}^{k} -  \Ga_{\al,\beta}^{k} \wedge
\bigl( \Si_{\ga,\de}^l- \Si_{\al,\beta}^k \wedge \Si_{\ga,\de}^l
\bigr)
\eeqq
In the limit $l\to\infty$, $ \Si_{\ga,\de}^l\to 0$  by Assumption \eqref{test_H1'}, thus
\be\label{2nd_rhs_1}
\lim_{l\to\infty}\Ga_{\al,\beta}^{k} \wedge
\bigl( \Si_{\ga,\de}^l- \Si_{\al,\beta}^k \wedge \Si_{\ga,\de}^l
\bigr) =0
\ee
This gives the correct left marginal for all $k>0$, 
\be\label{correct-left-marg}
\sum_{l'\ge 0} G_{\al,\beta;\, \ga,\de}^{k;\, l'}=
G_{\al,\beta;\, \ga,\de}^{k;\, 0}+
\sum_{l'> 0} G_{\al,\beta;\, \ga,\de}^{k;\, l'}= \Ga_{\al,\beta}^{k}
\ee
The right marginal can be treated in a similar way using \eqref{cou2-1} and \eqref{cou-alt2}. We get
\be\label{2nd_rhs_2}
\lim_{k\to\infty}\bigr(\Si_{\al,\beta}^k- \Si_{\al,\beta}^k \wedge \Si_{\ga,\de}^l\bigr)\wedge \Ga_{\ga,\de}^l =0
\ee
thus, for all $l>0$,
\be\label{correct-right-marg}
\sum_{k'\ge 0} G_{\al,\beta;\, \ga,\de}^{k';\, l}= 
G_{\al,\beta;\, \ga,\de}^{0;\, l}+
\sum_{k'> 0} G_{\al,\beta;\, \ga,\de}^{k';\, l}=
\Ga_{\ga,\de}^{l}
\ee 
\eprm

Proofs of Propositions \ref{altprinc2} and  \ref{altprinc} rely on the following lemmas.

\bl\label{lemma-altprin2a}
The coupling rates defined by \eqref{cou2}--\eqref{cou2-3} satisfy the recursion relations \eqref{coupl2}--\eqref{coupl5}.
\el

\bpr
Equations \eqref{coupl3}--\eqref{coupl4} are verified by  \eqref{cou2}--\eqref{cou2-2} as a direct consequence of \eqref{correct-left-marg}--\eqref{correct-right-marg} (and  \eqref{cou2-3} is identical to
\eqref{coupl5}).
Equation  \eqref{coupl2} requires a partial resummation of the coupling rates :
using \eqref{cou-alt1}, \eqref{2nd_rhs_1} on the one hand and
\eqref{cou-alt2}, \eqref{2nd_rhs_2} on the other hand gives
\be\label{2nd_rhs_3}
\sum_{l'>l} G_{\al,\beta;\, \ga,\de}^{k;\, l'} =
\,\Ga_{\al,\beta}^{k}\wedge 
\bigl( \Si_{\ga,\de}^{l} - \Si_{\al,\beta}^{k} \wedge\Si_{\ga,\de}^{l} \bigr)
\ee
for $k>0$ and $l\ge 0$, and
\be\label{2nd_rhs_4}
\sum_{k'>k} G_{\al,\beta;\, \ga,\de}^{k';\, l} =
\bigl( \Si_{\al,\beta}^{k} - \Si_{\al,\beta}^{k} \wedge\Si_{\ga,\de}^{l} \bigr) 
\wedge  \Ga_{\ga,\de}^{l}
\ee
for $k\ge 0$ and $l>0$. 
Both right hand sides above appear in  \eqref{cou2}; substituting them by the respective left hand side of \eqref{2nd_rhs_3} and \eqref{2nd_rhs_4}  in \eqref{cou2}  gives \eqref{coupl2}.
\epr
\bl\label{lemma-altprin2l}
For all $(\al,\beta, \ga,\de)\in X^4$, any solution of \eqref{coupl2}--\eqref{coupl5} satisfies for all $(k,l)\in \N\times \N$,
\beq\label{lemma-altprin2la}
\Si_{\al,\beta}^{k} \ge \Si_{\ga,\de}^{l} \Longrightarrow \forall k'\le k,\forall l'>l, G_{\al,\beta;\, \ga,\de}^{k';\, l'} =0 \\
\label{lemma-altprin2lb}
\Si_{\al,\beta}^{k} < \Si_{\ga,\de}^{l} \Longrightarrow  \forall k'> k,\forall l'\le l, 
G_{\al,\beta;\, \ga,\de}^{k';\, l'} =0 
\eeq
\el

\bpr
Inserting \eqref{coupl3}--\eqref{coupl4} into \eqref{coupl2} writes
\beq
\bigl( \sum_{0\le l'\le l} G_{\al,\beta;\, \ga,\de}^{k;\, l'} \bigr)
 \wedge 
\bigl( \sum_{0\le k'\le k} G_{\al,\beta;\, \ga,\de}^{k';\, l} \bigr)
=G_{\al,\beta;\, \ga,\de}^{k;\, l} \nonumber
\eeq
Substracting $G_{\al,\beta;\, \ga,\de}^{k;\, l}$ on both sides gives
\be\label{lemma-altprin2ld}
\bigl( \sum_{0\le k'<k} G_{\al,\beta;\, \ga,\de}^{k';\, l} \bigr)
\wedge \bigl( \sum_{0\le l'<l} G_{\al,\beta;\, \ga,\de}^{k;\, l'} \bigr)=0 
\ee
On the other hand, we have
\beq
\Si_{\al,\beta}^{k} - \Si_{\ga,\de}^{l} 
&=& \sum_{k'> k \, ; \, l'\ge 0 } G_{\al,\beta;\, \ga,\de}^{k';\, l'} 
- \sum_{k'\ge 0 \, ; \,  l' > l } G_{\al,\beta;\, \ga,\de}^{k';\, l'}\nonumber\\
&=& \sum_{k'> k  \, ; \,  l'\le l } G_{\al,\beta;\, \ga,\de}^{k';\, l'} 
- \sum_{k'\le k  \, ; \,  l' > l } G_{\al,\beta;\, \ga,\de}^{k';\, l'}
\label{lemma-altprin2lc}
\eeq

Let $\Si_{\al,\beta}^{k} \ge \Si_{\ga,\de}^{l} $. Suppose that  the first sum in \eqref{lemma-altprin2lc} is zero;
by non-negativity of the coupling rates, all terms in the second sum in \eqref{lemma-altprin2lc} are zero, which gives  \eqref{lemma-altprin2la}.
Now suppose that the first sum in \eqref{lemma-altprin2lc} is not zero. Then  there 
are $k^*>k$, $l^*\le l$ such that $G_{\al,\beta;\, \ga,\de}^{k^*;\, l^*} > 0$.  Thus, for all $l''> l$,
we have
\be
\sum_{0\le l'<l''} G_{\al,\beta;\, \ga,\de}^{k^*;\, l'} >0\nonumber
\ee
and by \eqref{lemma-altprin2ld} applied to each  $l''$
\be
\sum_{0\le k'<k^*} G_{\al,\beta;\, \ga,\de}^{k';\, l''} =0\nonumber
\ee
By non-negativity of the coupling rates, all the elements in the above sums are zero; in particular,
since $k< k^*$, \eqref{lemma-altprin2la} holds.

We skip a similar derivation for  \eqref{lemma-altprin2lb}.
\epr

\bl\label{lemma-altprin2b}
The rates defined by \eqref{altprinc2-sola}--\eqref{altprinc2-solb} are the unique solution of the recursion relations \eqref{coupl2}--\eqref{coupl5}.
\el

\bpr
It consists in an explicit resolution of  \eqref{coupl2}--\eqref{coupl5}. We first prove that any solution is zero outside $\mathcal P_{\al,\beta;\, \ga,\de}$, that is \eqref{altprinc2-solb}.
Let $(k,l)\notin\mathcal P_{\al,\beta;\, \ga,\de}$ and define $i=k+l$ and $(k_i,l_i)$ as in \eqref{eq2.40}. 
Notice that by construction $k_i + l_i =i$ (hence $k\not= k_i$, $l\not= l_i$), $\{k_0,k_1,\cdots,k_i\}=\{0,1,\cdots,k_i\}$ and 
$\{l_0,l_1,\cdots,l_i\}=\{0,1,\cdots,l_i\}$. 

Suppose first that $k< k_i$. Let $i'$ denote the greatest element of   $\{j\ge 0 : k_j=k\}$. Hence $k_{i'+1}= k_{i'}+1$ and by   \eqref{eq2.40}, we have
$\Si_{\al,\beta}^{k_{i'}} \ge \Si_{\ga,\de}^{l_{i'}}$.  Since $k=k_{i'}$ and $l> l_i\ge l_{i'}$,
$G_{\al,\beta;\, \ga,\de}^{k;\, l} =0$ by \eqref{lemma-altprin2la}.

Suppose now $k> k_i$. Then $l<l_i$ and we denote by  $i''$  the greatest element of $\{j\ge0: l_j=l\}$. Hence $l_{i''+1}= l_{i''}+1$ and by  \eqref{eq2.40}, we have
$\Si_{\al,\beta}^{k_{i''}} < \Si_{\ga,\de}^{l_{i''}}$. Since $k> k_i\ge k_{i''}$ and $l= l_{i''}$,  
$G_{\al,\beta;\, \ga,\de}^{k;\, l} =0$ by \eqref{lemma-altprin2lb}.

We now are able to compute explicitely the values of the rates on $\mathcal P_{\al,\beta;\, \ga,\de}$.

Take $i\ge 0$ and $(k_i,l_i)$ as in \eqref{eq2.40}. Suppose first that $\Si_{\al,\beta}^{k_i} \ge \Si_{\ga,\de}^{l_i}$.
Then $k_j>k_i$ for all $j>i$ and thus, using  \eqref{altprinc2-solb},
\beq\nonumber
\sum_{j>i} G_{\al,\beta;\, \ga,\de}^{k_j;\, l_j}
= \sum_{k>k_i} \sum_{l\ge 0 } G_{\al,\beta;\, \ga,\de}^{k;\, l}
= \sum_{k>k_i} \Ga_{\al,\beta}^{k}
= \Si_{\al,\beta}^{k_i}
\eeq

Suppose now that $\Si_{\al,\beta}^{k_i} < \Si_{\ga,\de}^{l_i}$.
Then $l_j>l_i$ for all $j>i$ and we have
\beq\nonumber
\sum_{j>i} G_{\al,\beta;\, \ga,\de}^{k_j;\, l_j}
= \sum_{l>l_i} \sum_{k\ge 0 } G_{\al,\beta;\, \ga,\de}^{k;\, l}
= \sum_{l>l_i} \Ga_{\al,\beta}^{k}
=\Si_{\ga,\de}^{l_i}
\eeq
Putting both cases together gives for all $i\ge 0$
\beq
\sum_{j>i} G_{\al,\beta;\, \ga,\de}^{k_j;\, l_j} =
\Si_{\al,\beta}^{k_i} \vee  \Si_{\ga,\de}^{l_i}\nonumber
\eeq
These relations lead to the expression \eqref{altprinc2-sola}: for all $i>0$,
\beq
G_{\al,\beta;\, \ga,\de}^{k_i;\, l_i} &=&
\sum_{j>i-1} G_{\al,\beta;\, \ga,\de}^{k_j;\, l_j}
- \sum_{j>i} G_{\al,\beta;\, \ga,\de}^{k_j;\, l_j}\nonumber\\
&=&
\bigl(\Si_{\al,\beta}^{k_{i-1}} \vee 
 \Si_{\ga,\de}^{l_{i-1}}\bigr)
- \bigl(\Si_{\al,\beta}^{k_i} \vee  \Si_{\ga,\de}^{l_i}\bigr)\nonumber
\eeq
\epr

\noindent {\bf Proof of Propositions \ref{altprinc2} and  \ref{altprinc}.}

\bprm
From Lemma \ref{lemma-altprin2a} the coupling rates defined by \eqref{cou2}--\eqref{cou2-3}
verify the recursion relations \eqref{coupl2}--\eqref{coupl5}, which, by Lemma \ref{lemma-altprin2b},
have a unique solution given by \eqref{altprinc2-sola}--\eqref{altprinc2-solb}. Therefore
expressions \eqref{cou2}--\eqref{cou2-3} and  \eqref{altprinc2-sola}--\eqref{altprinc2-solb} are identical.
\eprm

\noindent {\bf Proof of Proposition \ref{main}.}

\bprm
Using conditions  \eqref{inc2}--\eqref{dec2}, we now prove that the process is attractive.
We need to show that $G_{\al,\beta;\, \ga,\de}^{k;\, l} =0$ whenever
$(\al,\beta) \le (\ga,\de)$ and $(\al-k,\beta+k) \not\le (\ga-l,\de+l)$.
This last condition splits into two possible cases
\beq
l> \ga-\al +k\label{inq-1}\\
 k>\de -  \beta +l\label{inq-2}
\eeq
In the first case, using \eqref{inq-1}, non-negativity of the 
$\Ga$'s and  \eqref{dec2}, we write
\beqq
\,\Si_{\ga,\de}^{l-1} = \sum_{l'\ge l} \,\Ga_{\ga,\de}^{l'}
\le \sum_{l'> \ga-\al +k} \,\Ga_{\ga,\de}^{l'}
\le \sum_{k'> k}\,\Ga_{\al,\beta}^{k'} = \,\Si_{\al,\beta}^{k}
\eeqq
Since $\Si_{\ga,\de}^{l}\le\Si_{\ga,\de}^{l-1}$, one has both $\Si_{\ga,\de}^{l}\le\Si_{\al,\beta}^{k}$ and $\Si_{\ga,\de}^{l-1}\le\Si_{\al,\beta}^{k}$ 
and the two terms appearing between parentheses in expressions
\eqref{cou-alt1} for $k\not= 0$ or \eqref{cou-alt3} for $k=0$ are equal to zero. Hence
$G_{\al,\beta;\, \ga,\de}^{k;\, l}= 0.$

In the second case \eqref{inq-2}, one writes in a similar way, using \eqref{inc2}
\beqq
\Si_{\al,\beta}^k\le\Si_{\al,\beta}^{k-1} = \sum_{k'\ge k} \,\Ga_{\al,\beta}^{k'}
\le \sum_{k'> \de-\beta +l} \,\Ga_{\al,\beta}^{k'}
\le \sum_{l'> l}\,\Ga_{\ga,\de}^{l'} = \,\Si_{\ga,\de}^{l}
\eeqq
Inserting these inequalities into the expression for the coupling rate, that is
\eqref{cou-alt2} for $l > 0$ or \eqref{cou-alt4} for $l=0$,
one gets again $G_{\al,\beta;\, \ga,\de}^{k;\, l}= 0.$
 
In conclusion, when $(\al,\beta)\le(\ga,\de)$, we have
\[
G_{\al,\beta ;\, \ga,\de}^{k ;\, l} = 0 \hbox{ if }
l-k \not\in \{-(\de-\beta),\cdots,\ga-\al\}
\]
The case $(\al,\beta)\ge(\ga,\de)$ is treated 
by symmetry using Remark \ref{cou2-symm},
which yields \eqref{4.10}.
\eprm

\noindent {\bf Proof of Theorem \ref{discrep}.}

\bprm
We first prove that for all $(\al,\beta,\ga,\de)\in X^4$, the coupling rate
$G_{\al,\beta;\, \ga,\de}^{k;\, l}$ is zero if $k-l$ is outside some range depending only on $\al-\ga$ and $\beta-\de$.

{}From Proposition \ref{main}, we already know that
if $(\al,\beta),(\ga,\de)$ are  ordered, then \eqref{4.10} is satisfied.
We now show similar restrictions for non ordered pairs, for which
equations \eqref{inc2}--\eqref{dec2} lead also, indirectly, to  inequalities between partial sums of rates. 
Suppose for instance that  $\al>\ga$, $\beta<\de$. 
We insert an intermediate pair of values and write 
\beqq
(\al,\beta)\ge(\ga,\beta) \quad\hbox{ and }\quad (\ga,\de)\ge(\ga,\beta)
\eeqq
We apply equations \eqref{inc2}, \eqref{dec2} on both pairs of values and get two sets of 
inequalities valid for all $k\ge 0$,
\beqq
\sum_{k'>k}\,\Ga_{\ga,\beta}^{k'} \le \sum_{l'>k} \,\Ga_{\al,\beta}^{l'} \quad &;& \quad
\sum_{l'>\al-\ga+k}\,\Ga_{\al,\beta}^{l'} \le \sum_{k'>k} \,\Ga_{\ga,\beta}^{k'}\\
\sum_{k'>\de-\beta+k}\,\Ga_{\ga,\beta}^{k'} \le \sum_{l'>k} \,\Ga_{\ga,\de}^{l'} \quad &;& \quad
\sum_{l'>k}\,\Ga_{\ga,\de}^{l'} \le \sum_{k'>k} \,\Ga_{\ga,\beta}^{k'}
\eeqq
Combining them by pairs in order to eliminate the intermediate values $(\ga,\beta)$ gives
\beqq
\sum_{l'>k}\,\Ga_{\ga,\de}^{l'} &\le& \sum_{k'>k} \,\Ga_{\al,\beta}^{k'}\\
\sum_{k'>\de-\beta+\al-\ga+k}\,\Ga_{\al,\beta}^{k'} &\le& \sum_{l'>k} \,\Ga_{\ga,\de}^{l'}
\eeqq
which implies
\beq\label{in-disc1}
\Si_{\ga,\de}^l \le \Si_{\ga,\de}^{l-1} &\le& \Si_{\al,\beta}^{k} \hbox{ for all } l>k\\
\label{in-disc2}
\Si_{\al,\beta}^k \le \Si_{\al,\beta}^{k-1} &\le& \Si_{\ga,\de}^{l} \hbox{ for all } k>\de-\beta+\al-\ga+l
\eeq
Inserting \eqref{in-disc1} in \eqref{cou-alt1} (or \eqref{cou-alt3} if $k=0$)
and  \eqref{in-disc2} in \eqref{cou-alt2} (or \eqref{cou-alt4} if $l=0$), one gets
\be\label{4.14}
G_{\al,\beta ;\, \ga,\de}^{k ;\, l} = 0 \hbox{ if } \al>\ga, \beta<\de
\hbox{ and } k-l \not\in \{0,\cdots,\al-\ga+\de-\beta\}
\ee
We have also by symmetry through Remark \ref{cou2-symm}
\be\label{4.15}
G_{\al,\beta ;\, \ga,\de}^{k ;\, l} = 0 \hbox{ if } \al<\ga, \beta>\de
\hbox{ and } l-k \not\in \{0,\cdots,\ga-\al+\beta-\de\}
\ee
Thus,  putting together \eqref{4.10}, \eqref{4.14}, \eqref{4.15}, the coupling rates $G_{\al,\beta ;\, \ga,\de}^{k ;\, l}$ are nonzero only if the increments satisfy 
\beq\label{k-l...}
\begin{cases}
l-k \in \{-(\de - \beta) ,\cdots, (\ga - \al)\} &
\mbox{ if } \al\le \ga \mbox{ and } \beta \le \de\\
l-k \in \{-(\al-\ga),\cdots,(\beta-\de)\} &
\mbox{ if }  \al\ge\ga \mbox{ and }  \beta \ge \de\\
l-k \in \{-(\al-\ga)-(\de - \beta),\cdots,0\}  &
\mbox{ if }  \al>\ga \mbox{ and }  \beta < \de\\
l-k \in \{0,\cdots,(\ga - \al)+(\beta-\de)\}  &
\mbox{ if }  \al<\ga \mbox{ and }  \beta > \de
\end{cases}
\eeq
All four cases can be written together as the single statement
\eqref{crnz} or equivalently
\be\label{crnz2}
\bigl|l - k +\frac{1}{2} (\al-\ga) -\frac{1}{2} (\beta- \de) \bigr| \le
\frac{1}{2} \bigl|\al-\ga\bigr|+\frac{1}{2} \bigl|\beta- \de\bigr|
\ee
These restrictions induce in turn a control on the sign of $\Delta (\al,\beta, \ga,\de,k,l)$ through the 
inequality:
For all $a,b,x$ in $\R$ such that $|x| \le |a| +|b|$, 
\be\label{ineq+}
\bigl[a + b+x\bigr]^+ + \bigl[a + b - x\bigr]^+ \le 2 [a]^+ +  2 [b]^+
\ee
Inequality \eqref{ineq+} is trivial if its left hand side  is equal to 0.  When exactly one of the two quantities on its left hand side is nonzero, \eqref{ineq+} follows directly from the bound on $x$; otherwise, the value of the left hand side is independent of $x$ and trivially bounded by the right hand side.

We now apply \eqref{ineq+},
choosing (in view of \eqref{crnz2})
\be\label{ineq+abx}
a= \frac{1}{2} (\al-\ga) \qquad b = \frac{1}{2} (\beta- \de)
\qquad x= l - k +\frac{1}{2} (\al-\ga) -\frac{1}{2} (\beta- \de)
\ee
we get
\be\label{ineq++abx}
[(\al-k)-(\ga-l)]^+ + [(\beta+k)-(\de+l)]^+
\le[\al-\ga]^+ + [\beta-\de]^+
\ee
Hence, recalling Definition \eqref{Delta_neg} of $\Delta (\al,\beta, \ga,\de,k,l)$, 
 and \eqref{Delta_sym},
\be\label{Delta_neg-bis}
G_{\al,\beta ;\, \ga,\de}^{k ;\, l}\ne 0 \Lra
 \Delta (\al,\beta, \ga,\de,k,l)=\Delta (\beta, \al,\de,\ga,l,k)\le 0
\ee
To go from \eqref{Delta_neg-bis} to \eqref{gen_negatif},
notice first that 
 for $k=l$, $\Delta(\xi(x),\xi(y),\zeta(x),\zeta(y),k,l)=0$.
We have, since $f_{x,y}^+(\xi,\zeta)=f_{y,x}^+(\xi,\zeta)$, and by \eqref{Delta_sym}, 
\beq\nonumber
\overline{\calL}_{x,y}f_{x,y}^+(\xi,\zeta) &=&
\sum_{\al,\beta\in X} \sum_{\ga,\de \in X} \chi_{x,y}^{\al,\beta}(\xi)
\chi_{x,y}^{\ga,\de}(\zeta)\times\\\label{L_xy_1}
&&\sum_{k\not=l}
\Bigl( G_{\al,\beta;\, \ga,\de}^{k;\, l}(y-x)
\Delta(\al,\beta,\ga,\de,k,l)
+ G_{\beta,\al;\, \de,\ga}^{l;\, k} (x-y) 
\Delta (\beta, \al,\de,\ga,l,k)\Bigr)  \\\nonumber &=&
\sum_{\al,\beta\in X} \sum_{\ga,\de \in X} \chi_{x,y}^{\al,\beta}(\xi)
\chi_{x,y}^{\ga,\de}(\zeta)\times\\\label{L_xy_2}
&&\sum_{k\not=l}
\Bigl( G_{\al,\beta;\, \ga,\de}^{k;\, l}(y-x)
+ G_{\beta,\al;\, \de,\ga}^{l;\, k} (x-y) \Bigr)  \Delta(\al,\beta,\ga,\de,k,l)
\eeq

 By \eqref{Delta_neg-bis}, each term of the sum in the right hand side
 of \eqref{L_xy_1} is non-negative.
Since by  \eqref{Delta+=Delta-}, 
$\overline{\calL}_{x,y}f_{x,y}^+(\xi,\zeta)=
\overline{\calL}_{x,y}f_{x,y}^-(\xi,\zeta)$,
 \eqref{gen_negatif} is satisfied.
 
To derive \eqref{Delta_neg-detail}, we first check that equality  holds in \eqref{ineq+}
 either for $a+b=\pm (|a|+|b|)$ or for $x= \pm (|a|+|b|)$. Recalling \eqref{ineq+abx}, the first case corresponds to ordered pairs of values, 
$(\al,\beta)\le(\ga,\de)$ or  $(\al,\beta)\ge(\ga,\de)$, for all values of $k$ and $l$. The other one gives two new possibilities for non ordered pairs
of values: either $l-k = 0$ or $l - k = -(\al-\ga) + (\beta - \de)$.
Notice that in this last case, the values of the discrepancies 
at sites $x$ and $y$ are exchanged. 
\eprm

\noindent {\bf Proof of Lemma \ref{lem:noteod}.}

\bprm
Let $(x,y)\in S^2$,
$(\al,\beta,\ga,\de)\in X^4$, non-negative $k,l$ such that  $G_{\al,\beta;\, \ga,\de}^{k;\, l}(y-x) >0$.
Either $(\al,\beta),(\ga,\de)$ are  ordered and \eqref{noteod} reduces to the expression  \eqref{4.10} for attractiveness; or $(\al,\beta),(\ga,\de)$ are not ordered and \eqref{noteod} is Definition \ref{idsc-exchange} of absence of exchanges of discrepancies.
\eprm

\noindent {\bf Proof of Theorem \ref{macrostab-ppv}.}

\bprm
We fix $(\xi_0,\zeta_0)\in{\bf X}^2$
with $\sum_{y\in\Z}[\xi_0(y)+\zeta_0(y)]<+\infty$.  
We prove \eqref{eq:macrostab} by checking that 
\be\label{(3.43)}
\overline{\calL} S(\xi_t,\zeta_t) \le 0 \hbox{ for all } t\ge 0
\ee
We have
\beq
\overline{\calL} S(\xi_t,\zeta_t) 
&=& \sum_{x,y \in \Z,|x-y|=1} \sum_{\al,\beta\in X} \sum_{\ga,\de \in X} \chi_{x,y}^{\al,\beta}(\xi_t) \chi_{x,y}^{\ga,\de}(\zeta_t)
\sum_{k,l} G_{\al,\beta;\, \ga,\de}^{k;\, l}(y-x)\times \nonumber\\
&&\quad \bigl(\sup_{z\in\Z} \abs{s_z(S_{x,y}^k \xi_t,S_{x,y}^l \zeta_t)} - \sup_{z\in\Z} \abs{s_z(\xi_t,\zeta_t)}\bigr)\label{3.34}
\eeq
We note that
\be\label{id_hors_xy}
\forall z\notin [x\wedge y ,x\vee y), \,\,s_z(S_{x,y}^k \xi_t,S_{x,y}^l \zeta_t) = s_z(\xi_t,\zeta_t)
\ee
because, for $z < x\wedge y$, jumps involve only sites $x,y$ and because,  for $z\ge x\vee y$, the dynamics is conservative (cf. \eqref{Delta+=Delta-}). Due to the restriction to nearest neighbor interactions, $|y-x|=1$, so that $s_z$ may change only for $z=x\wedge y$, 
\be\begin{cases}
 s_z(S_{x,y}^k \xi_t,S_{x,y}^l \zeta_t) 
 &= s_z( \xi_t,\zeta_t) + l - k \hbox{ if } z=x=y-1\label{post-jump}\\
s_z(S_{x,y}^k \xi_t,S_{x,y}^l \zeta_t) 
&= s_z( \xi_t,\zeta_t) + k - l \hbox{ if } z=y=x-1
\end{cases}
\ee
and  we have
\beq\label{(3.38)}
\overline{\calL} S(\xi_t,\zeta_t) &=&
\sum_{x \in \Z} \sum_{\al,\beta\in X} \sum_{\ga,\de \in X} \chi_{x,x+1}^{\al,\beta}(\xi_t) \chi_{x,x+1}^{\ga,\de}(\zeta_t)
\sum_{k,l} \bigl(G_{\al,\beta;\, \ga,\de}^{k;\, l}(1) + G_{\beta,\al;\, \de,\ga}^{l;\, k}(-1)\bigr)\times\nonumber\\
&&\quad \bigl( \abs{s_x(\xi_t,\zeta_t)+l -k}\vee\sup_{z\not=x} \abs{s_z(\xi_t,\zeta_t)} \ - S(\xi_t,\zeta_t)\bigr)
\nonumber\\
&\le&
\sum_{x \in \Z} \sum_{\al,\beta\in X} \sum_{\ga,\de \in X} \chi_{x,x+1}^{\al,\beta}(\xi_t) \chi_{x,x+1}^{\ga,\de}(\zeta_t)
\sum_{k,l} \bigl(G_{\al,\beta;\, \ga,\de}^{k;\, l}(1) + G_{\beta,\al;\, \de,\ga}^{l;\, k}(-1)\bigr)\times\nonumber\\
&&\quad \bigl[ \abs{s_x(\xi_t,\zeta_t)+l -k}\ - S(\xi_t,\zeta_t)\bigr]^+
\eeq
Now we prove that if the coupling is increasing and does not allow exchanges of discrepancies, all terms in the last line
of \eqref{(3.38)} are zero. Let $(\xi_t,\zeta_t)$ be such that 
$\xi_t(x)=\al$,  $\xi_t(x+1)=\beta$, $\zeta_t(x)=\ga$,  $\zeta_t(x+1)=\de$. For all transition rates with  $G_{\al,\beta;\, \ga,\de}^{k;\, l}(1) + G_{\beta,\al;\, \de,\ga}^{l;\, k}(-1) >0$,
 we apply Lemma \ref{lem:noteod}: Indeed, either $G_{\al,\beta;\, \ga,\de}^{k;\, l}(1)>0$
hence
we have \eqref{noteod}, or $G_{\beta,\al;\, \de,\ga}^{l;\, k}(-1) >0$, which implies inequalities
\eqref{noteod} again (by symmetry the latter stay the same when exchanging $\al$ with $\beta$,
$\ga$ with $\de$, $k$ with $l$). They have the
 equivalent formulation
\be\label{noteod-equiv}
0 \wedge (\ga - \al) \wedge (\beta -\de) \le l-k \le 0 \vee (\ga - \al) \vee (\beta -\de)
\ee
Recalling \eqref{def:sx} we have 
\be\label{sx:valeurs}
 s_{x-1}(\xi_t,\zeta_t) = s_x(\xi_t,\zeta_t) + \ga - \al, \qquad 
  s_{x+1}(\xi_t,\zeta_t) = s_x(\xi_t,\zeta_t) + \beta -\de 
 \ee
Thus, by adding $s_x(\xi_t,\zeta_t)$ to each term of  \eqref{noteod-equiv} we get
\[
s_x(\xi_t,\zeta_t) \wedge s_{x-1}(\xi_t,\zeta_t) \wedge s_{x+1}(\xi_t,\zeta_t) \le s_x(\xi_t,\zeta_t)+ l-k 
\le s_x(\xi_t,\zeta_t) \vee  s_{x-1}(\xi_t,\zeta_t)  \vee s_{x+1}(\xi_t,\zeta_t)
\]
 which implies 
\be\label{(3.42)}
\abs{s_x(\xi_t,\zeta_t)+ l-k } \le 
\abs{s_x(\xi_t,\zeta_t)} \vee  \abs{s_{x-1}(\xi_t,\zeta_t)}  \vee \abs{s_{x+1}(\xi_t,\zeta_t)} \le S(\xi_t,\zeta_t)
\ee
and finally, inserting  \eqref{(3.42)} in  \eqref{(3.38)}, we get \eqref{(3.43)}.
\eprm

\noindent {\bf Proof of Remark \ref{rk:macrostab}.}

\bprm
We exhibit two counter-examples for which \eqref{eq:macrostab} is wrong: in 
{\em (i)}, interactions are not only nearest neighbor; in {\em (ii)}, \eqref{noteod} is not fulfilled. For both cases,  let $(\xi_0,\zeta_0)$ be such that for $x<y$, $\xi_0(x)=\al$,  $\xi_0(y)=\beta$, $\zeta_0(x)=\ga$,  $\zeta_0(y)=\de$ with $(\al,\beta),(\ga,\de)$ not ordered, and 
$\xi_0(z)=\zeta_0(z)=0$ for $z\notin[x-1,y]$. Let the first coupled transition occur at time $t>0$ between sites $x,y$ at rate $G_{\al,\beta;\, \ga,\de}^{k;\, l}(y-x)$.

{\em (i)}
Assume that $y=x+2$, $\xi_0(x+1)=\widetilde\al$,   $\zeta_0(x+1)=\widetilde\ga$, 
$\xi_0(x-1)=\zeta_0(x-1)=0$, with $\ga>\al,\beta>\de$,
$\ga-\al=\beta-\de$,
$\widetilde\al-\widetilde\ga=\ga-\al+1$, and that $l-k+1=\ga-\al$ in $G_{\al,\beta;\, \ga,\de}^{k;\, l}(2)>0$. Therefore property \eqref{id_hors_xy} is still valid, and \eqref{noteod-equiv} is satisfied. An analogous computation to \eqref{3.34}--\eqref{(3.38)} yields
\beqq
\sup_{z\in\Z} \abs{s_z(\xi_{t^-},\zeta_{t^-})}&=&\sup(\ga-\al,1,\ga-\al+1)\\
\sup_{z\in\Z} \abs{s_z(S_{x,x+2}^k \xi_{t^-},S_{x,x+2}^l 
\zeta_{t^-})}&=&\sup(1,\ga-\al+2,\ga-\al+1)\\
\overline{\calL} S(\xi_t,\zeta_t) 
\ge G_{\al,\beta;\, \ga,\de}^{k;\, l}(2)\bigl(\sup_{z\in\Z} \abs{s_z(S_{x,x+2}^k \xi_{t^-},S_{x,x+2}^l 
\zeta_{t^-})} &- &\sup_{z\in\Z} \abs{s_z(\xi_{t^-},\zeta_{t^-})}\bigr)>0\label{3.41}
\eeqq
thus \eqref{(3.43)} is wrong.

{\em (ii)} Assume that
 $y=x+1$, $\xi_0(x-1)=\ga$, $\zeta_0(x-1)=\al$, and that in $G_{\al,\beta;\, \ga,\de}^{k;\, l}(1)$,
\be\label{noteod:contrex}
0<\beta-\de \le \ga-\al < l-k \le (\ga-\al)+(\beta-\de)
\ee
so that \eqref{noteod} is not satisfied
while \eqref{crnz} is. We have by \eqref{id_hors_xy}, \eqref{post-jump}, \eqref{sx:valeurs}
\beqq
S(\xi_{t^-},\zeta_{t^-})&=&s_{x-1}(\xi_t,\zeta_t) =s_{x-1}(\xi_{t^-},\zeta_{t^-}) =  \ga - \al\\
s_{x+1}(\xi_t,\zeta_t)&=&s_{x+1}(\xi_{t^-},\zeta_{t^-}) =  \beta -\de\\
s_x(\xi_{t^-},\zeta_{t^-})&=&0;\quad
s_x(\xi_t,\zeta_t)= l-k=S(\xi_t,\zeta_t)
 \eeqq
so that \eqref{eq:macrostab} fails at time $t$ because of \eqref{noteod:contrex}.
\eprm

\noindent {\bf Proof of Proposition \ref{macrostab:cns}.}

\bprm
Let $(\al,\beta,\ga,\de)\in X^4$ and $k,l$ non-negative. 
Assuming \eqref{cs-sum_l}--\eqref{cs-sum_k}, we show that if $k,l$
do not satisfy \eqref{noteod}, then $G_{\al,\beta;\, \ga,\de}^{k;\, l}=0$.

If  $l>K$, \eqref{cs-sum_k} implies
that $\Si_{\al,\beta}^k \ge \Si_{\ga,\de}^{l-1}$  and thus 
$\Si_{\al,\beta}^k -\Si_{\al,\beta}^k\wedge\Si_{\ga,\de}^l \ge \Ga_{\ga,\de}^l$.
Therefore  $G_{\al,\beta;\, \ga,\de}^{k;\, l}=0$ by \eqref{cou2} (if $k>0$) or by  \eqref{cou2-1} (if $k=0$).

If  $k>L$, \eqref{cs-sum_l} implies
that $\Si_{\ga,\de}^l \ge \Si_{\al,\beta}^{k-1}$  and thus 
$\Si_{\ga,\de}^l -\Si_{\al,\beta}^k\wedge\Si_{\ga,\de}^l\ge \Ga_{\al,\beta}^k$.
Therefore  $G_{\al,\beta;\, \ga,\de}^{k;\, l}=0$ by \eqref{cou2} (if $l>0$) or by  \eqref{cou2-2} (if $l=0$).
\eprm

\section {Coupling and attractiveness on examples}\label{application_coupling}

In this section, we apply results of Section \ref{sec:2} to  examples,  first the classical ones, SEP-ZRP-MP,
then  a stick process (StP), 
and finally a two species exclusion model (S$_2$EP). 

\subsection{SEP-ZRP-MP} \label{sec:SEP-ZRP-MP-description}

According to notations in Sections \ref{sec:1} and \ref{sec:2}, let 
$p(x,y)=p(y-x)$ be an irreducible translation
invariant probability transition on $\Z^d$;  set the transition rates
$\Ga_{\al,\beta}^k(y-x)$
equal to 0 unless $k=1$, and write $\Ga_{\al,\beta}^1(y-x)=p(y-x)b(\al,\beta)$.
We specify below (cf. \eqref{oleinik-ea})  the necessary 
conditions on the rates and the proper reduction of the state space (when needed) which ensure that \eqref{gene} is 
the generator of a well defined Markov process  
(see \cite{L}, \cite{A}, \cite{C}) with a one parameter family of 
invariant and translation invariant product probability measures, 
denoted by $\{\mu_\rho\}_\rho$. The parameter 
$\rho$ is the average particles' density per site.
We assume that the transition rates satisfy 
condition ${\bf(A)}$ of Section \ref{sec:1}.

{\em (a) MP: The misanthropes process} (\cite{C}). 
There, $X=\N$,  $S=\Z^d$. We assume that the transition 
rates $b(\al,\beta),\al,\beta\in\N$
are Lipschitz functions w.r.t. the first variable, with a Lipschitz 
constant uniform in the second variable; the state space is reduced to
\be\label{existence-MP}
\Omega_0=\{\eta\in X^S; \sum_{x\in S}a(x)\eta(x)<+\infty\}
\ee
where $a$ is a function which satisfies $\sum_{y\in S}p(x,y)a(y)\le M a(x)$ for all $x\in S$, 
for some constant $M$ (see \cite{A}, whose construction for ZRP extends for MP to these conditions, weaker than in \cite[(1.7)]{C}). 

We quote \cite[(2.3), (2.4)]{C}, which  imply the existence of  
product probability measures $\{\mu_\rho\}_\rho$, with 
$\rho\in[0,+\infty)$: 
\beqq
&(M1)& \quad b(0,.)=0 \hbox{ and }b(\al,\beta)>0 \hbox{ for }\al>0  \hbox{ and }\beta\geq 0
 \hbox{  (irreducibility on $X^S$) } \\
 &(M2)&
 \begin{cases}
 \bigl(\forall z\in\Z^d,\,p(z)=p(-z)\bigr) \hbox{ or } b(\al,\beta)-b(\beta,\al)=b(\al,0)-b(\beta,0)\cr
  \displaystyle{
\frac{b(\al,\beta)}{b(\beta+1,\al-1)}=\frac{b(\al,0)b(1,\beta)}{b(\beta+1,0)b(1,\al-1)}}
\end{cases}
\eeqq
{\em (b) ZRP: The zero-range process} (\cite{A}) is a  
misanthropes process for which the rates depend on the 
first coordinate only, that is, $b(\al,\beta)=g(\al)$
(for more general existence conditions in the totally 
asymmetric case, we refer to \cite{BRSS}).
Condition $(M1)$ reduces  to $g(0)=0$ and 
\be\label{zrp-ir}
g(\al) >0 \hbox{ for all } \al>0
\ee
Condition $(M2)$ is verified in all cases and  $\rho\in[0,\lim_\al g(\al))$.

{\em (c) SEP: The simple exclusion process} (\cite{L1}, \cite{L}) 
has state space $\{0,1\}^{\Z^d}$. The only nonzero jump 
rate is $\Ga_{1,0}^1(y-x)=p(y-x)b(1,0)=p(y-x)$. 
Irreducibility for $p(x,y)$ is relaxed to
\be\label{pt_irred}
\forall x,y\in\Z^d,\forall t>0,\,p_t(x,y)+p_t(y,x)>0, 
\ee
where $p_t(x,y)$ are the transition probabilities for 
the continuous time Markov chain associated to $p(x,y)$.
The invariant product probability measures are Bernoulli, with 
$\rho=\mu_\rho(\eta(x)=1)\in[0,1]$.

In those three examples, \eqref{inc2}--\eqref{dec2} reduce to Condition 
${\bf(A)}$, which thus insures attractiveness. 
For ZRP, ${\bf(A)}$ reads: $g(\al)$ is a non-decreasing 
function of $\al$. For SEP, ${\bf(A)}$ is empty and does not impose anything.

 The coupling rates' formulas \eqref{cou2}--\eqref{cou2-1} reduce to  those of basic coupling,
 since the only non-zero possibility for $k$ and $l$ is $1$:
\beq\begin{cases}\label{cou-misa}
G_{\al,\beta;\, \ga,\de}^{1;\, 1} =& 
\Ga_{\al,\beta}^{1}\wedge\Ga_{\ga,\de}^{1}\\
G_{\al,\beta;\, \ga,\de}^{0;\, 1}=& 
\Ga_{\ga,\de}^{1}  -  \Ga_{\al,\beta}^{1}  \wedge 
\Ga_{\ga,\de}^{1}\\
G_{\al,\beta;\, \ga,\de}^{1;\, 0}=& 
\Ga_{\al,\beta}^{1} - \Ga_{\al,\beta}^{1} \wedge \Ga_{\ga,\de}^{1}
\end{cases}\eeq
In Section \ref{sec:IcapS_ex} we revisit the conditions previously stated to determine extremal invariant and translation invariant probability measures for SEP-ZRP-MP.

\subsection{A stick process (StP) } \label{sec:StP-description}
This example is motivated by its connections with a generalization of the discrete 
Hammersley-Aldous-Diaconis (HAD) process, a continuous-time 
Markov process  taking values in a subset of $\mathcal{X}=\{0,1\}^\Z$
(see  \cite{FM}).
Let $p(x,y)=p(y-x)$ be an irreducible nearest-neighbor translation
invariant probability transition on $\Z$,
with $p(1)+p(-1)=1$.
For each empty site $j$, the closest particle to its left (resp. right) 
jumps to site $j$ with rate $p(1)$ (resp. $p(-1)$). The formal generator of the generalized HAD process is
\be\label{a1}
  L_Hf(\widetilde\eta) = \sum_{j\in\Z}\sum_{z=\pm 1}p(z) [f(S_{x,x_z}^1\widetilde\eta) - f(\widetilde\eta)]
\ee
where, for all $\widetilde\eta\in \mathcal{X}$, $x\in \Z$ and $z=\pm 1$, 
\be\label{a2}
 x_z = x_{z}(\widetilde\eta, x )= z \max\{ y < z x :\, \widetilde\eta(zy ) = 1\}
 \ee
The existence of this dynamics is proved through a graphical construction, and to exclude the possibility of particles
escaping immediately to infinity, 
the state space is restricted to
\[
\widetilde{\mathcal{X}}=\left\{
\widetilde\eta\in{\mathcal{X}}:
\lim_{m\to+\infty}
m^{-1/2}\sum_{j=-m}^{-1}\widetilde\eta(j)=\lim_{m\to+\infty}
m^{-1/2}\sum_{j=1}^m\widetilde\eta(j)=\infty
\right\}
\]
Notice that particles cannot jump over each other, so they keep their
relative order. We denote by $(z_i(t),i\in\Z,t\ge 0)$ their positions,
with $z_0(0)$ the position of the first $\widetilde\eta$-particle initially to the right of (or at) the origin.

We study here a {\sl generalized stick process} (StP)  which is a discrete version with a
generalization to nearest neighbor jumps of the {\sl stick process}
defined in \cite{Se}
(for a totally asymmetric case).  It is  defined on the state space 
\[\Om_0 =\{\eta\in\N^\Z; \lim_{n\to-\infty}n^{-2}\sum_{x=n}^{-1}\eta(x)=
\lim_{n\to+\infty}n^{-2}\sum_{x=1}^{n}\eta(x)=0 \}\] 
 so that $\Om_0\subset X^S$ with $X=\N,S=\Z$;
it describes translation invariant nearest-neighbor jumps. Its transition rates are
\beq\label{HAD-rates}
\begin{cases}
\Ga_{\al,\beta}^k (z)=p(z){\bf 1}_{\{k\le\al\}}
&\hbox{ for }z=\pm 1,\\ 
\Ga_{\al,\beta}^k (z)=0&\hbox{ otherwise}
\end{cases}
\eeq
They satisfy \eqref{test_H1'}.

Similarly to the original models (see \cite{Se}), 
the generalized stick process is in bijection with 
the generalized discrete HAD process seen from a tagged particle:
If we assume that at time $0$, $\widetilde\eta_0(0)=1$ so that
$z_0(0)=0$, we have for all $t\ge 0,i\in\Z$,
\be\label{HAD-stick}
\eta_t(i)=z_{i+1}(t)-z_i(t)
\ee
In other words, holes (resp. positions of particles) of the HAD are 
turned into particles (resp. sites) for the stick model.
 
The geometric product probability measures $\{\mu_\rho,\,\rho\in[0,+\infty)\}$
with parameter $\rho(1+\rho)^{-1}$ 
are invariant for the generalized stick process ($\rho$
is the average particles' density per site). 
\bp\label{stick-coupl-att}
The stick process is attractive. The rates of the coupling generator $\overline\calL$ are, for all $(\al,\beta, \ga,\de)\in\N^4$, positive $k,\,l$, $z=\pm 1$,
\beq\label{cou2-HAD}
\hbox{        }&& G_{\al,\beta;\, \ga,\de}^{k;\, l}(z)=
p(z){\bf 1}_{\{k\le\al,l\le\ga,\al-\ga=k-l\}}\\
\hbox{        }&& G_{\al,\beta;\, \ga,\de}^{0;\, l}(z)=
p(z){\bf 1}_{\{l\le\ga-\al\}}\quad\mbox{ where }\ga>\al\label{cou2-1-HAD}\\
\hbox{        }&& G_{\al,\beta;\, \ga,\de}^{k;\, 0}(z)=
p(z){\bf 1}_{\{k\le\al-\ga\}}\quad\mbox{ where }\al>\ga\label{cou2-2-HAD}
\eeq
This increasing coupling does not allow exchanges of discrepancies. 
\ep
\bpr
Inequalities \eqref{inc2}--\eqref{dec2} are satisfied
for the rates \eqref{HAD-rates} of StP, which is thus attractive. 
The coupling rates formulas \eqref{cou2}--\eqref{cou2-2} write here
\eqref{cou2-HAD}--\eqref{cou2-2-HAD}. Finally, recalling Theorem \ref{discrep}, notice that the case $\Delta(\al,\beta,\ga,\de,k,l)= 0$, $(\al,\beta)$, $(\ga,\de)$ not ordered (line $2$ of \eqref{Delta_neg-detail}) is not possible; more generally  
 (cf Definition \ref{idsc-exchange}) exchanges of discrepancies are impossible
 since $|k-l|\le|\al-\ga|$ and thus $\al-\ga+l-k$ cannot change sign.
\epr
\br\label{originalHAD}
The rates \eqref{cou2-HAD}--\eqref{cou2-2-HAD} are very different from the basic coupling rates. For $k,l$ positive, $k=l$ in $G_{\al,\beta;\, \ga,\de}^{k;\, l}(z)$
is possible if and only if $\al=\ga$, and 
\[
G_{\al,\beta;\, \al,\de}^{k;\, k}(z)=p(z){\bf 1}_{\{k\le\al\}}
=\Ga_{\al,\beta}^{k}(z)=\Ga_{\al,\de}^{k}(z)
\]
\er
\subsection{The two species exclusion model (S$_2$EP)}\label{sec:SOS-description} 
This one dimensional model with charge
conservation  was introduced and studied in 
\cite{CDFG} (see also \cite{TS1}, \cite{TS2} for a full account on conservation laws in this model). Its state space is 
$\{-1,0,1\}^{\Z}$, with the  interpretation that 
a site on $\Z$ is either empty or occupied by at most one charge, positive or negative.
 A transition consists in a
simultaneous change of the charges on two
nearest-neighbor sites,  such that the total charge is conserved:
in a transition, the value at a site can change by 
$+1,\,-1,\,+2$ or $-2$, therefore the admissible values for 
$k$ are $1$ and $2$. 
Since in \cite{CDFG} the model is interpreted as a 
Solid-on-Solid interface, conventions there are different 
from the present ones, and the rates $\Ga_{\al,\beta}^k(1)$ 
(resp. $\Ga_{\al,\beta}^k (-1)$) here are equal to their 
$\Ga_{\al,\beta}^{-k}$ (resp. $\Ga_{\beta,\al}^k$). 
There are possibly ten different transition rates, namely:
$\Ga_{0,-1}^{1}(z)$, $\Ga_{1,-1}^{2}(z)$,
$\Ga_{1,-1}^{1}(z)$, $\Ga_{0,0}^{1}(z)$, $\Ga_{1,0}^{1}(z)$, $z=\pm 1$. We do not consider the case 
$\Ga_{0,0}^{1}(z)=\Ga_{1,-1}^{1}(z)=0$,  where there is a 
second conserved 
quantity (the number of particles involved in each transition).
We assume, in order to avoid degeneracies, that 
\beq\label{non-deg}
\begin{cases}
\Ga_{1,-1}^{1}(1)+\Ga_{1,-1}^1(-1)>0\\ 
\Ga_{0,0}^{1}(1)+\Ga_{0,0}^1(-1)>0
\end{cases}
\eeq
Under the condition
\beq\label{prodnoncons}
\frac{\Ga_{0,0}^1(-1) \Ga_{1,-1}^{1}(1) - \Ga_{0,0}^{1}(1) \Ga_{1,-1}^{1}(-1)}
{\Ga_{0,0}^1(1)+\Ga_{0,0}^{1}(-1)} &=&\nonumber\\
\sum_{z=\pm 1} &z&\bigl( \Ga_{1,-1}^{2}(z)+\Ga_{1,-1}^{1}(z)
- \Ga_{1,0}^{1}(z)-\Ga_{0,-1}^1(z)\bigr)
\eeq
the process has a one-parameter family of stationary 
product probability measures $\{\mu_\rho,\,\rho\in[-1,1]\}$, where
$\mu_1 = \de_{\underline 1}$ and
$\mu_{-1} = \de_{\underline {-1}}$  are the Dirac 
measures concentrating respectively on $\eta\equiv 1$ and 
$\eta\equiv -1$.  
See Section \ref{sec:6} below for more details.

\bp\label{prop:sos-exch}
1) The two species exclusion model is attractive 
if and only if the rates satisfy, for $z=\pm 1$,
\be
\label{attr-3}
\Ga_{1,-1}^{2}(z)\vee\Ga_{0,0}^{1}(z) \leq 
\Ga_{0,-1}^{1}(z)\wedge\Ga_{1,0}^{1}(z)\leq 
\Ga_{0,-1}^{1}(z)\vee\Ga_{1,0}^{1}(z) \leq
\Ga_{1,-1}^{1}(z)+\Ga_{1,-1}^{2}(z)
\ee
2) The increasing coupling for the two-species model does not allow
exchanges of  discrepancies if and only if, for $z=\pm 1$,
\be\label{sos-exch}
\Ga_{1,-1}^{2}(z)\le\Ga_{0,0}^{1}(z) 
\ee
\ep
\bpr
1. Inequalities  \eqref{attr-3} are just a compact form of
\eqref{inc2}--\eqref{dec2} for this model.

2. By Definition \ref{idsc-exchange}, we have exchanges of discrepancies for $(x,y)\in S^2$, $(\al,\beta,\ga,\de)\in X^4$ with $(\al-\ga) (\beta-\de) <0$, non-negative $k,l$ with $|k-l|>|\al-\ga|\vee |\beta-\de|$.
Since $|\al-\ga|\vee |\beta-\de|\ge 1$, the only possibility is $|k-l|=2$
hence $(k,l)\in\{(2,0),(0,2)\}$. Also $y-x=z$ has to belong to $\{-1,1\}$. For $\al<\ga,\beta>\de$ the only possible positive coupling rate is 
\[
G_{0,0;\,1,-1}^{0;\, 2}(z)=\Ga_{1,-1}^{2}(z)-\Ga_{0,0}^{1}(z)
\]
and for $\al>\ga,\beta<\de$ it is 
\[
G_{1,-1;\,0,0}^{2;\, 0}(z)=\Ga_{1,-1}^{2}(z)-\Ga_{0,0}^{1}(z)
\]
\epr

The rates of the coupling process are here rather different from those of basic coupling. Their explicit form is derived in Appendix (Section \ref{sec:A1}). 
\br\label{Fritz}
Hydrodynamic limits of non attractive 
two species exclusion models have been studied in \cite{FT} (where the dynamics has two conserved quantities), and in
\cite[Section 6]{FN}, where the rates are
$\Ga_{1,-1}^{2}(1)=2+\si,\Ga_{1,-1}^{2}(-1)=\si,\Ga_{0,-1}^{1}(1)
=\Ga_{1,0}^{1}(1)=1+\si,\Ga_{0,-1}^{1}(-1)=\Ga_{1,0}^{1}(-1)=\si,
\Ga_{0,0}^{1}(1)=\beta+\si,\Ga_{0,0}^{1}(-1)=\si,\Ga_{1,-1}^{1}(1)
=\si,\Ga_{1,-1}^{1}(-1)=\beta+\si$ (for suitable positive parameters $\beta,\si$), which do not satisfy  \eqref{attr-3}.
\er
We determine in Section \ref{sec:IcapS_ex} $\left(\mathcal {I}\cap\mathcal{S}\right)_e$ and in Section \ref{sec:6} the hydrodynamic limit for StP and attractive S$_2$EP 
under Assumption \eqref{sos-exch}.

\section{Invariant measures: Determination of 
$\left(\mathcal {I}\cap\mathcal{S}\right)_e$.}\label{sec:IcapS}

The main result in this  Section, Theorem \ref{th:(i cap s)_e}, is
the determination of the extremal invariant 
and translation invariant probability measures of
the process $(\eta_t)_{t\ge 0}$ with generator \eqref{gene}, assuming it is attractive.
This result, which does not depend on the existence of product invariant probability measures,
was already known in some cases (for instance SEP). 
The proof's skeleton is classical (and goes back to \cite{L1}), 
but the derivation of its main step (\eqref{mesure_ordonne} below) in a rather general setting is new; 
it sheds some light on the  relevant properties of the model, and will enable (see Section \ref{sec:IcapS_ex})
to relax the assumptions on the rates on the examples ZRP-MP. This
proof is linked with Section \ref{sec:2.3}.
\subsection{Preliminaries and statement of Theorem \ref{th:(i cap s)_e}}\label{prel-stat}
We first give {\em irreducibility conditions 
for the coupled process} $(\xi_t,\zeta_t)_{t\ge 0}$ of Proposition 
\ref{princ}, which are required assumptions for
Theorem \ref{th:(i cap s)_e}. Their 
goal is to ascertain that if a coupled
configuration contains a pair of discrepancies of opposite signs, there is a positive probability (for the coupled evolution) that it evolves into a locally  ordered coupled configuration. Whenever basic coupling can be used  ($k\le 1$), such event can be described as a motion of discrepancies towards neighboring sites where they eventually merge; as a consequence, irreducibility conditions are more intuitive and can be stated directly in terms of the marginal rates. Here more involved events may occur, such as (partial or total) exchanges of discrepancies (see Definition \ref{idsc-exchange}) and irreducibility conditions need to be expressed in terms of the coupling rates.

\bd\label{def:edge}
A pair of sites $(x,y)\in S^2$ is {\rm an edge for $\Ga$}, and we write
$(x,y)\in E_\Ga$,  if 
\be\label{G-neighbors}
\sum_{k}\sum_{\al,\beta \in X}\Bigl[\Ga_{\al,\beta}^k(y-x) + \Ga_{\beta,\al}^k (x-y)
\Bigr]>0
\ee 
\ed
\br\label{rk:edge1}
 Because the left hand side of  \eqref{G-neighbors}
is symmetric and the rates are translation invariant, we have
\[
(x,y)\in  E_\Ga \Llra (y,x)\in  E_\Ga 
 \Llra (0,y-x)\in  E_\Ga  
\]
\er
\br\label{rk:edge2}
Due to the definitions \eqref{cou2}--\eqref{cou2-2}
of the coupling rates,
the set of edges $ E_\Ga$ for $\Ga$  coincides with the set $E_G$ of edges for $G$, defined as 
\[
E_G=\Bigl\{ (x,y)\in S^2 : 
\sum_{k,l}\sum_{\al,\beta \in X}\sum_{\ga,\de\in X}\Bigl[G_{\al,\beta;\, \ga,\de}^{k;\, l}(y-x) + G_{\beta,\al;\, \de,\ga}^{l;\, k} (x-y)
\Bigr]>0
\Bigr\}
\]

\er
\bd\label{def:path_sites}
 For any pair of sites $(x,y)\in S^2$, the distance $d(x,y)$ is the number of edges in the shortest
{\rm path  of sites} $p$ between $x$ and $y$, 
\beqq
d(x,y) &=&\inf\bigl\{n:  \exists\, p\in \calP^n(x,y)\bigr\}\\
\calP^n(x,y)&=&\bigl\{p=(x_0,\cdots,x_n)\in S^{n+1}: x_0=x, x_n=y, \forall\,1\le i\le n,(x_{i-1},x_i)\in E_\Ga \bigr\}
\eeqq
\ed
\bd\label{def:finite_path}
For any edge $(x,y)\in E_\Ga$, a {\rm finite path of coupled transitions on $(x,y)$}   starting from $(\al',\beta',\ga',\de') \in X^4$ and
ending at $(\al'',\beta'',\ga'',\de'') \in X^4$ is
a finite sequence $\{(\al_r,\beta_r,\ga_r,\de_r)\}_{0\le r\le u}\subset(X^4)^{u+1}$, for $u\in \N\setminus \{0\}$, such that 
 $(\al_0,\beta_0,\ga_0,\de_0)=(\al',\beta',\ga',\de')$, $(\al_u,\beta_u,\ga_u,\de_u)=(\al'',\beta'',\ga'',\de'')$, and   
for all $0\le r\le u-1$, 
\beqq
\al_r- \al_{r+1}   = \beta_{r+1} - \beta_r =: k_r,  \qquad \ga_r -  \ga_{r+1} =  \de_{r+1} - \de_r =: l_r, \hbox{   and}\\
\hbox{either } (k_r\ge 0, l_r \ge 0,\, G_{ \al_r ,\beta_r;\, \ga_r,\de_r}^{k_r;\, l_r}(y-x) >0)
\\
\hbox{               or }  (k_r\le 0, l_r \le 0,\, G_{\beta_r,\al_r;\, \de_r,\ga_r}^{-k_r;\,- l_r} (x-y) > 0)
\eeqq
\ed
\br\label{rk:finite_path}
This notion is the natural `building block' to construct an event in which the width of discrepancies decreases. It plays the same role as a single coupled jump in the basic coupling construction. 
\er
We also define the set 
\[
X^4_D = \{(\al,\beta,\ga,\de)\in X^4 :
(\al<\ga,\beta>\de) \,\hbox{ or }(\al>\ga,\beta<\de)\}
\]
of values
corresponding to 
couples of discrepancies of opposite signs (cf. Definition \ref{def-disc}) located on two sites $x,y$ of $S$, that is 
$\xi(x)=\al,\xi(y)=\beta,\zeta(x)=\ga,\zeta(y)=\de$.

In order to cope with `strongly' asymmetric processes (in the sense that $\sum_k\Ga_{\al;\beta}^k(y-x)=0$ for some $(x,y)\in E_\Ga$ and $(\al,\beta)\in X^2$), we introduce the following notions:
\bd\label{def:wedge}
An {\rm oriented wedge} is an ordered triple $(x_0,x_1,x_2)\in X^3$ such that\hfill\break
$(x_0,x_1)\in E_\Ga$, $(x_1,x_2)\in E_\Ga$ and $x_0\not= x_2$.
\ed
\bd\label{def:W-path}
Let $W_\Ga$ be a set of oriented wedges. For any pair of sites $(x,y)\in S^2$, a {\rm $W_\Ga$-path} from $x$ to $y$ is a set of oriented wedges
$p_{W_\Ga} = (w_1,\cdots,w_{n-1})$, $n\ge 2$,  such that $w_i=(x_0^i,x_1^i,x_2^i)$ for all $1\le i\le n-1$,
$x_1^i=x_0^{i+1}$, $x_2^i=x_1^{i+1}$ for all $1\le i\le n-2$ and $x^1_0=x$, $x_2^{n-1}=y$. We denote by 
$\calP_{W_\Ga}^n(x,y)$ the set of all $W_\Ga$-pathes with $x$, $y$ as endpoints and $n-1$ wedges.
\ed
\bd\label{def:W-connected}
The set of sites $S$ is  {\rm $W_\Ga$-connected}  if $(S,E_\Ga)$ is a connected graph and for all $(x,y)\in S^2$ such that $d(x,y)\ge 2$, there exists a $W_\Ga$-path
from $x$ to $y$ or from $y$ to $x$.
\ed
\bd\label{def:d_W}
If $S$ is $W_\Ga$-connected, for all $(x,y)\in S^2$,  we define the quantity $d_{W_\Ga}(x,y)$ as
\beqq
d_{W_\Ga}(x,y)=
\begin{cases}
 d(x,y)& \text{ if } d(x,y)\le 1\\
\inf\bigl\{n:  \exists\, p\in \calP_{W_\Ga}^n(x,y)\bigr\} & \text{ otherwise }
 \end{cases}
 \eeqq
\ed
\br\label{rk:d-and-d_W}
$d_{W_\Ga}$ is a pseudo-distance since triangular inequality may not hold.
If $W_\Ga$ is taken as the set of all oriented wedges, $d(\cdot,\cdot)$ and $d_{W_\Ga}(\cdot,\cdot)$ coincide. In general, we have only
$d_{W_\Ga}(x,y)\ge d(x,y)$ since the existence of a $W_\Ga$-path with $n-1$ wedges,
 $p_{W_\Ga} = (w_1,\cdots,w_{n-1})$,  $w_i=(x_0^i,x_1^i,x_2^i)$ for all $1\le i\le n-1$,
implies the existence of a path of sites $p=(x_0^1,x_1^1,\cdots,x_1^{n-1},x_2^{n-1})$ with $n+1$ elements and the same endpoints. See in Examples \ref{ex:IC-SEP} cases where $d_{W_\Ga}(x,y)> d(x,y)$.
\er

The {\bf irreducibility condition (IC)} for the coupled process is: 

{\sl 
$\quad\,\,\,\, (o)$   $(S, E_\Ga)$ forms a connected lattice.

 For all $(x,y)\in E_\Ga$, 

$\quad\,\,\,\, (a)$  
for any couple of discrepancies 
$(\al,\beta,\ga,\de)\in X^4_D$, there is a finite path of coupled transitions on $(x,y)$ starting from $(\al,\beta, \ga,\de)$ along which  $f_{x,y}^+$ decreases;

$\quad\,\,\,\, (b)$ let $X_R(y-x)= X_L(x-y)$ be the set of values $\ve\in X$ such that 
for any discrepancy with values $(\al,\ga)$ located at $x$,  there is a finite path of coupled transitions on $(x,y)$ starting from
$(\al,\ve,\ga,\ve)$ which ends with a discrepancy on $y$; then either
$X_R(y-x) = X$ or $X_L(y-x) = X$.

\noindent
When $(b)$ is not satisfied, we replace $(b)$ by $(b')$: 

There exists a set of oriented wedges $W_\Ga$ such that

$\quad\,\,\,\, (b_0')$ $S$ is $W_\Ga$-connected;

$\quad\,\,\,\, (b_1')$ for all $w= (x_0,x_1,x_2)\in W_\Ga$, 
$X_R(x_1-x_0)\bigcup X_R(x_1-x_2) = X$;

$\quad\,\,\,\, (b_2')$ for all $w= (x_0,x_1,x_2)\in W_\Ga$, 
all $ \ve_1\notin X_R(x_1-x_0)$ and  all
$\ve_2\in X_R(x_2-x_1)$, there is a finite path of coupled transitions on  $(x_1,x_2)$ 
from $(\ve_1,\ve_2,\ve_1,\ve_2)$ to $(\ve_3,\ve_4,\ve_3,\ve_4)$ such that $\ve_3\in X_R(x_1-x_0)$.

}

\bt\label{th:(i cap s)_e}
If the process $(\eta_t)_{t\ge 0}$
with generator \eqref{gene} is attractive
and satisfies Assumption ${\bf (IC)}$, then 

 1) if the state space of $(\eta_t)_{t\ge 0}$
is compact, that is $X=\{a,\cdots,b\}$
for $(a,b)\in\Z^2$, we have ${(\ii\cap\s)}_e=\{\mu_\rho,\,\rho\in{\mathcal R}\}$, 
where $\mathcal R$ is a closed subset of $[a,b]$ containing 
$\{a,b\}$, and $\mu_\rho$ is a translation invariant probability 
measure on $\Om$ with $\mu_\rho[\eta(0)]=\rho$; the measures 
$\mu_\rho$ are stochastically ordered, that is,
$\mu_\rho\leq\mu_{\rho'}$ if $\rho\leq\rho'$;

 2)  if $(\eta_t)_{t\ge 0}$ possesses a one parameter family
$\{\mu_\rho\}_\rho$ of product invariant and translation 
invariant probability measures, we have ${(\ii\cap\s)}_e=\{\mu_\rho\}_\rho$.
\et

\subsection{Proof of Theorem \ref{th:(i cap s)_e}}\label{proof_th(i cap s)_e}
 We consider  
  case (2). For
 (1), we refer to \cite[Proposition 3.1]{BGRS2}, 
 which deals with a misanthrope type 
model. The only difference to cope 
with here is 
\eqref{mesure_ordonne} below, the difficult part common to both cases.

$\bullet$ {\em The method.}

Let $\mu\in{(\ii\cap\s)}_e$, and $\overline\nu$ be an 
extremal invariant and translation invariant probability measure for the coupled process $(\xi_t,\zeta_t)_{t\ge 0}$ 
with marginals $\mu$ and $\mu_\rho$, for some  value $\rho$ 
of the parameter (see
\cite[Lemma 4.4]{A} or \cite[Proposition VIII.2.14]{L}). 
We claim that
\be\label{mesure_ordonne}
\forall (x,y)\in S^2,\qquad
\overline\nu\{(\xi,\zeta): \xi(x)>\zeta(x) \,\,
\hbox{ and }\xi(y)<\zeta(y)\}=0
\ee
Then $\overline\nu\{(\xi,\zeta): \xi\leq\zeta \,\,
\hbox{ or }\xi\geq\zeta\}=1$ 
(see \cite[Corollary 4.8]{A}, \cite[Lemma VIII.3.2(b)]{L}),
so that $\mu=\mu_{\rho_0}$, where 
$\rho_0=\sup\{\rho, \mu\geq\mu_{\rho}\}$
(see \cite[Proposition 5.5]{A}, \cite[Proposition VIII.2.13 and 
Theorem VIII.3.9]{L}), which concludes the proof.

We are left to  prove 
 \eqref{mesure_ordonne},
that is, there cannot be discrepancies of opposite signs
under the coupled
measure $\overline\nu$, which we can also write
 \beq\label{mesure_ordonne-bis}
\forall (x,y)\in S^2, &&\forall (\al,\beta,\ga,\de) \in X^4_D
\mbox{  with  } \al>\ga,\beta<\de,\\\nonumber
&&\overline\nu\{(\xi,\zeta): \xi(x)=\al,\zeta(x)=\ga,\xi(y)=\beta,\zeta(y)=\de\}=0
\eeq
{}For this, we show that in a coupled evolution where the 
initial configuration has  at least on a pair of sites $(x,y)\in S^2$ a couple of discrepancies of opposite signs 
$(\al,\beta,\ga,\de) \in X^4_D$, then  with a positive probability,
the width of the positive discrepancy decreases to 0 in a 
(succession of) coupled transition(s). The latter exists
by Assumption ${\bf (IC)}$. 

We prove \eqref{mesure_ordonne-bis} by induction on 
the distance $d(x,y) =n$  (resp. $d_{W_\Ga}(x,y)=n$ under Assumptions ${\bf (IC},b')$) between the  discrepancies 
of opposite signs. 

To clarify the presentation,
we turn some claims into lemmas,  whose proofs are
postponed to next subsection.

$\bullet$ {\em The case $d(x,y)=1$.}

 By Theorem \ref{discrep},  the function $f_{x,y}^+=\psi_{x}^++\psi_y^+$ with 
$\psi_z^+(\xi,\zeta)=[\xi(z)-\zeta(z)]^+$
defined in \eqref{f_xy} satisfies \eqref{gen_negatif}, that is
$\overline{\calL}_{x,y}f_{x,y}^+(\xi,\zeta)\leq 0$.
Since $\overline\nu$ is invariant 
and translation invariant for the coupled process, 
\beqq
0 &= &\int \overline{\calL}\psi_0^+ d\overline\nu 
=\int \sum_{u\in S} \overline{\calL}_{u,0}\psi_0^+ d\overline\nu +  \int \sum_{v\in S} \overline{\calL}_{0,v}\psi_0^+ d\overline\nu \\
&=& \int \sum_{v\in S} \bigl( \overline{\calL}_{-v,0}\psi_0^+  + \overline{\calL}_{0,v}\psi_0^+ \bigr)d\overline\nu 
=\int  \sum_{v\in S}  \overline{\calL}_{0,v} \bigl(\psi_v^+  + \psi_0^+ \bigr) d\overline\nu \\
&=& \sum_{v\in S}  \int   \overline{\calL}_{0,v}  f_{0,v}^+ d\overline\nu \label{int_gen_zero}
\eeqq
Since each term is non-positive by \eqref{Delta_neg-detail}, $\int   \overline{\calL}_{0,v}  f_{0,v}^+ d\overline\nu =0$
for any site $v$, and by translation invariance we also get $\int   \overline{\calL}_{x,y}  f_{x,y}^+ d\overline\nu =0$. 
Using \eqref{L_xy_2}, we obtain 
\beq\nonumber
0&=& \overline\nu 
\{\xi(x)=\al,\zeta(x)=\ga,\xi(y)=\beta,\zeta(y)=\de\}\times \\
\label{vrai-disc-dec}
&&\sum_{k-l\notin\{0,(\al-\ga)+(\de-\beta)\}} \Bigl( G_{\al,\beta;\, \ga,\de}^{k;\, l}(y-x)
+ G_{\beta,\al;\, \de,\ga}^{l;\, k} (x-y) \Bigr)  \Delta(\al,\beta,\ga,\de,k,l)
\eeq
We conclude  with 
\bl\label{lem:n=1}
Under Assumption ${\bf (IC},a)$, equation
\eqref{vrai-disc-dec}
implies \eqref{mesure_ordonne-bis} for $d(x,y)=1$.
\el
$\bullet$ {\em The first steps to $n\ge 2$.}

 We suppose \eqref{mesure_ordonne-bis} satisfied for $d(x,y)\le n-1$ (resp.
 for $d_{W_\Ga}(x,y)\le n-1$ under Assumptions ${\bf (IC},b')$). 
Consider any path of sites $p=(x_0,\cdots,x_n)\in 
\calP^n(x,y)$, (resp. $W_\Ga$-path $p_{W_\Ga}=((x_0,x_1,x_2),(x_1,x_2,x_3),\cdots,(x_{n-2},x_{n-1},x_n))\in \calP_{W_\Ga}^n(x,y)$). For all $1\leq i\leq n-1$, by the induction hypothesis,
\beq\nonumber
\overline\nu\{(\xi,\zeta)&:& \xi(x_0)>\zeta(x_0),
 \xi(x_n)<\zeta(x_n),\xi(x_i)>\zeta(x_i)\}\\
\qquad&+&
 \overline\nu\{(\xi,\zeta): \xi(x_0)>\zeta(x_0),\xi(x_n)<\zeta(x_n),
 \xi(x_i)<\zeta(x_i)\}=0\label{first-step-intermediaire}
 \eeq
Therefore if for $(\ve_1,\cdots,\ve_{n-1})\in X^{n-1}$, we define the event
\beq\label{E_p}
{\mathcal E}_p(\ve_1,\cdots,\ve_{n-1}) =  \{(\xi,\zeta)&:&
\xi(x_0)=\al>\zeta(x_0)=\ga,\xi(x_n)=\beta<\zeta(x_n)=\de;\\\nonumber
&&(\forall\, 1\le i\le n-1,\,\xi(x_i)=\zeta(x_i)=\ve_i)\}
\eeq 
we now have
\beqq\label{detail-induction_milieu}
\overline\nu\{(\xi,\zeta):\xi(x_0)=\al>\zeta(x_0)=\ga&,&\xi(x_n)=\beta<\zeta(x_n)=\de\}\\
&=&\sum_{(\ve_1,\cdots,\ve_{n-1})\in X^{n-1}} 
\overline\nu({\mathcal E}_p(\ve_1,\cdots,\ve_{n-1}))
\eeqq
We show that the left hand side is zero by proving the existence of 
a path of sites 
$p\in \calP^n(x,y)$ such that for all  $(\ve_1,\cdots,\ve_{n-1})\in X^{n-1}$,
\be\label{milieu2}
\overline\nu({\mathcal E}_p(\ve_1,\cdots,\ve_{n-1})) = 0
\ee 
We fix arbitrary $(\ve_1,\cdots,\ve_{n-1})\in X^{n-1}$ and   $p=(x_0,\cdots,x_n)\in \calP^n(x,y)$ (resp. $p_{W_\Ga}=((x_0,x_1,x_2),(x_1,x_2,x_3),\cdots,(x_{n-2},x_{n-1},x_n))\in \calP_{W_\Ga}^n(x,y)$). 
\bl\label{lem:n_ge2}
Equation \eqref{milieu2}  is satisfied for $(\ve_1,\cdots,\ve_{n-1})\in X^{n-1}$ if either 
(a) $\ve_1\in X_R(x_1-x_0)$, or
(b) $\ve_{n-1}\in X_R(x_{n-1}-x_n)$.
\el
If moreover either  
$X_R(x_1-x_0)=X$ or $X_R(x_{n-1}-x_n)=X$ 
Lemma \ref{lem:n_ge2}(a) or Lemma \ref{lem:n_ge2}(b) implies that \eqref{milieu2} holds for any $(\ve_1,\cdots,\ve_{n-1})\in X^{n-1}$, 
and we are done. Otherwise, we have both $X_R(x_1-x_0)\ne X$ and $X_R(x_{n-1}-x_n)\ne X$.  Some more work is required, depending on which assumption we work
with:

$\bullet$ {\em Assumption ${\bf (IC}, b)$ is satisfied}.

Thus $X_R(x_0-x_1)= X_R(x_n-x_{n-1})= X$,
and we construct a new path of sites $p'=(x_0',\cdots,x_n')$ where 
\beq\label{path-b}
\begin{cases}
x_0'=x_0=x,\,x_n'=x_n=y\\
x_1'= x_0+y-x_{n-1}\\
x_{n-1}'= y+x_0-x_1\\
x'_{j+1}-x'_j=x_{j+1}-x_j&\hbox{for all }1\le j < n-1\\
\end{cases}
\eeq
By translation invariance (cf. Remark \ref{rk:edge1}),  $p'\in \calP^n(x,y)$. 
Since $X_R(x_1'-x)=X_R(y-x_{n-1})=X$ and $X_R(x'_{n-1}-y)=X_R(x-x_1)= X$, both cases of Lemma \ref{lem:n_ge2}  apply
for the path $p'$,  and
$\overline\nu({\mathcal E}_{p'}(\ve_1,\cdots,\ve_{n-1})) = 0$
 for all $(\ve_1,\cdots,\ve_{n-1})\in X^{n-1}$.
 
 $\bullet$ {\em Assumptions ${\bf (IC},b')$ are satisfied.}

We consider the path  of wedges $p_{W_\Ga}=((x_0,x_1,x_2),(x_1,x_2,x_3),\cdots,(x_{n-2},x_{n-1},x_n))\in
\calP_{W_\Ga}^n(x,y)$ and the induced path of sites $p=(x_0,x_1,\cdots,x_n)\in\calP_n(x,y)$.
Using ${\bf (IC},b_1')$, we have 
\be\label{b-onbothends}
X_R(x_1-x_0)\cup X_R(x_1-x_2)=X
\ee
When $n=2$,  we conclude by Lemma \ref{lem:n_ge2}. 

 When $n\ge 3$, we assume that 
\be\label{assume-n3bis}
\ve_1\notin X_R(x_1-x_0),\qquad\ve_{n-1}\notin X_R(x_{n-1}-x_n)
\ee
 (otherwise Lemma \ref{lem:n_ge2} implies \eqref{milieu2}). Then by \eqref{b-onbothends},
\be\label{by-b-onbothends}
\ve_1\notin X_R(x_1-x_0),\qquad\ve_{n-1}\in X_R(x_{n-1}-x_{n-2})
 \ee
\bl\label{lem:n=3-bis}
Under Assumption ${\bf (IC}, b_2')$,
Equation \eqref{milieu2}  is satisfied   $\forall (\ve_1,\cdots,\ve_{n-1})\in X^{n-1}$ such that 
$\ve_1\notin X_R(x_1-x_0)$ and $\ve_{n-1}\in X_R(x_{n-1}-x_{n-2})$.
\el 
By \eqref{by-b-onbothends}, Lemma \ref{lem:n=3-bis} exhausts the remaining cases for $n\ge 3$ and $\overline\nu({\mathcal E}_{p}(\ve_1,\cdots,\ve_{n-1})) = 0$
 for all $(\ve_1,\cdots,\ve_{n-1})\in X^{n-1}$.
 
 Theorem \ref{th:(i cap s)_e} is proven. 
\QED

\subsection{Proofs of  Lemmas}\label{proofs_lemmas}
All three  proofs share a common part that we state as another lemma.
\bl\label{lemma:common}
Let $(\widetilde x,\widetilde y)\in E_\Ga$ and a finite path of coupled transitions $\{(\al_r,\beta_r,\ga_r,\de_r)\}_{0\le r \le u}$
on $(\widetilde x,\widetilde y)$. Define a family of cylinder indicator functions $\{\phi_r\}_{0\le r \le u}$ on coupled configurations such
that for all $(\xi,\zeta)\in \Om^2$,
\beq\begin{cases}\label{def:phi-r}
\forall 0\le r\le u, 
&\phi_r(\xi,\zeta) \le \chi_{\widetilde x,\widetilde y}^{\al_r,\beta_r}(\xi) \chi_{\widetilde x,\widetilde y}^{\ga_r,\de_r}(\zeta)\\
\forall  0\le r\le u-1,  &
\phi_r(\xi,\zeta) = 
\begin{cases}
\phi_{r+1}(S_{\widetilde x,\widetilde y}^{k_r}\xi,S_{\widetilde x,\widetilde y}^{l_r}\zeta)&\text{ if } k_r+l_r >0\\
\phi_{r+1}(S_{\widetilde y,\widetilde x}^{-k_r}\xi,S_{\widetilde y,\widetilde x}^{-l_r}\zeta)&\text{ if } k_r+l_r <0
\end{cases}
\end{cases}
\eeq
Then $\int\phi_v d\overline\nu =0$ for $v=u-1$ or for $v=u$
implies $\int\phi_0 d\overline\nu =0$.
\el 
\bpr
The key idea is: 
if there is a succession of coupled transitions from an event
of $\overline\nu$-probability 0 to another event,
then the latter must also be of 
$\overline\nu$-probability 0.

We make a downwards induction on $0\le r\le v$. We start the induction with the hypothesis $\int\phi_v d\overline\nu =0$. Then we suppose 
that for $0\le r\le v-1$ we have $\int\phi_{r+1} d\overline\nu=0$ and  write 
 \beqq\nonumber
0 &=& \int \overline \calL \phi_{r+1} d\overline\nu\\
\nonumber
&=& \sum_{x',y'} \sum_{\al',\beta'} \sum_{\ga',\de'} \int \chi_{x',y'}^{\al',\beta'}(\xi)  \chi_{x',y'}^{\ga',\de'}(\zeta) \times\\\nonumber
&&\qquad \sum_{k,l} \bigl( G_{\al',\beta';\, \ga',\de'}^{k;\, l}(y'-x') 
\bigl[\phi_{r+1}(S_{x',y'}^k\xi,S_{x',y'}^l\zeta) - \phi_{r+1}(\xi,\zeta)\bigr]\bigr) d\overline\nu(\xi,\zeta)\\
\nonumber
&=& \sum_{x',y'} \sum_{\al',\beta'} \sum_{\ga',\de'} \int \chi_{x',y'}^{\al',\beta'}(\xi)  \chi_{x',y'}^{\ga',\de'}(\zeta) \times\\
&&\qquad \sum_{k,l}  G_{\al',\beta';\, \ga',\de'}^{k;\, l}(y'-x') 
\phi_{r+1}(S_{x',y'}^k\xi,S_{x',y'}^l\zeta)  d\overline\nu(\xi,\zeta)\label{calculdebase}
\eeqq
The integral over negative terms in the second line is zero as a function integrated
on an event of $\overline\nu$-probability 0 (by the induction hypothesis). The third line is a sum of nonnegative terms, so each of them is separately zero, in particular the term corresponding to $k_r,l_r$ given in Definition \ref{def:finite_path}, that is either 
\[
\begin{array}{l} k_r \ge 0,l_r \ge 0, 
x'=\widetilde x, y'=\widetilde y, k=k_r, l=l_r, \al'=\al_r, \beta'=\beta_r, \ga'=\ga_r, \de'=\de_r,\\
G_{\al_r,\beta_r;\, \ga_r,\de_r}^{k_r;\, l_r}(\widetilde y-\widetilde x) > 0;\qquad 
\chi_{\widetilde x,\widetilde y}^{\al_r,\beta_r}(\xi) \chi_{\widetilde x,\widetilde y}^{\ga_r,\de_r}(\zeta)\phi_{r+1} (S_{\widetilde x,\widetilde y}^{k_r}\xi,S_{\widetilde x,\widetilde y}^{l_r}\zeta)= \phi_{r}(\xi,\zeta)\\
\end{array}\]
or 
\[
\begin{array}{l}
k_r \le 0, l_r \le 0, x'=\widetilde y, y'=\widetilde x, k=-k_r, l=-l_r, \al'=\beta_r, \beta'=\al_r, \ga'=\de_r, \de'=\ga_r,\\
 G_{\beta_r,\al_r;\, \de_r,\ga_r}^{-k_r;\, -l_r}(\widetilde x-\widetilde y) > 0;\qquad 
 \chi_{\widetilde x,\widetilde y}^{\al_r,\beta_r}(\xi) \chi_{\widetilde x,\widetilde y}^{\ga_r,\de_r}(\zeta)
\phi_{r+1} (S_{\widetilde y,\widetilde x}^{-k_r}\xi,S_{\widetilde y,\widetilde x}^{-l_r}\zeta)= \phi_{r}(\xi,\zeta)\\
\end{array}\]
In both cases, this yields
\[
\int\phi_{r} d\overline\nu=0
\]
The  induction  is proven and Lemma \ref{lemma:common} follows.
\epr

\noindent
{\bf Proof of Lemma \ref{lem:n=1}.}
 
\noindent
We have fixed
 an arbitrary element $(\al,\beta,\ga,\de)$ of $X^4_D$. 
By Assumption ${\bf (IC},a)$, for any coupled configuration 
$(\xi,\zeta)$ such that $\xi(x)=\al$, $\xi(y)=\beta$, $\zeta(x)=\ga$, 
$\zeta(y)=\de$, there is a finite path of coupled transitions
$\{(\al_r,\beta_r,\ga_r,\de_r)\}_{0\le r \le u}$ on $(x,y)$ starting from  $(\al_0,\beta_0,\ga_0,\de_0)=(\al,\beta,\ga,\de)$ such that 
\be\label{u-correct}
f_{x,y}^+(S_{x,y}^{K_{u-1}}\xi,S_{x,y}^{L_{u-1}}\zeta) <  f_{x,y}^+(\xi,\zeta)
\ee
where
$K_{s}=\sum_{r=0}^{s} k_r,\,L_{s}=\sum_{r=0}^{s} l_r$ for $0\le s\le u-1$.
Without loss of generality, we  
assume $u$ to be the first step of the path 
of coupled transitions in which $f_{x,y}^+$ decreases, and,
by Theorem \ref{discrep},
$f_{x,y}^+(S_{x,y}^{K_r}\xi,S_{x,y}^{L_r}\zeta) =  f_{x,y}^+(\xi,\zeta)$
for all $0\le r < u-1$, so that 
$(\al_r,\beta_r,\ga_r,\de_r)\in X^4_D$, and equality \eqref{vrai-disc-dec}
is valid for $(\al_r,\beta_r,\ga_r,\de_r)$.
For $0\le r \le u-1$, we use Lemma \ref{lemma:common} for
\[
\varphi_r(\xi,\zeta) = \chi_{x,y}^{\al_r,\beta_r}(\xi) \chi_{x,y}^{\ga_r,\de_r}(\zeta)
\]
By our assumption \eqref{u-correct} on $u$ and Definition \ref{def:finite_path}, for $r=u-1$, we have either 
\beqq
k_r \ge 0,l_r \ge 0&&\,\mbox{and  }\qquad
 G_{\al_r,\beta_r;\, \ga_r,\de_r}^{k_r;\, l_r}(y-x) > 0, \\
\Delta(\al_r,\beta_r,\ga_r,\de_r,k_r,l_r) &=&f_{x,y}^+(S_{x,y}^{K_r}\xi,S_{x,y}^{L_r}\zeta) - f_{x,y}^+(S_{x,y}^{K_{r-1}}\xi,S_{x,y}^{L_{r-1}}\zeta)  
<0
\eeqq
or
\beqq
k_r\le 0,l_r \le 0&&\,\mbox{and  }\qquad
G_{\beta_r,\al_r;\, \de_r,\ga_r}^{-k_r;\, -l_r}(x-y) > 0, \\
\Delta(\beta_r,\al_r,\de_r,\al_r,-k_r,-l_r) &=&f_{x,y}^+(S_{x,y}^{K_r}\xi,S_{x,y}^{L_r}\zeta) - f_{x,y}^+(S_{x,y}^{K_{r-1}}\xi,S_{x,y}^{L_{r-1}}\zeta)
<0
\eeqq
In both cases, the sum in \eqref{vrai-disc-dec} applied to
$(\al_r,\beta_r,\ga_r,\de_r)$
contains a  negative term, thus  is  negative by Theorem \ref{discrep}. Hence, \eqref{vrai-disc-dec} applied to
$(\al_r,\beta_r,\ga_r,\de_r)$ gives 
\beqq
\int \varphi_{u-1} d\overline\nu = \overline\nu\{(\xi,\zeta):\,\xi(x)=\al_{u-1},\zeta(x)=\ga_{u-1},\xi(y)=\beta_{u-1},\zeta(y)=\de_{u-1}\} = 0
\eeqq
For $u> 1$, by Lemma \ref{lemma:common} with $v=u-1$ we get  
\beqq
\int \varphi_{0} d\overline\nu = \overline\nu\{(\xi,\zeta):\,\xi(x)=\al,\zeta(x)=\ga,\xi(y)=\beta,\zeta(y)=\de\} = 0
\eeqq
\QED

\noindent
{\bf Proof of Lemma \ref{lem:n_ge2}.}

\noindent
We only derive 
Lemma \ref{lem:n_ge2}(a), supposing that $\ve_1\in X_R(x_1-x)$ (The proof of Lemma \ref{lem:n_ge2}(b), when $\ve_{n-1}\in X_R(x_{n-1}-y)$,
is similar and omitted). 
Then, there is a finite path of coupled transitions  on $(x,x_1)$, 
$\{(\widetilde\al_r,\widetilde\beta_r,\widetilde\ga_r,\widetilde\de_r)\}_{0\le r \le u}$ such that
$ (\widetilde\al_0,\widetilde\beta_0,\widetilde\ga_0,\widetilde\de_0)=(\al,\ve_1,\ga,\ve_1)$, $\widetilde\beta_r=\widetilde\de_r$ 
for all $1\le r < u$ and $\widetilde\beta_u>\widetilde\de_u$ 
(by attractiveness $(\al,\ve_1)\ge(\ga,\ve_1)$ implies that
$(\widetilde\al_r,\widetilde\beta_r)\ge(\widetilde\ga_r,\widetilde\de_r)$).

For $0\le r \le u$, we define the cylinder indicator functions 
\beqq
\varphi^{(n)}_r(\xi,\zeta) &=&{\bf 1}\{(\xi,\zeta):\xi(x)=\widetilde\al_r,\zeta(x)=\widetilde\ga_r,\xi(x_1)=\widetilde\beta_r,\zeta(x_1)=\widetilde\de_r;\\ 
&&(\xi(x_i)=\zeta(x_i)=\ve_i,\forall\, 2\le i\le n-1 );\;
\xi(y)=\beta, \zeta(y)=\de\}
\eeqq
These functions fulfill  Conditions \eqref{def:phi-r} of Lemma \ref{lemma:common} for the finite path of coupled transitions 
$\{(\widetilde\al_r,\widetilde\beta_r,\widetilde\ga_r,\widetilde\de_r)\}_{0\le r \le u}$  on $(x,x_1)$.
In addition, we have
\beqq
\varphi^{(n)}_u(\xi,\zeta)\le  {\bf 1}\{(\xi,\zeta):\xi(x_1)< \zeta(x_1),\xi(y)>\zeta(y)\}
\eeqq
Since $d(x_1,y)=n-1$ (resp. $d_{W_\Ga}(x_1,y)=n-1$), 
\[
\overline\nu\{(\xi,\zeta): \xi(x_1)<\zeta(x_1),
 \xi(y)>\zeta(y)\}= 0
 \]
by the induction hypothesis (we assumed \eqref{mesure_ordonne-bis}
satisfied for $d(x,y)\le n-1$ under Assumption ${\bf (IC},b)$, resp.
 for $d_{W_\Ga}(x,y)\le n-1$ under Assumptions ${\bf (IC},b')$). Hence $\int \varphi^{(n)}_u d\overline\nu =0$. Applying Lemma \ref{lemma:common}
with $v=u$ implies then
\beqq
0 = \int \varphi^{(n)}_0 d\overline\nu=
 \overline\nu({\mathcal E}_p(\ve_1,\cdots,\ve_{n-1}) )
\eeqq
\QED

\noindent
{\bf Proof of Lemma \ref{lem:n=3-bis}.}

\noindent
We define 
\[
j_0=\inf\{j:1\le j\le n-1,\ve_j\in X_R(x_j-x_{j-1})\}
\]
 By \eqref{by-b-onbothends}, $j_0$ exists and
$1<j_0$. 
Using ${\bf (IC},b'_2)$, we make a downwards induction on 
$1\le m\le j_0-1$ to construct 
a sequence of finite pathes of coupled transitions
on $(x_m,x_{m+1})$,
$\{(\widetilde\ve_{m}^{(m,r)},\widetilde\ve_{m+1}^{(m,r)},\widetilde\ve_{m}^{(m,r)},\widetilde\ve_{m+1}^{(m,r)})\}_{0\le r \le u_m}$ 
such that 
\beqq
\begin{cases}
\widetilde\ve_{j_0-1}^{(j_0-1,0)}=\ve_{j_0-1}&\notin X_R(x_{j_0-1}-x_{j_0-2})\\
\widetilde\ve_{j_0}^{(j_0-1,0)}=\ve_{j_0}&\in X_R(x_{j_0}-x_{j_0-1})\\
\text{ and }\\
\widetilde\ve_{j_0-1}^{(j_0-1,u_{j_0-1})}&\in X_R(x_{j_0-1}-x_{j_0-2})\\
\end{cases}
\eeqq
 starts the induction for $m=j_0-1$ (thanks to the definition of $j_0$), and for each step $1\le m< j_0-1$ 
\beq\label{lem3-ind1}
\begin{cases}
\widetilde\ve_m^{(m,0)}=\ve_m&\notin X_R(x_m-x_{m-1})\\
\widetilde\ve_{m+1}^{(m,0)}=
\widetilde\ve_{m+1}^{(m+1,u_{m+1})}&\in X_R(x_{m+1}-x_{m})\\
\text{ and }\\
\widetilde\ve_{m}^{(m,u_m)}&\in X_R(x_m-x_{m-1})
\end{cases}
\eeq
For all $1\le m\le  j_0-1$ and all $0\le r \le u_m$, we define the cylinder indicator function
\beqq
\widetilde\varphi^{m}_r(\xi,\zeta) &=&{\bf 1}\{(\xi,\zeta):\xi(x)=\al,\zeta(x)=\ga, 
(\xi(x_i)=\zeta(x_i)=\ve_i,\,\forall\, 1\le i\le m-1);\\
&&\xi(x_m)=\zeta(x_m)=\widetilde\ve_m^{(m,r)},
\xi(x_{m+1})=\zeta(x_{m+1})=\widetilde\ve_{m+1}^{(m,r)},\\
&&(\xi(x_i)=\zeta(x_i)=\widetilde\ve_i^{(i-1,u_{i-1})},\,\forall\, m+2\le i\le j_0);\\
&&(\xi(x_i)=\zeta(x_i)=\ve_i,\,\forall\, j_0+1\le i\le n-1);
\xi(y)=\beta,\zeta(y)=\de\}
\eeqq
(with the expressions between parentheses possibly empty according to the values of $m$ and $j_0$)
which verify \eqref{def:phi-r} and in addition 
\beq\label{phi-it}
\widetilde\varphi^{m}_0(\xi,\zeta) = \widetilde\varphi^{m+1}_{u_{m+1}}(\xi,\zeta)
\eeq
for all $1\le m<  j_0-1$. By \eqref{lem3-ind1} for $m=1$,
we can now use Lemma \ref{lem:n_ge2}(a) to get
$\int \widetilde\varphi^{1}_{u_{1}} d\overline\nu =0$, then we apply Lemma \ref{lemma:common} for $v=u_1$ which gives 
$\int \widetilde\varphi^{1}_0 d\overline\nu =0$. This is the first step ($m=1$) of another induction to prove that $\int \widetilde\varphi^{m}_0 d\overline\nu =0$ for all $1\le m\le j_0-1$. To go from $m$ to $m+1$, we use \eqref{phi-it} then 
another application of Lemma \ref{lemma:common}
to $\{\widetilde\varphi^{m}_r\}_{0\le r\le u_m}$ for $v=u_m$.
For $m=  j_0-1$, we get 
 \beqq
0 = \int \widetilde\varphi^{j_0-1}_{0} d\overline\nu = \overline\nu({\mathcal E}_p(\ve_1,\cdots,\ve_{n-1}))
\eeqq
The Lemma is proven.
\QED

\section{ 
Irreducibility Condition and invariant measures on examples.
}\label{sec:IcapS_ex}

In this Section, we determine what  Assumption ${\bf (IC)}$ imposes on the rates of
the various examples of Section \ref{application_coupling}, 
so that Theorem \ref{th:(i cap s)_e} applies: For ZRP-MP,   Assumption ${\bf (IC)}$
 enables to extend the previously known results to probability transitions $p(\cdot,\cdot)$ such that 
 \be\label{qt_irred}
 q(x,y)=\frac{p(x,y) + p(y,x)}2\qquad\hbox{ satisfies \eqref{pt_irred}}
 \ee
 For SEP, Assumption ${\bf (IC})$ is proven equivalent to \eqref{pt_irred}.
 Results on Stick process (StP) and on the two-species exclusion process (S$_2$EP)
are new. 

\subsection{SEP}\label{SEP:IcapS}
 Theorem \ref{th:(i cap s)_e} for SEP is equivalent to \cite[Theorem 3.9(a)]{L}. Indeed
\bp\label{prop:IC-SEP}
For SEP, Condition \eqref{pt_irred}  for $p(.,.)$ is equivalent to Assumption ${\bf (IC)}$. 
\ep

\bpr
We assume Condition \eqref{pt_irred}, and check whether Assumption ${\bf (IC})$ is satisfied. The lattice being connected
by \eqref{pt_irred},
${\bf (IC}, o)$ holds; 

There exists two pairs of discrepancies of opposite signs
\beqq
X^4_D=\{(0,1,1,0),(1,0,0,1)\}
\eeqq
By symmetry and Remark \ref{cou2-symm}, we can restrict the analysis to the first element in $X^4_D$. 
By definition, for any edge $(x,y)\in E_\Ga$, we have:
\beqq
0 < G_{0,1,1,0}^{0;\, 1}(y-x)+G_{1,0,0,1}^{1;\, 0}(x-y)= p(y-x) +p(x-y)
\eeqq
thus at least one of the two terms in the last sum is non zero.

\noindent
$\bullet$ ${\bf (IC}, a)$. A path of coupled transition on $(x,y)$ starting from $(0,1,1,0)$ along which $f^+_{x,y}$ decreases  is given by 
\beqq
\begin{cases}
\{ (0,1,1,0), (0,1,0,1)\} & \hbox{ if }p(y-x) >0\cr
\{ (0,1,1,0), (1,0,1,0)\} & \hbox{ if } p(x-y) >0\cr
\end{cases}
\eeqq
so that Assumption ${\bf (IC}, a)$ is always satisfied.

\noindent
$\bullet$ ${\bf (IC}, b)$, ${\bf (IC}, b')$. The value of $X_R(y-x)$ can be read from (cf. \eqref{cou-misa})
\beqq
G_{0,0,1,0}^{0;\, 1}(y-x)&=& G_{1,0,0,0}^{1;\, 0}(y-x) = p(y-x)\\
G_{1,1,1,0}^{0;\, 1}(x-y)&=& G_{1,1,1,0}^{1;\, 0}(x-y) = p(x-y)
\eeqq
which gives
\beq\label{XR-SEP}
\begin{cases}
0 \in X_R(y-x) & \hbox{ if  and only if  } p(y-x) >0\cr
1 \in X_R(y-x) & \hbox{ if  and only if  } p(x-y) >0\cr
\end{cases}
\eeq
Thus Assumption ${\bf (IC}, b)$ requires 
\be\label{ICb-SEP}
\forall (x,y)\in E_\Ga,\,p(y-x)>0 \hbox{ and } p(x-y)>0
\ee
If this fails, we take
 \beq\label{SEP-WGa}
 W_\Ga=\{(x_0,x_1,x_2)\in S^3&:&
 p(x_1-x_0)p(x_2-x_1)>0\nonumber\\
&&\hbox{ and either }p(x_1-x_2)=0 \hbox{ or } p(x_0-x_1)p(x_1-x_2)>0\}
 \eeq
Thus \eqref{pt_irred} implies that for any pair $(x,y)\in S^2$, there is a $W_\Ga$-path from either $x$ to $y$, or from $y$ to $x$ and   ${\bf (IC}, b'_0)$ holds. 

Let $(x_0,x_1,x_2)\in W_\Ga$. By \eqref{XR-SEP}, $0\in X_R(x_1-x_0)$, $1\in X_R(x_1-x_2)$ and ${\bf (IC}, b'_1)$ holds.
Now suppose $X_R(x_1-x_0)\not= X$; then $p(x_0-x_1)=0$, $p(x_1-x_2)=0$ and $X_R(x_1-x_0)=X_R(x_2-x_1)=\{0\}$. Since $p(x_2-x_1)>0$, there 
is a finite path of coupled transitions from $(1,0,1,0)$ to $(0,1,0,1)$ on $(x_1,x_2)$ and ${\bf (IC}, b'_2)$ holds.
Assumption ${\bf (IC})$ is thus satisfied.

For the converse, we assume  ${\bf (IC})$. Assumption ${\bf (IC}, o)$ implies Condition \eqref{qt_irred}. 
By \eqref{ICb-SEP}, ${\bf (IC}, b)$ implies then \eqref{pt_irred}. If  Assumptions ${\bf (IC}, b')$ hold for some $W_\Ga$, let $(x_0,x_1,x_2)\in W_\Ga$ be such that $1\notin X_R(x_1-x_0)$. Assumption ${\bf (IC}, b_1')$
implies $1\in X_R(x_1-x_2)$, hence $p(x_2-x_1)>0$ thus $0\in X_R(x_2-x_1)$ by \eqref{XR-SEP};
then Assumption ${\bf (IC}, b_2')$ implies that $1\notin X_R(x_2-x_1)$
(hence $p(x_1-x_2)=0$ by \eqref{XR-SEP}),
since there is no possible coupled transition starting from $(1,1,1,1)$. Similarly, if
$0\notin X_R(x_1-x_0)$ (hence $p(x_1-x_0)=0$ by \eqref{XR-SEP}), by ${\bf (IC}, b_1')$
we have $0\in X_R(x_1-x_2)$ and $p(x_1-x_2)>0$ by \eqref{XR-SEP};
then by ${\bf (IC}, b_2')$, $0\notin X_R(x_2-x_1)$
(hence $p(x_2-x_1)=0$ by \eqref{XR-SEP}, thus $1\notin X_R(x_1-x_2)$ and by ${\bf (IC}, b_1')$
we have $1\in X_R(x_1-x_0)$ thus $p(x_1-x_0)>0$ by \eqref{XR-SEP}),
since there is no possible transition starting from $(0,0,0,0)$. Therefore
 the only possible choice for $W_\Ga$ is \eqref{SEP-WGa}.
 
 This implies \eqref{pt_irred}.
\epr

\bex\label{ex:IC-SEP}
Let $S=\Z^2$ be endowed with its canonical basis $(e_1,e_2)$.
We denote by $\underline 0$ the origin.

1. Let $p(e_1)=p_1,p(-e_1)=q_1,p(e_2)=p_2,p(-e_2)=q_2$, with 
$p_1,q_1,p_2,q_2$ all distinct, $p_1+q_1+p_2+q_2=1$. There is an edge in $E_\Ga$ between $\underline 0$ and its four neighoring sites hence Assumption ${\bf (IC}, o)$ holds, and  $p(y-x)p(x-y)>0$ for all $(x,y)\in E_\Ga$ so that Assumption ${\bf (IC}, b)$ is satisfied.

2. Let $p(2e_1)=p_1,p(-e_1)=q_1,p(e_2)=p_2$, with 
 $p_1+q_1+p_2=1$. 
 \beqq
 E_\Ga&=&\{(\underline 0,2e_1),(\underline 0,-e_1),(\underline 0,e_2)\}\\
W_\Ga&=&\{(\underline 0,-e_1,e_1),(\underline 0,2e_1,2e_1+e_2),(2e_1,e_1,\underline 0),(e_1-e_2,e_1,\underline 0),
(e_1-e_2,-e_2,\underline 0)\}
 \eeqq
 Notice first that the distances $d$ and $d_{W_\Ga}$ are distinct: 
 for $x=\underline 0,y=4e_1-e_2$ we have $d(x,y)=3,d_{W_\Ga}(x,y)=5$,
 corresponding to the following path of sites and  $W_\Ga$-path
 \beqq
 p&=&(\underline 0,2e_1,2e_1-e_2,4e_1-e_2)\\ 
p_{W_\Ga}&=&((\underline 0,e_1,2e_1),(e_1,2e_1,2e_1-e_2),(2e_1,2e_1-e_2,3e_1-e_2),\\&&\quad(2e_1-e_2,3e_1-e_2,4e_1-e_2))
  \eeqq
  There is an edge in $E_\Ga$ between $\underline 0$ and its four neighoring sites hence Assumption ${\bf (IC}, o)$ holds.
  There exists a  $W_\Ga$-path between $\underline 0$ and each of its 4 second neighbors, $e_1+e_2,-e_1+e_2,-e_1-e_2,e_1-e_2$ so that ${\bf (IC}, b'_0)$ holds. By \eqref{XR-SEP}, ${\bf (IC}, b'_1)$ is satisfied, and so is ${\bf (IC}, b'_2)$ (by inspection of all elements of $W_\Ga$).
  
\eex

\subsection{ZRP}\label{ZRP:IcapS}
Applied to this case, Theorem \ref{th:(i cap s)_e} generalizes  \cite[Theorem 1.9]{A}, since
\bp\label{prop:IC-ZRP}
For ZRP, Assumption ${\bf (IC)}$ is equivalent to \eqref{qt_irred} and \eqref{zrp-ir}.
\ep
\bpr
Because the transition rates have a product form and  $g(\cdot)$ is non decreasing (by attractiveness), for all $(x,y)\in E_\Ga$  necessarily $p(x,y) + p(y,x)>0$, and for any $\ga>0$ the existence of a coupled transition on $(x,y)$ from $(0,\ga,\ga,0) \in X_D^4$ reads
\beqq
\bigl(  p(x,y) + p(y,x)\bigr) g(\ga) > 0
\eeqq
that is Assumption ${\bf (IC}, a)$ implies \eqref{zrp-ir},
 the usual irreducibility condition on $g(.)$.  We now verify 
 that  \eqref{qt_irred} and \eqref{zrp-ir} imply Assumption ${\bf (IC)}$.
 
\noindent
$\bullet$ ${\bf (IC}, a)$. Consider $(x,y)\in E_\Ga$ and a pair of discrepancies $(\al,\beta,\ga,\de)\in X_D^4$ on $(x,y)$ such that $\al< \ga,\beta>\de$ (the other case being treated by reversing the roles of the two marginals).  A path of coupled transitions on $(x,y)$ starting from $(\al,\beta,\ga,\de)$ along which $f_{x,y}^+$ decreases (by 1) is given by either
\beq\label{zrp-a-path1}
\bigl\{(\al,\beta,\ga,\de),\cdots,(0,\beta+\al,\ga-\al,\de+\al),(0,\beta+\al,\ga-\al-1,\de+\al+1)\bigr\}
\eeq
when $p(y-x) >0$ (with rates $p(y-x) $ times $g(\al),g(\al-1),\cdots,g(1),g(\ga-\al)$), or by
\beq\label{zrp-a-path2}
\bigl\{(\al,\beta,\ga,\de),\cdots,(\al+\de,\beta-\de,\ga+\de,0),(\al+\de+1,\beta-\de-1,\ga+\de,0)\bigr\}
\eeq
when  $p(x-y) >0$ (with rates $p(x-y) $ times $g(\de),g(\de-1),\cdots,g(1),g(\beta-\de)$). 

\noindent
$\bullet$ ${\bf (IC}, b)$. Consider $\al < \ga$ and $\ve \ge 0$. Suppose $p(y-x) >0$; then a path of coupled transitions on $(x,y)$ starting from $(\al,\ve ,\ga,\ve )$ which creates a discrepancy on $y$
is given by
\beq\label{zrp-b-path}
\bigl\{(\al ,\ve ,\ga ,\ve ),\cdots,(0,\ve +\al ,\ga -\al ,\ve +\al ),(0,\ve +\al ,\ga -\al -1,\ve +\al +1)\bigr\}
\eeq
(with rates $p(y-x) $ times $g(\al),g(\al-1),\cdots,g(1),g(\ga-\al)$),
which leads to $X_R(y-x)=X$. 
\epr

 \cite[Lemma 4.6]{A} constructs
the path of coupled transitions \eqref{zrp-a-path1} (resp. \eqref{zrp-a-path2})  to go from $(\al,\beta,\ga,\de)$ 
to $(0,\beta+\al,\ga-\al,\de+\al)$ when $p(y-x)>0$ 
(resp. to $(\al+\de,\beta-\de,\ga+\de,0)$ when $p(x-y)>0$). 
Lemma \ref{lem:n_ge2} corresponds to \cite[Lemma 4.7]{A}.

\subsection{MP}\label{MP:IcapS}
This is the third example with single particle jumps. 
Applied to this case, Theorem \ref{th:(i cap s)_e} generalizes \cite[Theorem 2.13]{C}: we do not require product invariant probability measures, that is \cite[(2.3), (2.4)]{C}, and we extend \cite[(1.7)]{C}.
\bp\label{prop:IC-MP}
For MP, Assumption ${\bf (IC)}$ is equivalent to either (i) $b(\ga,\al) > 0 $ for all $\ga >0$, $\al\ge 0$, and $p(\cdot,\cdot)$ satisfies \eqref{qt_irred}; or (ii) for all $(x,y)\in E_\Ga$, $p(x,y)p(y,x)>0$ and $b(\cdot,\cdot)$ may take some zero values
 along with
\be\label{positive-MP}
b(\ga,\al) > 0 \qquad \hbox{ for all } \ga >\al\ge 0
\ee
\ep
In \cite{C}, $(i)$ and $(ii)$ have to be simultaneously valid. 

\bpr
As for ZRP, necessarily $p(x,y) + p(y,x)>0$  for all $(x,y)\in E_\Ga$. 
For any $\ga>\al$ the existence of a coupled transition on $(x,y)$ from $(\al,\ga,\ga,\al) \in X_D^4$ reads
(we cannot reduce ourselves here to $\al=0$ since $b(\cdot,\cdot)$ depends on $\al,\ga$, and not only on  $\al$ like $g(.)$ for ZRP) 
\beqq
\bigl(  p(x,y) &+& p(y,x)\bigr)\bigl( b(\al,\ga)\wedge b(\ga,\al) +[b(\ga,\al)-b(\al,\ga)\wedge b(\ga,\al)]^+
\\&+&
[b(\al,\ga)-b(\al,\ga)\wedge b(\ga,\al)]^+\bigr)
=\bigl(  p(x,y) + p(y,x)\bigr) b(\ga,\al) > 0
\eeqq
and thus \eqref{positive-MP} is valid.
On the other hand, suppose that there is $(x,y)\in E_\Ga$ such that $p(x-y)=0$ and that $b(\ga_0,\de_0)=0$ for some pair $(\ga_0,\de_0)$
such that $0<\ga_0 \le\de_0$. By attractiveness, $ b(\al,\beta)=0$ for all $(\al,\beta)$ such that $\al<\ga_0$ and $\beta>\de_0$ and there is no path of coupled transitions on $(x,y)$ starting from
$(\al,\beta,\ga_0,\de_0) \in X^4_D$. 
Hence Assumption ${\bf (IC)}$ excludes having simultaneously  $p(x-y)=0$ for some $(x,y)\in E_\Ga$ 
and nontrivial zeros in $b(\cdot,\cdot)$.  We thus have the two  (non mutually exclusive) cases $(i)$ and $(ii)$ stated above.

Case $(i)$ follows exactly the same lines as for ZRP.
For case $(ii)$, the construction of the correct pathes of coupled transitions on $(x,y)\in E_\Ga$ depends on the values $\al_0,\beta_0$
for which $b(\al_0,\beta_0)=0$. To explain how to proceed, we show that a Misanthrope process with rates  $b(\cdot,\cdot)$ such that $b(\ga,\de)>0$ for all $\ga>\de$ and $b(\ga,\de)=0$ for all $\ga\le\de$
 satisfies Assumption ${\bf (IC)}$.

\noindent
$\bullet$ ${\bf (IC}, a)$. Let $(x,y)\in E_\Ga$ and $(\al,\beta,\ga,\de)\in X_D^4$ on $(x,y)$
such that $\al< \ga,\beta>\de$. Suppose first that $\ga>\de$ and let $r$ be the smallest integer such that $\ga-r\le\de+r$. Observe that
$\beta+r > \de +r \ge \ga-r> \al - r$ and thus $ \al - r +1 \le \beta+r -1$ while by definition of $r$, $ \ga - r +1 > \de+r -1$. Now choose $r'\le r-1$ to be the smallest integer verifying both $ \al - r' \le \beta+r'$ and $ \ga - r' > \de+r'$; a path of coupled transitions on $(x,y)$ starting from $(\al,\beta,\ga,\de)$ along which $f_{x,y}^+$ decreases (by 1) is given by
\beqq\label{mp-a-path1}
\bigl\{(\al,\beta,\ga,\de),\cdots,(\al-r',\beta+r',\ga-r',\de+r'),(\al-r',\beta+r',\ga-r'-1,\de+r'+1)\bigr\}
\eeqq
(with rates $p(x,y) $ times $b(\al,\beta),b(\al-1,\beta+1),\cdots,b(\al-r'+1,\beta+r'-1),b(\ga-r',\de+r')$).

\noindent
Suppose now that $\ga\le\de$.  Take $r$ as the smallest integer such that $\ga+r\ge \de-r$. Hence $r< (\de-\ga+1)/2$ and since $\beta>\de>\ga>\al$, $(\beta -r)-(\al +r)\ge (\de-r) - (\ga+ r) +2 \ge 1$. Thus $\beta -r > \al +r$ and a  path of coupled transitions on $(x,y)$ starting from $(\al,\beta,\ga,\de)$ along which $f_{x,y}^+$ decreases (by 1) is given by
\beqq\label{mp-a-path2}
\bigl\{(\al,\beta,\ga,\de),\cdots,(\al+r,\beta-r,\ga+r,\de-r),(\al+r+1,\beta-r-1,\ga+r,\de-r)\bigr\}
\eeqq
(with rates $p(y,x) $ times $b(\de,\ga),b(\de-1,\ga+1),\cdots,b(\de-r+1,\ga+r-1),b(\beta-r,\al+r)$).

\noindent
$\bullet$ ${\bf (IC}, b)$. Consider $\al < \ga$ and $\ve \ge 0$. There are five possibilities and  for each one we give a path of coupled transitions on $(x,y)$ starting from $(\al,\ve ,\ga,\ve )$ which creates a discrepancy on $y$. 

\noindent
1) $\ve \le \al$, $r_0 = \inf\{r > 0 :\ve + 2 r\ge \al\}$ and $\ga > \ve + 2 r_0$ :
\beqq\label{mp-b-path1}
\bigl\{(\al ,\ve ,\ga ,\ve ),\cdots,(\al - r_0,\ve + r_0 ,\ga - r_0 ,\ve +r_0 ),
(\al - r_0,\ve + r_0 ,\ga -r_0 -1,\ve +r_0 +1)\bigr\}
\eeqq
(with rates $p(x,y) $ times $b(\al,\ve),b(\al-1,\ve+1),\cdots,b(\al-r_0+1,\ve+r_0-1),b(\ga-r_0,\ve+r_0)$).

\noindent
2) $\ve \le \al$ 
and $\ga= \ve + 2 r_0 =\al +1$ :
\beqq\label{mp-b-path2}
\bigl\{(\al ,\ve ,\ga ,\ve ),\cdots,(\al - r_0,\ve + r_0 ,\ga - r_0 ,\ve +r_0 ),
(\al - r_0+1,\ve + r_0-1 ,\ga -r_0 ,\ve +r_0 )\bigr\}
\eeqq
(with rates $p(x,y) $ times $b(\al,\ve),b(\al-1,\ve+1),\cdots,b(\al-r_0+1,\ve+r_0-1)$,
then $p(y,x) $ times $b(\ve+r_0,\al-r_0)$).

\noindent
3) $\al < \ve < \ga$ :
\beqq\label{mp-b-path3}
\bigl\{(\al ,\ve ,\ga ,\ve ),(\al ,\ve ,\ga - 1 ,\ve +1)\bigr\}
\eeqq
(with rate $p(x,y) $ times $b(\ga,\ve)$).

\noindent
4) $\ga \le \ve$, $r_1 = \inf\{r > 0 :\ve  - 2 r\le \ga\}$ and $\al= \ve - 2 r_1 =\ga -1$ :
\beqq\label{mp-b-path4}
\bigl\{(\al ,\ve ,\ga ,\ve ),\cdots,(\al + r_1,\ve - r_1,\ga + r_1,\ve -r_1),
(\al + r_1,\ve - r_1,\ga +r_1-1 ,\ve -r_1+1 )\bigr\}
\eeqq
(with rates $p(y,x) $ times $b(\ve,\ga),b(\ve-1,\ga+1),\cdots,b(\ve-r_1+1,\ga+r_1-1)$,
then $p(x,y) $ times $b(\ga+r_1,\ve-r_1)$).

\noindent
5) $\ga \le \ve$ 
 and $\al< \ve - 2 r_1$ :
\beqq\label{mp-b-path5}
\bigl\{(\al ,\ve ,\ga ,\ve ),\cdots,(\al + r_1,\ve - r_1,\ga + r_1,\ve -r_1),
(\al + r_1+1,\ve - r_1-1 ,\ga +r_1,\ve -r_1 )\bigr\}
\eeqq
(with rates $p(y,x) $ times $b(\ve,\ga),b(\ve-1,\ga+1),\cdots,b(\ve-r_1+1,\ga+r_1-1),
b(\ve-r_1,\al + r_1)$).

This leads to $X_R(y-x)=X$. 

Assumptions ${\bf (IC}, a, b)$ are thus satisfied.
\epr

\subsection{StP}\label{StP:IcapS}
For all $(x,y)\in E_\Ga$ (so that $|y-x|=1$), $(\al,\beta)\in X^2$ with $\al\not=0,\beta\not=0$, by \eqref{HAD-rates} and the definition of $E_\Ga$,
\[\sum_{k}\bigl(\Ga_{\al,\beta}^k (1)+\Ga_{\beta,\al}^k (-1)\bigr)=p(1)\al+
p(-1)\beta>0\]
therefore necessarily $p(1)+p(-1)>0$ and Assumption ${\bf (IC}, o)$ holds. We now check that Assumptions ${\bf (IC}, a, b)$ are satisfied.

\noindent
$\bullet$ ${\bf (IC}, a)$. A path of coupled transition on $(x,y)$ starting from $(\al,\beta,\ga,\de)\in X^4_D$ with for instance 
$\al<\ga,\beta>\de$, along which $f^+_{x,y}$ decreases  is given by
(cf. \eqref{cou2-HAD}--\eqref{cou2-2-HAD}) 
\beqq
\begin{cases}
\bigl\{ (\al,\beta,\ga,\de), (\al,\beta,\al,\de+\ga-\al)\bigr\} & \hbox{ if }p(1) >0\quad\hbox{ (with rate }G_{\al,\beta;\,\ga,\de}^{0;\, \ga-\al}(1)=p(1))\cr
\bigl\{ (\al,\beta,\ga,\de), (\al+\beta-\de,\de,\ga,\de)\bigr\} & \hbox{ if } p(-1) >0\quad\hbox{ (with rate }G_{\beta,\al;\,\de,\ga}^{\beta-\de;\, 0}(-1)=p(-1))\cr
\end{cases}
\eeqq
$\bullet$ ${\bf (IC}, b)$. Consider $\al < \ga$ and $\ve \ge 0$. Suppose $p(y-x)=p(1) >0$; then a path of coupled transitions on $(x,y)$ starting from $(\al,\ve,\ga,\ve)$ which creates a discrepancy on $y$ is given by
\beqq\label{stp-b-path-r}
\bigl\{(\al ,\ve ,\ga ,\ve ),(\al ,\ve ,\al ,\ve +\ga -\al)\bigr\}\quad\hbox{ (with rate }G_{\al,\ve;\,\ga,\ve}^{0;\, \ga-\al}(1)=p(1))
\eeqq
which leads to $X_R(1)=X$.
When $p(x-y)=p(-1) >0$, a path of coupled transitions on $(x,y)$ starting from $(\ve,\al,\ve,\ga)$ which creates a discrepancy on $x$ is given by
\beqq\label{stp-b-path-l}
\bigl\{(\ve,\al,\ve,\ga ),(\ve,\al,\al ,\ve +\ga -\al)\bigr\}\quad\hbox{ (with rate }G_{\al,\ve;\,\ga,\ve}^{0;\, \ga-\al}(-1)=p(-1))
\eeqq
which leads to $X_L(1)=X$.
\subsection{S$_2$EP}\label{S2EP:IcapS} 
We first give the general conditions on the rates
under which Assumption {\bf (IC)} is satisfied. They are obtained by direct inspection of all possible cases, we rely on Table \ref{table:soscplrts}.
 To illustrate this,   we present three examples which fulfill Assumption {\bf (IC)} in different manners. 
 Notice first that since the dynamics is nearest neighbor and non degenerated by \eqref{non-deg}, 
 ${\bf (IC}, o)$ is always satisfied.

\noindent
$\bullet$ ${\bf (IC}, a)$ 
\beqq
X^4_D&=&\{(-1,0,0,-1),(-1,0,1,-1),(-1,1,0,0),(-1,1,0,-1),
(-1,1,1,0),\\
&&(-1,1,1,-1),(0,0,1,-1),(0,1,1,0),(0,1,1,-1),
(0,-1,-1,0),(0,-1,-1,1),\\
&&(0,0,-1,1),(1,-1,-1,0),
(1,-1,-1,1),(1,-1,0,0),(1,-1,0,1),(1,0,-1,1),\\
&&(1,0,0,1)\}
\eeqq
The following expressions need to be positive for $z=\pm 1$. They correspond to coupled transitions on $(0,z)$ from  $(\al,\beta,\ga,\de)\in X^4_D$ with 
$\al<\ga,\beta>\de$ for which $f^+_{0,z}$ decreases
\beqq
G_{-1,0;\,0 ,-1}^{0;\, 1}(z) + G_{0,-1;\,-1 ,0}^{1;\, 0}(-z)&= &\Ga_{0,-1}^{1}(z) + \Ga_{0,-1}^{1}(-z) \\
G_{-1,1;\,1 ,-1}^{0;\, 1}(z) + G_{-1,1;\,1 ,-1}^{0;\, 2}(z) &+& G_{1,-1;\,-1 ,1}^{1;\, 0}(-z) + G_{1,-1;\,-1 ,1}^{2;\, 0}(-z)\\
&=&\Ga_{1,-1}^{1}(z) + \Ga_{1,-1}^{2}(z)+\Ga_{1,-1}^{1}(-z)+\Ga_{1,-1}^{2}(-z)\\
G_{0,1;\,1 ,0}^{0;\, 1}(z)  + G_{1,0;\,0 ,1}^{1;\, 0}(-z) &=&
\Ga_{1,0}^{1}(z) + \Ga_{1,0}^{1}(-z)
\eeqq
\beqq
G_{-1,0;\,1 ,-1}^{0;\, 1}(z) + G_{-1,0;\,1 ,-1}^{0;\, 2}(z) + G_{0,-1;\,-1 ,1}^{1;\, 0}(-z) &=&
\Ga_{1,-1}^{1}(z) + \Ga_{1,-1}^{2}(z)+\Ga_{0,-1}^{1}(-z)\\
G_{0,1;\,1 ,-1}^{0;\, 1}(z) + G_{0,1;\,1 ,-1}^{0;\, 2}(z) + G_{1,0;\,-1 ,1}^{1;\, 0}(-z) &=&
\Ga_{1,-1}^{1}(z) + \Ga_{1,-1}^{2}(z)+\Ga_{1,0}^{1}(-z)\\
\eeqq
We need also that either
\beq
G_{0,0;\,1 ,-1}^{0;\, 1}(z) 
&+& G_{0,0;\,1 ,-1}^{1;\, 2}(z) +G_{0,0;\,-1 ,1}^{1;\, 0}(-z)\nonumber \\
&=& \Ga_{1,-1}^{1}(z)- [\Ga_{0,0}^{1}(z)-\Ga_{1,-1}^{2}(z)]^+ +\Ga_{0,0}^{1}(z)\wedge \Ga_{1,-1}^{2}(z) +\Ga_{0,0}^{1}(-z)
\label{sos-long}
\eeq
is positive for $z=\pm 1$ (this corresponds to a coupled transition on $(0,z)$ from  $(0,0,1,-1)$ for which $f^+_{0,z}$ decreases) or
\beq\label{sos-long-2}
G_{0,0;\,1 ,-1}^{1;\, 1}(-z) +G_{0,0;\,1 ,-1}^{0;\, 2}(-z)= |\Ga_{0,0}^{1}(-z)-\Ga_{1,-1}^{2}(-z)|
\eeq
and \eqref{sos-long} are simultaneously positive for $z=1$ or for $z=-1$.  There, $f^+_{0,z}$ decreases along paths
of coupled transitions on $(0,z)$ for which the first one induces an exchange of discrepancies (see Example \ref{ex:SOS-IC},$(iii)$ below). 

\noindent
$\bullet$ ${\bf (IC}, b)$,  ${\bf (IC}, b')$

Up to an exchange of marginals, there are three possible values $(\al,\ga)$ with $\al<\ga$ for a discrepancy on a given site: $(-1,0)$, $(-1,1)$, $(0,1)$. Therefore for 
each element $\ve\in\{-1,0,1\}$ and each site $z\in\{-1,1\}$ neighbor of the origin, a set of 3 inequalities has to be verified  by the coupling rates 
whenever $\ve$ belongs to $X_R(z)$ 
\beqq
-1\in X_R(z)
\Longleftrightarrow
\begin{cases}
G_{-1,-1;\,0 ,-1}^{0;\, 1}(z) >0\\
G_{-1,-1;\,1 ,-1}^{0;\, 1}(z) +G_{-1,-1;\,1 ,-1}^{0;\, 2}(z) >0\\
\begin{cases}
G_{0,-1;\,1 ,-1}^{0;\, 1}(z)+G_{0,-1;\,1 ,-1}^{0;\, 2}(z) +G_{0,-1;\,1 ,-1}^{1;\, 0}(z) +
G_{0,-1;\,1 ,-1}^{1;\, 2}(z)> 0\nonumber\\
\text{ or } \\
\begin{cases}
G_{0,-1;\,1 ,-1}^{1;\, 1}(z) >0 \\
G_{-1,0;\,0 ,0}^{0;\, 1}(z) +G_{0,-1;\,0 ,0}^{0;\, 1}(-z)+G_{0,-1;\,0 ,0}^{1;\, 0}(-z)>0
\end{cases}
\end{cases}
\end{cases}
\eeqq
 As previously, the 3 first lines correspond to coupled transitions which create a discrepancy on $z$,
and the combination of the fourth and fifth lines corresponds to a finite path of coupled transitions on $(0,z)$,
$\{(0,-1,1,-1),(-1,0,0,0),(-1,0,-1,1)\}$ which ends with a discrepancy on $z$.
\beqq
0\in X_R(z)
\Longleftrightarrow
\begin{cases}
\begin{cases}
G_{-1,0;\,0 ,0}^{0;\, 1}(z) +G_{0,-1;\,0 ,0}^{0;\, 1}(-z)+G_{0,-1;\,0 ,0}^{1;\, 0}(-z)>0\\
\text{ or }\\
\begin{cases}
G_{0,-1;\,0 ,0}^{1;\, 1}(-z)>0 \\
G_{0,-1;\,1 ,-1}^{0;\, 1}(z)+G_{0,-1;\,1 ,-1}^{0;\, 2}(z) +G_{0,-1;\,1 ,-1}^{1;\, 0}(z) +G_{0,-1;\,1 ,-1}^{1;\, 2}(z)> 0
\end{cases}
\end{cases}\\
G_{-1,0;\,1 ,0}^{0;\, 1}(z)+G_{0,-1;\,0 ,1}^{1;\, 0}(-z)>0\\
\begin{cases}
G_{0,0;\,1 ,0}^{0;\, 1}(z)+G_{0,0;\,1 ,0}^{1;\, 0}(z) +G_{0,0;\,0 ,1}^{1;\, 0}(-z) >0\\
\text{ or }\\
\begin{cases}
G_{0,0;\,1 ,0}^{1;\, 1}(z)>0 \\
G_{1,-1;\,1 ,0}^{1;\, 0}(-z)+G_{1,-1;\,1 ,0}^{2;\, 0}(-z) +G_{1,-1;\,1 ,0}^{0;\, 1}(-z) +G_{1,-1;\,1 ,0}^{2;\, 1}(-z)> 0
\end{cases}
\end{cases}
\end{cases}
\eeqq
\beqq
1\in X_R(z)
\Longleftrightarrow
\begin{cases}
\begin{cases}
G_{1,-1;\,1 ,0}^{1;\, 0}(-z)+G_{1,-1;\,1 ,0}^{2;\, 0}(-z) +G_{1,-1;\,1 ,0}^{0;\, 1}(-z) +G_{1,-1;\,1 ,0}^{2;\, 1}(-z)> 0\\
\text{ or } \\
\begin{cases}
G_{1,-1;\,1 ,0}^{1;\, 1}(-z) >0 \\
G_{0,0;\,1 ,0}^{0;\, 1}(z)+G_{0,0;\,1 ,0}^{1;\, 0}(z) +G_{0,0;\,0 ,1}^{1;\, 0}(-z) >0
\end{cases}
\end{cases}\\
G_{1,-1;\,1 ,1}^{1;\, 0}(-z) +G_{1,-1;\,1 ,1}^{2;\, 0}(-z) >0\\
G_{1,0;\,1 ,1}^{1;\, 0}(-z) >0
\end{cases}
\eeqq
Using the coupling rates' expressions given in Table \ref{table:soscplrts} as well as attractiveness conditions \eqref{attr-3} leads to following conditions on the rates:
\beqq
-1\in X_R(z)
\Longleftrightarrow
\begin{cases}
\Ga_{0,-1}^{1}(z) >0\\
\Ga_{1,-1}^{1}(z) +\Ga_{1,-1}^{2}(z) >0\\
\begin{cases}
(\Ga_{1,-1}^{1}(z) +\Ga_{1,-1}^{2}(z)-\Ga_{0,-1}^{1}(z)) +\Ga_{1,-1}^{2}(z)> 0\\
\text{ or }\\
 \Ga_{0,-1}^{1}(z)- \Ga_{1,-1}^{2}(z)>0\text{ and }
\Ga_{0,0}^{1}(z) +(\Ga_{0,-1}^{1}(-z)-\Ga_{0,0}^{1}(-z))>0
\end{cases}
\end{cases}
\eeqq
\beqq
0\in X_R(z)
\Longleftrightarrow
\begin{cases}
\begin{cases}
\Ga_{0,0}^{1}(z) +(\Ga_{0,-1}^{1}(-z)-\Ga_{0,0}^{1}(-z))>0\\
\text{ or } \\
\Ga_{0,0}^{1}(-z)>0\text{ and }
(\Ga_{1,-1}^{1}(z) +\Ga_{1,-1}^{2}(z)-\Ga_{0,-1}^{1}(z)) +\Ga_{1,-1}^{2}(z)> 0
\end{cases}\\
\Ga_{1,0}^{1}(z)+\Ga_{0,-1}^{1}(-z)>0\\
\begin{cases}
(\Ga_{1,0}^{1}(z)-\Ga_{0,0}^{1}(z))+\Ga_{0,0}^{1}(-z)>0\\
\text{ or } \\
\Ga_{0,0}^{1}(z)>0 \text{ and }
(\Ga_{1,-1}^{1}(-z)+\Ga_{1,-1}^{2}(-z)-\Ga_{1,0}^{1}(-z))+ \Ga_{1,-1}^{2}(-z) > 0
\end{cases}
\end{cases}
\eeqq
\beqq
1\in X_R(z)
\Longleftrightarrow
\begin{cases}
\begin{cases}
(\Ga_{1,-1}^{1}(-z)+\Ga_{1,-1}^{2}(-z)-\Ga_{1,0}^{1}(-z))+ \Ga_{1,-1}^{2}(-z) > 0\\
\text{ or }\\
 \Ga_{1,0}^{1}(-z)-\Ga_{1,-1}^{2}(-z) >0 \text{ and }
(\Ga_{1,0}^{1}(z)-\Ga_{0,0}^{1}(z))+\Ga_{0,0}^{1}(-z)>0
\end{cases}\\
\Ga_{1,-1}^{1}(-z)+ \Ga_{1,-1}^{2}(-z) >0\\
\Ga_{1,0}^{1}(-z) >0
\end{cases}
\eeqq

\bex\label{ex:SOS-IC}
We give three examples of coupling rates which satisfy
\eqref{attr-3} (that is correspond to an attractive process), 
and Assumption {\bf (IC)}.

{\em (i)} All the rates $\Ga_{\al,\beta}^{k}(z)$, 
$z=\pm 1$ are positive. Hence  ${\bf (IC}, a)$ holds and $X_R(z)=X$ for $z=\pm 1$, so that ${\bf (IC}, b)$ is satisfied.

{\em (ii)} All rates $\Ga_{\al,\beta}^{k}(-1)$ are equal to 0 and all rates $\Ga_{\al,\beta}^{1}(1)$ are positive (totally asymmetric case).  Here ${\bf (IC}, a)$ holds only through \eqref{sos-long} for $z=\pm 1$ if $\Ga_{0,0}^{1}(1)\le\Ga_{1,-1}^{2}(1)$,
and through the combination of  \eqref{sos-long}, \eqref{sos-long-2} if $0\le\Ga_{1,-1}^{2}(1)<\Ga_{0,0}^{1}(1)$. We always have $-1\in X_R(1),-1\notin X_R(-1)$ and $1\in X_R(-1),1\notin X_R(1)$, therefore ${\bf (IC},b)$ never holds. Here $W_\Ga=\{(x,x+1,x+2)\in\Z^3\}$. To have ${\bf (IC},b'_1)$,
 we need in addition
\beqq\label{SOS-ICb'1}
\begin{cases}
\Ga_{1,0}^{1}(1) >\Ga_{0,0}^{1}(1) \Longrightarrow X_R(1)\supset\{0\} \\
\text{or}\\
\Ga_{0,-1}^{1}(1) >\Ga_{0,0}^{1}(1) \Longrightarrow X_R(-1)\supset \{0\}\\
\end{cases}
\eeqq
In both cases, ${\bf (IC}, b'_2)$ is satisfied.

{\em (iii)} We choose:
\beqq\label{ex3_attr}
\begin{cases}
0=\Ga_{1,-1}^{1}(1)=\Ga_{0,0}^1(1),&
0<\Ga_{1,-1}^{2}(1)=\Ga_{0,-1}^{1}(1)=\Ga_{1,0}^{1}(1)\\
0=\Ga_{1,-1}^{2}(-1),&
0<\Ga_{0,0}^1(-1)=\Ga_{0,-1}^{1}(-1)=\Ga_{1,0}^{1}(-1)=\Ga_{1,-1}^{1}(-1)
\end{cases}
\eeqq
Here many coupling rates are equal to zero, and so is  \eqref{sos-long} for $z=-1$; we have ${\bf (IC}, a)$ satisfied through the combination of  \eqref{sos-long}, \eqref{sos-long-2} for $z=1$: the pair of discrepancies $(0,0,1,-1)$ on $(x,x+1)$
can disappear only through paths of coupled transitions involving an exchange of discrepancies: 
\[
\{(0,0,1,-1), (-1,1,0,0),(-1,1,-1,1)\}, \mbox{   or   } 
\{(0,0,1,-1), (0,0,-1,1),(-1,1,-1,1)\}
\]
By inspecting all possibilities, ${\bf (IC}, b)$ is satisfied. 
\eex

 For this model we can say more on invariant measures:
\bp
{}For the  attractive two species exclusion model, 
under Assumption ${\bf (IC)}$, $\mu_\rho\in{\mathcal I}_e$ 
(for $\rho\in[-1,1]$ under condition \eqref{prodnoncons}, 
or $\rho\in{\mathcal R}$ otherwise).
\ep
\bpr
It is similar to the one of \cite[Proposition 1.1]{B}, but, as  previously to derive \eqref{mesure_ordonne}, 
we have moreover to use Assumption ${\bf (IC)}$.
\epr

\section { Hydrodynamic limits for the stick process and the two species 
exclusion model.}\label{sec:6}
In this section, we derive almost-sure hydrodynamics for these two nearest-neighbor
one-dimensional models under Euler scaling following the 
constructive method of \cite{BGRS3}
(see also \cite{BGRS1}, \cite{BGRS2}, where hydrodynamics is proved in the usual sense), 
which does not require product invariant probability measures for the dynamics. 
We quote the result from  \cite{BGRS3} (after explaining a few preliminaries), then indicate the scheme of its derivation.
The fact that more than one particle can jump at a given time induces only one modification to the original proof. 

\subsection{The hydrodynamic limit result}\label{subsec:hydro-result}
A graphical construction of the models is  needed; 
it is explained as follows. Since for both models the rates 
are bounded, the construction, restricted to compact state spaces in \cite{BGRS3}, can be extended here.
For each $(x,z)\in\Z^2$, let $\{T_n^{x,z},n\geq
1\}$ be the arrival times of mutually independent rate
$\sup_{k\in\N;\al,\beta\in X}\Ga_{\al,\beta}^k(z)$ Poisson processes, let $\{U_n^{x,z},n\geq 1\}$
be mutually independent (and independent of the Poisson processes)
random variables, uniform on $[0,1]$. We denote by  $({\bf X},{\mathcal
F},\Prob)$ the probability space corresponding to these families of variables. 
At time $t=T_n^{x,z}$, the
configuration $\eta_{t^-}$ becomes $S_{x,x+z}^k\eta_{t^-}$ if
\[
U_n^{x,z}\le
\frac{\Ga_{\eta_{t^-}(x),\eta_{t^-}(x+z)}^k(z)}{\sup_{k\in\N;\al,\beta\in X}\Ga_{\al,\beta}^k(z)}
\] 
and stays unchanged otherwise. Let
also $({\bf X}_0,{\mathcal F}_0,\Prob_0)$ denote an `initial' probability space large enough to construct
random initial configurations $\eta_0=\eta_0(\omega_0)$ for
$\omega_0\in{\bf X}_0$. 

Let $N\in\N$ be the scaling parameter for the hydrodynamic limit. The empirical measure of a configuration $\eta$
viewed on scale $N$ is given by
\[
\alpha^N(\eta)(dx)=N^{-1}\sum_{y\in\Z}\eta(y)\delta_{y/N}(dx)
\]
It belongs to the set of positive, locally finite
measures on $\R$, equipped with the topology of vague convergence.
\bt\label{th:hydro} {\rm \cite[Theorem 2.1]{BGRS3}}
Let $(\eta^N_0,\,N\in\N)$ be a
sequence of $\Omega$-valued random variables on ${\bf X}_0$. Assume
there exists a measurable  bounded profile $u_0(.)$ on $\R$
such that
\be\label{initial_profile_vague}
\lim_{N\to\infty}\alpha^N(\eta^N_0)(dx)=
u_0(.)dx,\quad\Prob_0\mbox{-a.s.}\ee
that is,
\[
\lim_{N\to\infty}
\int_\R\psi(x)\alpha^N(\eta^N_0)(dx) =\int
\psi(x)u_0(x)dx,\quad\Prob_0\mbox{-a.s.} \]
for every continuous function $\psi$ on $\R$ with compact support.
 Let $(x,t)\mapsto u(x,t)$ denote the unique entropy
solution to the scalar conservation law
\be \label{hydro} \dt u+\dx[G(u)]=0 \ee
with initial condition $u_0$, where $G$ is a Lipschitz-continuous
flux function  (defined in \eqref{flux} below) determined by the $\Ga$'s.
Then,  with $\Prob_0\otimes\Prob$-probability one, the convergence
\be \label{later_profile}
\lim_{N\to\infty}\alpha^N(\eta_{Nt}(\eta^N_0(\omega_0),\omega))(dx)=u(.,t)dx
\ee
holds uniformly on all bounded time intervals. That is, for every
continuous function $\psi$ on $\R$ with compact support, the
convergence
\[
\lim_{N\to\infty}
\int_\R\psi(x)\alpha^N(\eta^N_{Nt})(dx) =\int \psi(x)u(.,t)dx\]
holds uniformly on all bounded time intervals.
\et
For $x\in\Z$, the {\em microscopic current} between sites $x$ and $x+1$ is defined by
\be
\label{micro_current}
J_{x,x+1}(\eta)=\sum_{k}k 
\bigl(\Ga_{\eta(x),\eta(x+1)}^k(1)
-\Ga_{\eta(x+1),\eta(x)}^k(-1)\bigr)
\ee
The  
{\em macroscopic flux} $G$ is defined by
\be \label{flux}
G(\rho)=\mu_\rho\left[J_{0,1}(\eta)\right]
\ee 
for $\rho\in\mathcal R$  (a set introduced in Theorem \ref{th:(i cap s)_e}), and by linear 
interpolation on the complement of $\mathcal R$
(which is at most a countable union of disjoint open intervals).

The first step to prove Theorem
\ref{th:hydro} works here unchanged: it
consists in first giving
a variational formula for the solution of the hydrodynamic 
equation \eqref{hydro} under Riemann initial condition
\be\label{eq:rie}
u_0(x)=\lambda {\bf 1}_{\{x<0\}}
+\rho {\bf 1}_{\{x\geq 0\}}
\ee
(where $\lambda,\rho$ belong to $\mathcal{R}$), and then deriving hydrodynamics under this initial profile. 
 
 The second step consists in extending the hydrodynamic limit   to any initial profile $u_0$ through an approximation scheme. Its key tools are a finite propagation property both at microscopic and macroscopic levels
(which work unchanged: for particles, we refer to  \cite[Lemma 3.1]{BGRS1} which does not rely on attractiveness) and the macroscopic stability property.
The latter requires here a different proof, that we derived in Proposition \ref{macrostab-ppv}. 

In the following subsections, we compute the involved quantities,
that is microscopic current, macroscopic flux, and check macroscopic stability
for both examples. We saw that the stick process possesses a one-parameter family of product probability measures, and that
 for the two species exclusion it is the case  under Condition \eqref{prodnoncons}, otherwise 
the invariant measures are not explicit (cf. Theorem \ref{th:(i cap s)_e}).
\subsection {Hydrodynamic limits for the stick process.}\label{sec:hydro-HAD}
{}For $x\in\Z$, setting $p(1)=p$ and $p(-1)= q$, 
the microscopic current between $x$ and $x+1$ is 
\beq\label{cur-HAD}
J_{x,x+1}(\eta)
&=&\sum_k k\bigl(p{\bf 1}_{\{k\le\eta(x)\}}-q{\bf 1}_{\{k\le\eta(x+1)\}}\bigr)\nonumber\\
&=& \frac{p}2\,\eta(x)(\eta(x)+1) -\frac{q}2 \,\eta(x+1)(\eta(x+1)+1)
\nonumber
\eeq
so that the macroscopic flux 
\be\label{flux-HAD}
G(\rho)=\frac{p-q}2\rho(1+\rho)
\ee
 is a convex function for $p-q>0$.

Proposition \ref{macrostab-ppv} holds here by Proposition \ref{stick-coupl-att}. 
\subsection {Hydrodynamic limits for the two species 
exclusion model.}\label{sec:hydro-SOS}
Under conditions \eqref{non-deg}--\eqref{prodnoncons}, the model admits a one-parameter
family of invariant product probability measures $\{\mu_\rho\}_{\rho\in (-1,1)}$ (cf. Section \ref{sec:SOS-description});
the marginals of $\mu_\rho$ can be written as (see \cite[Equations (3.6)--(3.8)]{CDFG}) 
\beq\label{marg_1}
\mu_\rho(\eta(x)=1)&=&\frac{cy}{1+cy+cy^{-1}}=\frac{\varphi(\rho)+\rho}{2}\\
\mu_\rho(\eta(x)=0)&=&\frac{1}{1+cy+cy^{-1}}=1-\varphi(\rho)\\
\mu_\rho(\eta(x)=-1)&=&\frac{cy^{-1}}{1+cy+cy^{-1}}=\frac{\varphi(\rho)-\rho}{2}\label{marg_-1}
\eeq
where $y\in (0,+\infty)$ is an auxiliary parameter related both to the mean charge, $\rho=\int \eta(x) d\mu_\rho(\eta)$
and to the mean squared charge, $\varphi(\rho)=\int \eta(x)^2 d\mu_\rho(\eta)$ as
\be\label{charge}
\rho=\frac{c(y-y^{-1})}{1+cy+cy^{-1}}\qquad ;\qquad
\varphi(\rho)=\frac{c(y+y^{-1})}{1+cy+cy^{-1}}
\ee
 The constant  $c$ depends on the jump rates
\be\label{cnoncons}
c=\sqrt{\frac{\Ga_{0,0}^1(1)+\Ga_{0,0}^{1}(-1)}
{\Ga_{1,-1}^{1}(1)+\Ga_{1,-1}^1(-1)}}
\ee
After elimination of $y$ in \eqref{charge},  $\varphi(\rho)$ appears as a function of $\rho\in(-1,1)$
\beq\label{phi_psi}
\varphi(\rho)=
\begin{cases}
\displaystyle{\frac{\psi(\rho)-4c^2}{1-4c^2}}
\quad\hbox{with }\psi(\rho)=\sqrt{4c^2+\rho^2 (1-4c^2)}
\qquad&\hbox{if } 4c^2\ne 1\cr
\qquad\qquad\quad\displaystyle{\frac{\rho^2+1}{2}}
&\hbox{if } 4c^2=1
\end{cases}
\eeq
Independently of \eqref{prodnoncons}, 
according to the values of the jump parameters, 
the scaling is either diffusive or Euler, and 
the model is studied either as a gradient model, 
or as an attractive one. In both cases, the microscopic current 
between $x$ and $x+1$ reads
\beq\label{SOS-cur}
J_{x,x+1}(\eta)
&=&\chi_{x,x+1}^{0,0}(\eta)\bigl(\Ga_{0,0}^1(1)
-\Ga_{0,0}^1(-1)\bigr)\nonumber\\
&&+\chi_{x,x+1}^{1,0}(\eta)\Ga_{1,0}^1(1)
-\chi_{x,x+1}^{0,1}(\eta)\Ga_{1,0}^1(-1)\nonumber\\
&&
+\chi_{x,x+1}^{0,-1}(\eta)\Ga_{0,-1}^1(1)
-\chi_{x,x+1}^{-1,0}(\eta)\Ga_{0,-1}^1(-1)\\
&&+\chi_{x,x+1}^{1,-1}(\eta)\bigl(\Ga_{1,-1}^1(1)
+2\Ga_{1,-1}^2(1)\bigr)
-\chi_{x,x+1}^{-1,1}(\eta)\bigl(\Ga_{1,-1}^1(-1)
+2\Ga_{1,-1}^2(-1)\bigr)\nonumber
\eeq
\subsubsection{ Hydrodynamic limits in the diffusive case.}
This case, for which we 
refer to \cite{KL}, \cite{V}, corresponds to the values of the 
jump rates under which the 
macroscopic flux computed  under Euler scaling 
in \eqref{SOS-flux} satisfies $G(\rho)\equiv 0$. 
We only derive the conditions under which the model 
is gradient:
\bl
The two species exclusion model is gradient 
if the rates satisfy
\be\label{eq:SOS-gradient}
\begin{cases}
\Ga_{0,0}^{1}(-1) =\Ga_{0,0}^{1}(1);& 
\Ga_{1,0}^{1}(-1) =\Ga_{1,0}^{1}(1) ;\quad
\Ga_{0,-1}^{1}(-1) = \Ga_{0,-1}^{1}(1) \\
\Ga_{1,-1}^{1}(-1)+2\Ga_{1,-1}^{2}(-1) 
&=\Ga_{1,-1}^{1}(1)+ 2 \Ga_{1,-1}^{2}(1) 
=\,\,\, \Ga_{0,-1}^{1}(1) + \Ga_{1,0}^{1}(1)
\end{cases} 
\ee
\el
\bpr
The model is gradient if there exists a function 
$q$ such that
\[
J_{x,x+1}(\eta)=(\tau_{x+1}-\tau_{x})q(\eta)
\]
Because the right hand side of \eqref{micro_current} depends only on
$\eta(x), \eta(x+1)$,
$J_{x,x+1}(\eta)$ has to be of the form
\beqq
g_1 \bigl(\chi_{x,x+1}^{1,0} (\eta) - \chi_{x,x+1}^{0,1} (\eta)\bigr)
+g_2 \bigl(\chi_{x,x+1}^{0,-1} (\eta) - \chi_{x,x+1}^{-1,0} (\eta)\bigr)
+g_3 \bigl(\chi_{x,x+1}^{1,-1} (\eta) - \chi_{x,x+1}^{-1,1} (\eta)\bigr)
\eeqq
with $g_1+g_2=g_3$.
This leads to \eqref{eq:SOS-gradient}, and 
\be\label{tauxq}
q(\eta)=-
[\Ga_{1,0}^{1}(1)+\Ga_{0,-1}^{1}(1)]  \frac{\eta(0)}{2}
+[\Ga_{0,-1}^{1}(1)-\Ga_{1,0}^{1}(1)]\frac{\eta(0)^2}{2}
\ee
We have used
 \[
{\bf 1}_{\{\eta(x)=0\}}=1-{\bf 1}_{\{\eta(x)=1\}}-{\bf 1}_{\{\eta(x)=-1\}};\,
{\bf 1}_{\{\eta(x)=1\}}=\eta(x)\frac{\eta(x)+1}{2};\,
{\bf 1}_{\{\eta(x)=-1\}}=\eta(x)\frac{\eta(x)-1}{2}
\]
to write in \eqref{SOS-cur}  the indicator functions 
as polynomials in $\eta(x),\eta(x+1)$.
\epr
\subsubsection { Hydrodynamic limit under Euler scaling
in the attractive case.}
Using expressions \eqref{SOS-cur} for the microscopic current 
and  \eqref{marg_1}--\eqref{marg_-1} for the product measures, we obtain the 
macroscopic flux as a function of the density $\rho$,
\beq\label{SOS-flux}
G(\rho)&=& ( \Ga_{0,0}^{1}(1) - \Ga_{0,0}^{1}(-1) ) 
(1-\varphi(\rho))^2\nonumber\\
&&+\frac{1}{2} ( \Ga_{1,0}^{1}(1) - \Ga_{1,0}^{1}(-1) ) 
(1- \varphi(\rho))(\varphi(\rho)+\rho)\nonumber\\
&&+\frac{1}{2} ( \Ga_{0,-1}^{1}(1) - \Ga_{0,-1}^{1}(-1) ) 
(1- \varphi(\rho))(\varphi(\rho)-\rho)\nonumber\\
&&+\frac{1}{4} \bigl( ( \Ga_{1,-1}^{1}(1) - \Ga_{1,-1}^{1}(-1)) +
2 ( \Ga_{1,-1}^{2}(1) - \Ga_{1,-1}^{2}(-1)) \bigr) 
(\varphi(\rho)^2-\rho^2)
\eeq
It is a nontrivial function, a priori 
neither concave nor convex in the parameter $\rho$.
Recalling \eqref{flux} and  \eqref{phi_psi}, its
second derivative reads
\beq
 G''(\rho) &=&
\frac{C_1 - C_2}{1-4c^2}
-\frac{2c^2 (2C_1 - (1+4c^2)C_2)}{(1-4c^2)\psi(\rho)^3} 
- \frac{C_3\rho}{\psi(\rho)} \Bigl(\frac{1}{2} +\frac{c^2}{ \psi(\rho)^2} \Bigr)
\eeq
where 
\beqq
C_1&=& 2(\Ga_{0,0}^{1}(1)-\Ga_{0,0}^{1}(-1))+ 
2 c^2 (\Ga_{1,-1}^{1}(1)-\Ga_{1,-1}^{1}(-1))
+4 c^2 (\Ga_{1,-1}^{2}(1)-\Ga_{1,-1}^{2}(-1))\\
C_2&=&(\Ga_{1,0}^{1}(1)-\Ga_{1,0}^{1}(-1))
+(\Ga_{0,-1}^{1}(1)-\Ga_{0,-1}^{1}(-1))\\
C_3&=&(\Ga_{1,0}^{1}(1)-\Ga_{1,0}^{1}(-1))
-(\Ga_{0,-1}^{1}(1)-\Ga_{0,-1}^{1}(-1))
\eeqq
According to the values of the jump rates, 
the flux function is either concave, or exhibits one 
or two inflexion points. This yields increasing or 
decreasing shocks, rarefaction fans as well as contact 
discontinuities for the unique entropy solution of the 
hydrodynamic equation \eqref{hydro}
with initial condition \eqref{eq:rie}.

Proposition \ref{macrostab-ppv} holds  under Condition \eqref{sos-exch}.

\noindent
{\bf Example.} We detail what happens for 
{\sl thermal bath dynamics}. These can be 
compared to the simulations described in 
\cite{CDFG}, though a discrete time 
Markov chain was used there. The model is used to describe the dynamics of a one dimensional Solid-on-Solid interface, that is a collection of heights on $\Z$ and a configuration $\eta$ describes only the height differences between neighboring sites. Jump rates are thus chosen according 
to the following rules:

$\bullet$ Detailed balance holds with respect to 
a formal Solid-on-Solid Hamiltonian.

\beqq
\Ga_{\eta(x),\eta(x+1)}^k(1) \exp\bigl\{H(S_{x,x+1}^k\eta) - H(\eta)\bigr\} = \Ga_{\eta(x+1)+k,\eta(x)-k}^k(-1)
\eeqq
for all $\eta$, all $x$ and all $k>0$, with
\beqq
\mathcal H(\eta) = J \sum_{x\in \Z} |\eta_(x)| -E \sum_{x\in \Z} x \eta(x)
\eeqq
There are two parameters: $J>0$ weights the height differences between neighboring sites, 
and $E>0$ is an external field which drives the interface.

$\bullet$ Rates are chosen invariant under the symmetry 
$x\to -x, \eta \to -\eta$.

$\bullet$ Condition \eqref{prodnoncons} for the existence of 
invariant product probability measures holds.

\begin{figure}[htbp]
\begin{center}
\includegraphics[scale=.4]{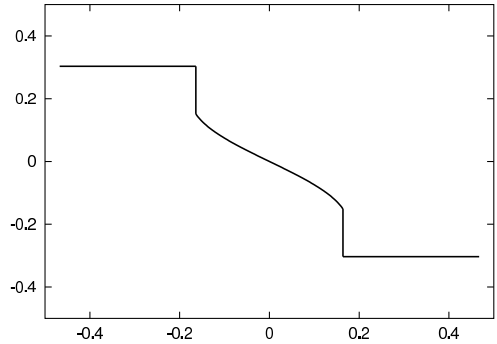}
\end{center}
\caption{Starting from an initial condition
$u_0(x)=\lambda ({\bf 1}_{\{x<0\}} 
- {\bf 1}_{\{x\ge 0\}})$, the hydrodynamic limit may
develop two shocks going in opposite directions at 
constant speeds and a rarefaction fan in-between. 
Here $c=.145$ and   $\lambda=.303$.}
\label{figure:2shocks}
\end{figure}

The above three conditions allow to write all jump rates in 
terms of two physically significant parameters $a= \exp(-E/2)$  and $b=\exp(-J)$:
\beq\begin{cases}\label{ga_bt}
\Ga_{1,0}^1(1)=\Ga_{0,-1}^1(1)= a v_2\\
\Ga_{1,0}^1(-1)=\Ga_{0,-1}^1(-1)= a^{-1} v_2\\
\Ga_{0,0}^1(1)=\Ga_{1,-1}^2(1)= a^2 v_0\\
\Ga_{0,0}^1(-1)=\Ga_{1,-1}^2(-1)= a^{-2} v_0\\
\Ga_{1,-1}^1(1)=\Ga_{1,-1}^1(-1)=b^{-2} v_0 
\end{cases}
\eeq
with 
\be
v_2=\frac{a+a^{-1}}{2} \quad;\quad v_0
=\frac{a^2+a^{-2}}{a^2+a^{-2}+b^{-2}}\nonumber
\ee
so that 
\be
c^2=\frac{b^2(a^{-2}+a^2)}{2}
\ee
The macroscopic flux and its second derivative read
\beqq
 G(\rho)&=& \frac{(a^{-2}-a^2)}{2} \Bigl(\rho^2 -\frac{\psi(\rho)}{1-4 c^2}\Bigr)\\
 G''(\rho)&=& (a^{-2}-a^2) \Bigl(1 -\frac{2c^2}{\psi(\rho)^3}\Bigr)
\eeqq
which yields no inflexion point if $c\in [1/4,1/\sqrt{2}]$. 
When $c>1/\sqrt{2}$ there are two inflexion points
near $\rho=\pm 1$; when $c<1/4$ the two inflexion points are 
near $\rho=0$. The situation is thus rather similar to that 
of the discrete time model studied heuristically in 
\cite{CDFG}. We have derived rigorously 
all the cases they got using finite size scaling analysis. The situation
in Figure \ref{figure:2shocks}  was beyond the reach
of their analysis: for $c<1/4$ and initial condition \eqref{eq:rie} 
with $\rho = -\lambda$ and $\lambda$ larger than the abscissa 
of the inflexion point, the system develops two shocks moving 
with opposite velocities and separated by a rarefaction fan.

\noindent 
{\bf Acknowledgments.} We are indebted to Fran\c cois Dunlop and
Pierre Collet for many stimulating exchanges at early
stages of this work. We thank Enrique Andjel for useful discussions, and Gunter Sch\"utz for pointing out relevant references. Part of this work was done during the authors'stay at Institut Henri Poincar\'e, Centre
Emile Borel (whose hospitality is acknowledged), for the semester
``Interacting Particle Systems, Statistical Mechanics and
Probability Theory''.
\bibliographystyle{amsalpha}

\section{Appendix}
\label{sec:A1}
In this part, we detail the derivation of coupling rates for the attractive two species exclusion model of Subsection \ref{sec:SOS-description}. 
In order to illustrate the coupling construction of Proposition \ref{princ}, 
we re-obtain the rates' values  by direct inspection, without using equations \eqref{cou2}--\eqref{cou2-3} but  relying on the following previously stated construction rules:
(o) marginals have to be recovered; (i) in coupled jumps, the 
same departure and arrival sites are taken for both processes;
 (ii) partial order on departure and arrival sites has to be preserved; 
(iii) discrepancies should not increase.
In some instances, these rules do not fully determine the coupling rates but give a family of solutions;
a consistent choice can be done by the additional rule: (iv)  the non-zero coupling rates  are located on a ``staircase-shaped path" in the $(k,l)$-quadrant (recall Proposition \ref{altprinc2}).
Rules (o - iv) single out the solution given by  \eqref{cou2}--\eqref{cou2-3}.

\begin{figure}[htbp]
\begin{center}
\includegraphics[scale=.4]{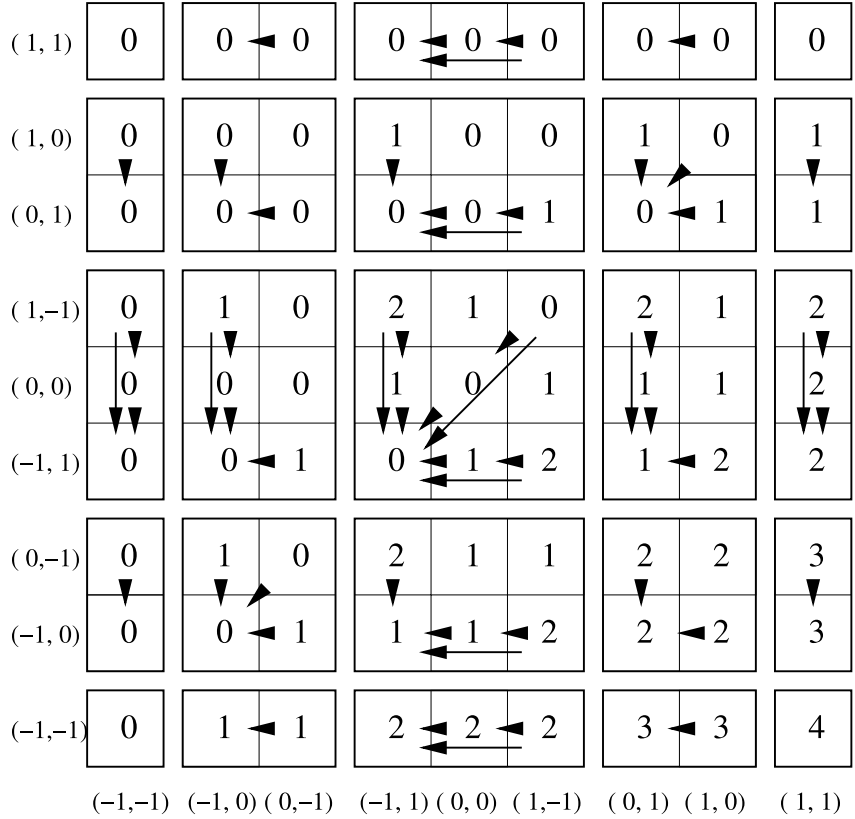}
\end{center}
\caption{ 
The coupling rates between two sites for the two species model.
Possible coupled jumps, represented by arrows, are subject to three restrictions:
arrows connect two cells in the same sub-array, point leftwards and/or downwards, cannot point to a larger value than the initial one. All trivial arrows have been represented, leaving four non trivial cases.}
\label{figure:sosattra1}
\end{figure}

By rule (i), we fix the same initial site $x$ and final site $y$ with $z=y-x=\pm 1$, consider only the local configuration $(\xi(x),\xi(y),\zeta(x),\zeta(y))$ as an element of  $X^4$, with $X=\{-1,0,1\}$, and describe a coupled jump as a transition on this reduced state space. Furthermore, since in the initial process, jump rates from $x$ to $y$ depend only on values at  these sites, one can look for a coupling process which has also this property so that the coupling construction can be given as if it were on  $X^4$.  In Figure \ref{figure:sosattra1}, we represent this reduced space as an $X^2\times X^2$ square array  with the values of $(\xi(x),\xi(y))$ (resp. $(\zeta(x),\zeta(y))$) as the first, horizontal (resp. second, vertical), coordinate from left to right on the horizontal axis, and from bottom to top on the vertical one. Before asking for the actual values of the coupling rates, note that three different factors limit the number of non zero coupling rates: 

(a) there
is charge conservation on each marginal. Since both $\xi(x)+\xi(y)$ and $\zeta(x)+\zeta(y)$ take value in $\{-2,\cdots,+2\}$, charge conservation splits the $X^2\times X^2$ square array into $25$ sub-arrays, and a transition takes place only between two elements of the same sub-array;

(b) 
only positive charges jump from $x$ to $y$ in both marginals  so that in a jump, by rule (i), the values at $x$ cannot increase;

(c)
the coupling should be increasing, reflecting that the initial process is attractive by rule (ii) and that by rule (iii), the discrepancies cannot grow. Both properties are 
related to  
the function (cf. \eqref{f_xy})
\[
f_{x,y}^+(\xi,\zeta)=[\xi(x)-\zeta(x)]^+
+[\xi(y)-\zeta(y)]^+
\]
Once fixed the values of the local charges $\xi(x)+\xi(y)$ and $\zeta(x)+\zeta(y)$, $f_{x,y}^+(\xi,\zeta)$
takes its extremal values on ordered pairs, namely,
\[
f_{x,y}^+(\xi,\zeta)=
\begin{cases} 0 &\quad \hbox{ if } (\xi(x),\xi(y)) \le (\zeta(x),\zeta(y))\\
 \xi(x)+\xi(y)-(\zeta(x)+\zeta(y))  &\quad \hbox{ if } (\xi(x),\xi(y)) \ge (\zeta(x),\zeta(y))
\end{cases}
\]
Rule (ii) requires that $f_{x,y}^+(\xi,\zeta)$ cannot escape its minimum by a jump between $x$ and $y$,
and rule (iii) extends this requirement to non ordered states by asking that $f_{x,y}^+(\xi,\zeta)$  is not increasing. 

These three restrictions can be depicted on the graphical construction of Figure \ref{figure:sosattra1}, where we represent a possible coupled transition as an arrow from a `cell' (initial values) to another one 
(final values).
By (a), an arrow can be drawn only between two cells of a same subarray. By (b), since configurations are ordered by increasing charge and increasing value, 
arrows have to point left and/or downwards; In order to use (c), we have written in each cell the value of $f_{x,y}^+(\xi,\zeta)$: arrows cannot go from a given value to a larger one.

\begin{figure}[htbp]
\begin{center}
\includegraphics[scale=.4]{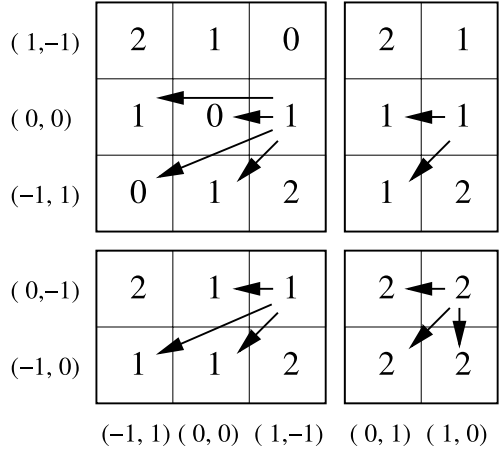}
\end{center}
\caption{Examples of non trivial cases}
\label{figure:sosattra2}
\end{figure}

All this already fixes  trivial rates for our coupling process, when only one transition from a cell is allowed, and we have indicated all of them on Figure \ref{figure:sosattra1}: these include non coupled $\xi$-jumps (horizontal arrows), non coupled $\zeta$-jumps (vertical arrows) or fully coupled jumps between identical configurations (arrows on the main diagonal
of each sub-array). We are thus left with
 only four types of coupled jumps, which we treat in more detail now, using the four examples represented in Figure \ref{figure:sosattra2}.

In the sub-array 
$A_1=\{(0,1),(1,0)\}\times\{(-1,0),(0,-1)\}$, all four pairs are ordered and the function $f_{x,y}^+(\xi,\zeta)$
is constant. 
Starting from the state $((1,0),(0,-1))$, three jumps are possible and by rule (o), the coupling rates are related through
\[
\begin{cases}
G_{1,0;\, 0,-1}^{1;\, 0}(z) =  \Ga_{1,0}^{1}(z) - G_{1,0;\, 0,-1}^{1;\, 1}(z) \\
G_{1,0;\, 0,-1}^{0;\, 1}(z) =  \Ga_{0,-1}^{1}(z) - G_{1,0;\, 0,-1}^{1;\, 1}(z)
\end{cases}
\]
Any choice for $G_{1,0;\, 0,-1}^{1;\, 1}(z)$ which preserves non-negativity is valid, so that 
\[ 
0 \le G_{1,0;\, 0,-1}^{1;\, 1}(z) \le \Ga_{1,0}^{1}(z) \wedge \Ga_{0,-1}^{1}(z)
\]
Here the only possible non-zero value for $k$ and $l$ is 
$1$. Rule (iv) imposes that either $G_{1,0;\, 0,-1}^{1;\, 0}(z)$ or $G_{1,0;\, 0,-1}^{0;\, 1}(z)$ is zero,
 hence
\[
G_{1,0;\, 0,-1}^{1;\, 1}(z) = \Ga_{1,0}^{1}(z) \wedge \Ga_{0,-1}^{1}(z)
\]
and the non-zero coupling rates are either on the $(k,l)$-path $\mathcal P_{1,0;\, 0,-1} =\{(1,0),(1,1)\}$
or $\mathcal P_{1,0;\, 0,-1} =\{(0,1),(1,1)\}$, depending on the value of the marginals.
Notice that the construction is here identical to the `basic coupling' one.

In rectangle 
$A_2=\{(0,1),(1,0)\}\times\{(-1,1),(0,0),(1,-1)\}$,  
we consider coupling rates from the initial state 
$((1,0), (0,0))$. Its coordinates being 
ordered (that is $(1,0)\ge (0,0)$), order preservation (rule (ii)) forbids a jump to 
$((1,0),(-1,1))$, thus
\beqq
G_{1,0;\, 0,0}^{0;\,1}(z)=0
\eeqq
The only choice to get the correct $\zeta$-marginal
(that is to satisfy rule (o)) is then
\beqq
G_{1,0;\,0,0}^{1;\,1} (z) &=&\Ga_{0,0}^{1}(z)
\eeqq
and the last rate follows
\beqq
G_{1,0;\, 0,0}^{1;\, 0} (z) = 
\Ga_{1,0}^{1}(z) - \Ga_{0,0}^{1}(z)
\eeqq
It has to be non-negative, which was
stated by \eqref{attr-3}. Notice that rule (iv) is automatically satisfied with
$\mathcal P_{1,0;\, 0,0} =\{(1,0),(1,1)\}$.

The situation in rectangle
$A_3=\{(-1,1),(0,0),(1,-1)\}\times\{(-1,0),(0,-1)\}$ is 
similar to the preceding one. We consider there the other type of non trivial initial state,
$((1,-1),(0,-1))$ whose coordinates are again ordered  $((1,-1)\ge(0,-1))$, and by rule (ii)
\[
G_{1,-1;\, 0,-1}^{2;\, 0}(z)=G_{1,-1}^{0;\, -1}(z)=0
\]
 Here the situation cannot be 
solved through basic coupling, since there is a marginal $\xi$-jump involving two positive charges.
There is only one solution for the coupling rates,
\beqq
G_{1,-1;\, 0,-1}^{2;\, 1}(z)&=&\Ga_{1,-1}^{2}(z)\\
G_{1,-1;\, 0,-1}^{1;\, 1}(z)&=& \Ga_{0,-1}^{1}(z) - \Ga_{1,-1}^{2}(z)\\
G_{1,-1;\, 0,-1}^{1;\, 0}(z) &=& \Ga_{1,-1}^{1}(z)+\Ga_{1,-1}^{2}(z) - \Ga_{0,-1}^{1}(z)
\eeqq
They are all non-negative by $\eqref{attr-3}$ and identical to the values given by \eqref{cou2}--\eqref{cou2-3}. Rule (iv) is again satisfied and $\mathcal P_{1,-1;\, 0,-1} =\{(1,0),(1,1),(2,1)\}$.

We find a last type of non trivial case in the larger square
\[
A_4=\{(-1,1),(0,0),(1,-1)\}\times\{(-1,1),(0,0),(1,-1)\}
\]
 and
we consider $((1,-1),(0,0))$ as initial state. Its coordinates  are not ordered so that, rather than rule (ii), selection rule (iii) may apply. This does not fix completely the coupling rates and we have
\beqq
G_{1,-1;\, 0,0}^{2;\,1}(z)&=&\Ga_{1,-1}^{2}(z) - G_{1,-1;\, 0,0}^{2;\,0}(z)\nonumber\\
G_{1,-1;\, 0,0}^{1;\,1}(z)&=&\Ga_{0,0}^{1}(z) + G_{1,-1;\, 0,0}^{2;\,0}(z) - \Ga_{1,-1}^{2}(z)
\nonumber\\
G_{1,-1;\, 0,0}^{1;\,0}(z)&=&\Ga_{1,-1}^{1}(z) + \Ga_{1,-1}^{2}(z)  
- \Ga_{0,0}^{1}(z) - G_{1,-1;\, 0,0}^{2;\,0}(z) 
\eeqq
where non-negativity is guaranteed by $\eqref{attr-3}$, and any choice for $G_{1,-1;\, 0,0}^{2;\,0}(z)$
in the range
\[
[\Ga_{1,-1}^{2}(z) - \Ga_{0,0}^{1}(z) ]^+ 
\le G_{1,-1;\, 0,0}^{2;\,0}(z)
\le \Ga_{1,-1}^{2}(z) \wedge
\bigl( \Ga_{1,-1}^{1}(z) + \Ga_{1,-1}^{2}(z) -  \Ga_{0,0}^{1}(z)\bigr)
\]
 is valid. Rule (iv) imposes to set it to its minimal value,
 \[ G_{1,-1;\, 0,0}^{2;\,0}(z) = [\Ga_{1,-1}^{2}(z) - \Ga_{0,0}^{1}(z) ]^+ \]
 and gives a path $\mathcal P_{1,-1;\, 0,0} =\{(1,0),(1,1),(2,1)\}$ or 
 $\mathcal P_{1,-1;\, 0,0} =\{(1,0),(2,0),(2,1)\}$ depending on the value of the marginals.
The solution given by \eqref{cou2}--\eqref{cou2-3} is again recovered.
 Note that when $\Ga_{1,-1}^{2}(z) \le \Ga_{0,0}^{1}(z)$, this choice is the only one which avoids an `exchange of discrepancies' (cf. Proposition \ref{prop:sos-exch}).

 We conclude that  
\eqref{cou2}--\eqref{cou2-3} lead, as in the spirit of `basic coupling', to changes as similar as possible in both marginals, which has in addition the property of minimizing  exchanges of discrepancies.

For the sake of completeness, all the coupling rates for this model are reported in Table \ref{table:soscplrts}; they are also useful to derive conditions on the rates in Subsection \ref{S2EP:IcapS}.

 \begin{sidewaystable}[ht]
 \centering
 \resizebox{21truecm}{!}{
 \begin{tabular}{|p{.5cm}p{.5cm}|p{.3cm}p{.3cm}|p{1.1cm}|p{1.1cm}|p{3.6cm}|p{1.2cm}|p{3.6cm}|p{3.6cm}|p{1.2cm}|p{3.5cm}|p{1.1cm}|}
 \hline
&\multicolumn{2}{c}{}&&\multicolumn{8}{c}{$\zeta(x)\quad\zeta(y)$}&\\
 \hline
\centerline{$\xi(x)$}&\centerline{$\xi(y)$}&\centerline{$k$}&\centerline{$l$}&
-1\;\;\;-1&-1\quad0&0\quad-1&-1\quad1&0\quad0&1\quad-1&0\quad 1&1\quad 0&1\quad 1\\[-2.0ex]
 \hline
 \centerline{-1}&\centerline{-1}&\centerline{0}&\centerline{1}&
&&$\Ga_{0,-1}^{1}$&&$\Ga_{0,0}^{1}$&$\Ga_{1,-1}^{1}$&&$\Ga_{1,0}^{1}$&\\[-2.0ex]
\centerline{-1}&\centerline{-1}&\centerline{0}&\centerline{2}&
&&&&&$\Ga_{1,-1}^{2}$&&&\\[-2.0ex]
\hline
 \centerline{-1}&\centerline{0}&\centerline{0}&\centerline{1}&
&&$\Ga_{0,-1}^{1}$&&$\Ga_{0,0}^{1}$&$\Ga_{1,-1}^{1}$&&$\Ga_{1,0}^{1}$&\\[-2.0ex]
\centerline{-1}&\centerline{0}&\centerline{0}&\centerline{2}&
&&&&&$\Ga_{1,-1}^{2}$&&&\\[-2.0ex]
\hline
\centerline{0}&\centerline{-1}&\centerline{0}&\centerline{1}&
&&&&&$\Ga_{1,-1}^{1}+\Ga_{1,-1}^{2}-\Ga_{0,-1}^{1}$&&$[\Ga_{1,0}^{1}-\Ga_{0,-1}^{1}]^+$&\\[-2.0ex]
\centerline{0}&\centerline{-1}&\centerline{1}&\centerline{0}&
$\Ga_{0,-1}^{1}$&$\Ga_{0,-1}^{1}$&&$\Ga_{0,-1}^{1}$&$\Ga_{0,-1}^{1}-\Ga_{0,0}^{1}$&&
$\Ga_{0,-1}^{1}$&$[\Ga_{0,-1}^{1}-\Ga_{0,-1}^{1}]^+$&$\Ga_{0,-1}^{1}$\\[-2.0ex]
\centerline{0}&\centerline{-1}&\centerline{1}&\centerline{1}&
&&$\Ga_{0,-1}^{1}$&&$\Ga_{0,0}^{1}$&$\Ga_{0,-1}^{1}-\Ga_{1,-1}^{2}$&&
$\Ga_{1,0}^{1}\wedge\Ga_{0,-1}^{1}$&\\[-2.0ex]
\centerline{0}&\centerline{-1}&\centerline{1}&\centerline{2}&
&&&&&$\Ga_{1,-1}^{2}$&&&\\[-2.0ex]
\hline
\centerline{-1}&\centerline{1}&\centerline{0}&\centerline{1}&
&&$\Ga_{0,-1}^{1}$&&$\Ga_{0,0}^{1}$&$\Ga_{1,-1}^{1}$&&$\Ga_{1,0}^{1}$&\\[-2.0ex]
\centerline{-1}&\centerline{1}&\centerline{0}&\centerline{2}&
&&&&&$\Ga_{1,-1}^{2}$&&&\\[-2.0ex]
\hline
\centerline{0}&\centerline{0}&\centerline{0}&\centerline{1}&
&&$\Ga_{0,-1}^{1}-\Ga_{0,0}^{1}$&&&$\Ga_{1,-1}^{1}-[\Ga_{0,0}^{1}-\Ga_{1,-1}^{2}]^+$&&
$\Ga_{1,0}^{1}-\Ga_{0,0}^{1}$&\\[-2.0ex]
\centerline{0}&\centerline{0}&\centerline{0}&\centerline{2}&
&&&&&$[\Ga_{1,-1}^{2}-\Ga_{0,0}^{1}]^+$&&&\\[-2.0ex]
\centerline{0}&\centerline{0}&\centerline{1}&\centerline{0}&
$\Ga_{0,0}^{1}$&$\Ga_{0,0}^{1}$&&$\Ga_{0,0}^{1}$&&&$\Ga_{0,0}^{1}$&&$\Ga_{0,0}^{1}$\\[-2.0ex]
\centerline{0}&\centerline{0}&\centerline{1}&\centerline{1}&
&&$\Ga_{0,0}^{1}$&&$\Ga_{0,0}^{1}$&$[\Ga_{0,0}^{1}-\Ga_{1,-1}^{2}]^+$&&$\Ga_{0,0}^{1}$&\\[-2.0ex]
\centerline{0}&\centerline{0}&\centerline{1}&\centerline{2}&
&&&&&$\Ga_{0,0}^{1}\wedge\Ga_{1,-1}^{2}$&&&\\[-2.0ex]
\hline
\centerline{1}&\centerline{-1}&\centerline{1}&\centerline{0}&
$\Ga_{1,-1}^{1}$&$\Ga_{1,-1}^{1}$&$\Ga_{1,-1}^{1}+\Ga_{1,-1}^{2}-\Ga_{0,-1}^{1}$&
$\Ga_{1,-1}^{1}$&$\Ga_{1,-1}^{1}-[\Ga_{0,0}^{1}-\Ga_{1,-1}^{2}]^+$&&$\Ga_{1,-1}^{1}$&
$\Ga_{1,-1}^{1}+\Ga_{1,-1}^{2}-\Ga_{1,0}^{1}$&$\Ga_{1,-1}^{1}$\\[-2.0ex]
\centerline{1}&\centerline{-1}&\centerline{1}&\centerline{1}&
&&$\Ga_{0,-1}^{1}-\Ga_{1,-1}^{2}$&&$[\Ga_{0,0}^{1}-\Ga_{1,-1}^{2}]^+$&$\Ga_{1,-1}^{1}$&&
$\Ga_{1,0}^{1}-\Ga_{1,-1}^{2}$&\\[-2.0ex]
\centerline{1}&\centerline{-1}&\centerline{2}&\centerline{0}&
$\Ga_{1,-1}^{2}$&$\Ga_{1,-1}^{2}$&&$\Ga_{1,-1}^{2}$&$[\Ga_{1,-1}^{2}-\Ga_{0,0}^{1}]^+$&&
$\Ga_{1,-1}^{2}$&&$\Ga_{1,-1}^{2}$\\[-2.0ex]
\centerline{1}&\centerline{-1}&\centerline{2}&\centerline{1}&
&&$\Ga_{1,-1}^{2}$&&$\Ga_{1,-1}^{2}\wedge\Ga_{0,0}^{1}$&&&$\Ga_{1,-1}^{2}$&\\[-2.0ex]
\centerline{1}&\centerline{-1}&\centerline{2}&\centerline{2}&
&&&&&$\Ga_{1,-1}^{2}$&&&\\[-2.0ex]
\hline
\centerline{0}&\centerline{1}&\centerline{0}&\centerline{1}&
&&$\Ga_{0,-1}^{1}$&&$\Ga_{0,0}^{1}$&$\Ga_{1,-1}^{1}$&&$\Ga_{1,0}^{1}$&\\[-2.0ex]
\centerline{0}&\centerline{1}&\centerline{0}&\centerline{2}&
&&&&&$\Ga_{1,-1}^{2}$&&&\\[-2.0ex]
\hline
\centerline{1}&\centerline{0}&\centerline{0}&\centerline{1}&
&&$[\Ga_{0,-1}^{1}-\Ga_{1,0}^{1}]^+$&&&$\Ga_{1,-1}^{1}+\Ga_{1,-1}^{2}-\Ga_{1,0}^{1}$&&&\\[-2.0ex]
\centerline{1}&\centerline{0}&\centerline{1}&\centerline{0}&
$\Ga_{1,0}^{1}$&$\Ga_{1,0}^{1}$&$[\Ga_{1,0}^{1}-\Ga_{0,-1}^{1}]^+$&$\Ga_{1,0}^{1}$&
$\Ga_{1,0}^{1}-\Ga_{0,0}^{1}$&&$\Ga_{1,0}^{1}$&&$\Ga_{1,0}^{1}$\\[-2.0ex]
\centerline{1}&\centerline{0}&\centerline{1}&\centerline{1}&
&&$\Ga_{1,0}^{1}\wedge\Ga_{0,-1}^{1}$&&$\Ga_{0,0}^{1}$&$\Ga_{1,0}^{1}-\Ga_{1,-1}^{2}$&&
$\Ga_{1,0}^{1}$&\\[-2.0ex]
\centerline{1}&\centerline{0}&\centerline{1}&\centerline{2}&
&&&&&$\Ga_{1,-1}^{2}$&&&\\[-2.0ex]
\hline
\centerline{1}&\centerline{1}&\centerline{0}&\centerline{1}&
&&$\Ga_{0,-1}^{1}$&&$\Ga_{0,0}^{1}$&$\Ga_{1,-1}^{1}$&&$\Ga_{1,0}^{1}$&\\[-2.0ex]
\centerline{1}&\centerline{1}&\centerline{0}&\centerline{2}&
&&&&&$\Ga_{1,-1}^{2}$&&&\\
\hline
\end{tabular}}
 \caption{non zero coupling rates for the two species exclusion model}\label{table:soscplrts}
\end{sidewaystable}

\end{document}